\documentclass[aps,prc,nofootinbib,twoside,twocolumn]{revtex4-2}
\usepackage{mathpazo}
\usepackage[T1]{fontenc}
\usepackage[latin9]{inputenc}
\usepackage[a4paper]{geometry}
\usepackage{color}
\usepackage{verbatim}
\usepackage{fancybox}
\usepackage{calc}
\usepackage{amsmath}
\usepackage{amssymb}
\usepackage{graphicx}
\PassOptionsToPackage{normalem}{ulem}
\usepackage{ulem}

\makeatletter
\def\@oddhead{\rightmark \hfill Revealing a deep connection between factorization and saturation \hfill \thepage}
\def\@evenhead{\thepage \hfill K. Werner \hfill}
\def\fnum@table{\tablename~{\bf\thetable}}
\def\fnum@figure{\figurename~{\bf\thefigure}}
\def\tablename{\footnotesize{\bf Table}}
\def\figurename{\footnotesize{\bf Figure}}

\usepackage{dcolumn}

\def\citet{\cite}

\makeatother
\begin{document}
\begin{comment}
created from version2b
\end{comment}

\title{Revealing a deep connection between factorization and saturation: New insight
into modeling high-energy proton-proton and nucleus-nucleus scattering
in the EPOS4 framework}

\author{K. Werner}

\affiliation{SUBATECH, Nantes University - IN2P3/CNRS - IMT Atlantique, Nantes, France }

\begin{abstract}
It is known that multiple partonic scatterings in high-energy proton-proton
($pp$) collisions must happen in parallel. However, a rigorous parallel
scattering formalism, taking energy sharing properly into account,
fails to reproduce factorization, which on the other hand
is the basis of almost all $pp$ event generators. 
In addition, binary scaling in nuclear scatterings is badly violated. 
These problems are usually ``solved'' by simply not considering strictly 
parallel scatterings, which is not a solution.
I will report on
new ideas (leading to EPOS4), which allow recovering perfectly factorization,
and also binary scaling in $AA$ collisions, in a rigorous unbiased parallel
scattering formalism. In this new approach, dynamical saturation scales
play a crucial role, and this seems to be the missing piece needed
to reconcile parallel scattering with factorization. From a practical
point of view, one can compute within the EPOS4 framework parton distribution
functions (EPOS PDFs) and use them to compute inclusive $pp$ cross sections.
So, for the first time, one may compute inclusive jet production (for
heavy or light flavors) at very high transverse momentum ($p_{t}$)
and at the same time in the same formalism study flow effects at low
$p_{t}$ in high-multiplicity $pp$ events, making EPOS4 a full-scale
``general purpose event generator''. I discuss applications, essentially
multiplicity dependencies (of particle ratios, mean $p_{t}$, charm
production) which are very strongly affected by the saturation issues
discussed in this paper. \\
\end{abstract}
\maketitle

%%#############################################################################
%%#############################################################################

\section{Some introductory remarks about factorization, parallel scattering,
and energy sharing \label{=======intro-factorization=======}}

%%#############################################################################
%%#############################################################################

Two major discoveries made it possible to reliably compute
cross sections in high-energy proton-proton ($pp$) scattering. There
is first of all the fact that the coupling constant $\alpha_{s}$
of strong interactions becomes weaker with increasing scale, referred
to as ``asymptotic freedom'' \cite{Gross:1973,Politzer:1973}, which
allows the use of perturbation theory to compute parton-parton cross
sections. The other crucial issue is called ``factorization'' \cite{Collins:1989,Ellis:1996},
which amounts to separating short- and long-distance physics at some
``factorization scale'' $\mu$, which allows one to write the inclusive
$pp$ cross section as a convolution of two parton distribution functions
(PDFs) and a (calculable) elementary parton-parton cross section.
The PDFs contain all the long-distance physics, below the scale $\mu$.
Factorization in connection with asymptotic freedom turned out to
be an extremely powerful concept, with numerous important applications.
Extended to collisions of two nuclei, composed of $A$ and $B$ nucleons,
factorization means that the cross section for rare processes is given
as $AB$ times the $pp$ cross section. This is usually referred to as
``binary scaling''.

Factorization is an impressive tool, being very useful when it comes
to studying inclusive particle production, but there are very interesting
cases not falling into this category, like high-multiplicity events
in proton-proton scattering in the TeV energy range, where a very
large number of parton-parton scatterings contribute. Such events
are particularly interesting, since the CMS Collaboration observed
long-range near-side angular correlations for the first time in high-multiplicity
proton-proton collisions \cite{CMS:2010ifv}, which was before considered
to be a strong signal for collectivity in heavy ion collisions. And
studying such high-multiplicity events (and multiplicity dependencies
of observables) goes much beyond the frame covered by factorization.
Here one needs an appropriate tool, able to deal with multiple scatterings.

The most important fact about multiple parton-parton scatterings is that
they must occur in parallel, and not sequentially, as I am going
to justify in the following. It is known that parton-parton scatterings
are preceded by a series of successive parton emissions according
to Dokshitzer-Gribov-Lipatov-Altarelli-Parisi (DGLAP) evolution equations \cite{GribovLipatov:1972,AltarelliParisi:1977,Dokshitzer:1977}.
In particular, the first emitted partons carry a large momentum corresponding
to a large $\gamma$ factor, so they are ``long-lived'' particles.
Correspondingly, the whole reaction takes a long time, which makes
it impossible to have two (or more) successive parton-parton scatterings.
Multiple scattering must therefore happen in parallel. In the case
of nucleus-nucleus scattering, the nucleon-nucleon collisions also
happen in parallel, and this is simply due to the fact that at very
high energies, the ``reaction time'' (the time it takes for the
two nuclei to pass through each other) is much shorter than the particle
formation time. So first all the interactions are realized (instantaneously)
and particle production comes later. One has a ``double parallel
scattering'' scenario: the nucleon-nucleon scatterings happen in
parallel, and for each nucleon-nucleon scattering, the parton-parton
collisions occur in parallel.

In the case of multiple scatterings, energy-momentum conservation is an
important issue. Of course, everybody agrees on that and all event
generators do conserve it. But this has to be seen in the light of
the underlying theory. Multiple scatterings have been incorporated
in an S-matrix approach in the Gribov-Regge (GR) theory a long time
ago \cite{Gribov:1967vfb,Gribov:1968jf,GribovLipatov:1972,Abramovsky:1973fm}.
All the scatterings ``are equal'' in the sense that there is no
sequence, nothing like ``a first scattering'' and a ``second scattering''
and so on. However, as discussed in \cite{bra90,abr92}, there is
an inconsistency: the ``energy-momentum sharing'' is simply not
taken into account. In a strictly parallel scenario, the initial energy-momentum
has to be shared among $n$ parallel scatterings and a projectile
and a target remnant in an unbiased way, in the following referred
to as ``rigorous parallel scattering scenario'', which
amounts to using integrands (for $pp$ scattering) like 
\begin{equation}
\prod_{i=1}^{n+2}f_{i}(p_{i})\times\delta(p_{\mathrm{initial}}-\sum_{i=1}^{n+2}p_{i})\label{delta}
\end{equation}
(with $p$ referring to four-momentum). In the case of $AA$ scattering,
one has products of $\delta$-functions of the type Eq. (\ref{delta}).
I insist on the fact that the theoretical basis (S-matrix theory)
on one hand, and the Monte Carlo realization in event generators on
the other hand, should deal with energy conservation in the same
way, i.e., they should be 100\% compatible with each other; this is also
what I mean by a ``rigorous parallel scattering scenario''. This
is usually not the case, as in some early work of the author \cite{Werner:1993-VENUS},
where the underlying theory has no energy sharing, but the Monte Carlo
realization does, and this is even in recent event generators the
usual method.

Employing the ``rigorous parallel scattering scenario'', one encounters
highly multidimensional integrals that cannot be separated. In \cite{Drescher:2000ha},
nevertheless energy sharing in the sense of Eq. (\ref{delta})
and its generalization to $AA$ scattering
could be implemented, and 
the technical difficulties (using Markov chain techniques) could be handled. I am
not aware of any other attempt in this direction.

\begin{figure}[h]
\centering{}\includegraphics[scale=0.22]
{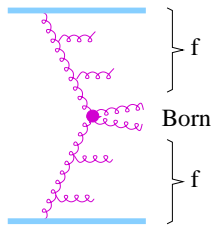}
\caption{Factorization formula. \label{factorization-formula}}
\end{figure}

Let me discuss the fundamental differences between the ``standard
QCD generators'' and the ``rigorous parallel scattering scenario''.
All ``standard QCD generators'' such as PYTHIA \cite{PYTHIA_2008}, HERWIG
\cite{HERWIG_2008}, or SHERPA \cite{SHERPA_2008}, take as starting
point the factorization formula, sketched in Fig. \ref{factorization-formula}.
In this plot and all the following ones, I show for simplicity only
gluons, in the real calculations all kinds of partons are considered.
The two light blue thick lines represent the projectile and the target
protons. The proton structure and the so-called space-like parton
cascade are taken care of by using parton distribution functions (PDFs)
$f$, which allows writing the jet cross section as a convolution
of these PDFs and an elementary QCD cross section for the Born process
in the middle. This formula (still based on Fig. \ref{factorization-formula})
serves as a probability distribution, which allows one to generate a sequence
of hard processes, which are ordered in ``hardness''. This is the
method to introduce multiple parton scattering. 

\begin{figure}[h]
\centering{}\includegraphics[scale=0.22]
{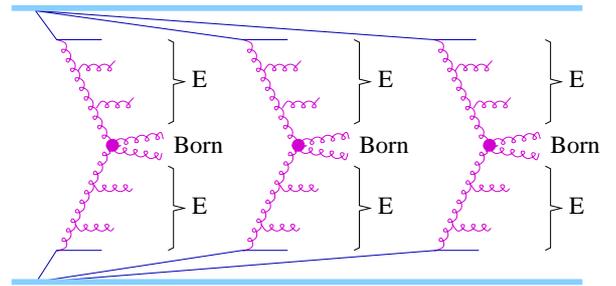}
\caption{Rigorous parallel scattering scenario, for $n=3$ parallel scatterings.
\label{rigorous parallel scattering}}
\end{figure}

In the ``rigorous parallel scattering scenario'', the starting point
is a multiple scattering diagram as shown in Fig. \ref{rigorous parallel scattering}
for $n=3$ scatterings, where the corresponding mathematical formula
contains in the case of $pp$ scattering a $\delta$ function as in Eq. (\ref{delta})
for energy-momentum conservation. Here one also considers parton evolutions
from both sides, but for each of the $n$ interactions, so one cannot
use the usual proton PDFs. Instead, one considers $2n$ evolutions,
starting always from the first perturbative parton on the projectile side
and the target side. But one is nevertheless able to define evolution
functions $E$, which are based on the same DGLAP partial differential
equations (see for example \cite{Collins:1989,Ellis:1996}), but in
our case, the initial condition is not a parton distribution $f(Q_{0}^{2},x)$
in the proton at some initial scale $Q_{0}^{2}$ , but a parton carrying
the full momentum fraction $x=1$. The Monte Carlo procedure to generate
partons will be done in two steps:
\begin{itemize}
\item Step 1: The multi-scattering formalism allows generating a number
of scatterings $n$ and in addition for each of the $n$ scatterings
its energy (expressed in terms of light-cone momentum fractions $x_{i}^{\pm}$),
with 100\% energy-momentum conservation (the cross section formulas
contain a $\delta$ function to assure it, so energy-momentum violating
configurations are never proposed). 
\item Step 2: With $n$ and all the $x_{i}^{\pm}$ known, one generates for
each of the $n$ scatterings the hard process based on a convolution
$E_{\mathrm{proj}}\otimes\mathrm{Born\otimes E_{targ}}$, and then
the parton emissions via backward evolution. 
\end{itemize}
The technical problems in the ``rigorous parallel scattering scenario''
can be handled, but there are conceptual problems. In the classical
GR approach, it is known that in the case of inclusive cross sections,
the multi-scattering contributions cancel, referred to as Abramovsky-Gribov-Kancheli (AGK) cancellations
\cite{Abramovsky:1973fm}, and considering a Pomeron to be a parton
ladder, one may deduce factorization and binary scaling in $AA$, as discussed
in more detail in Sec. \ref{=======factorization-without-energy-conservation=======}. 

In the parallel scattering scenario with energy-momentum sharing,
imposed in an unbiased fashion via a delta function as in Eq. (\ref{delta}),
one does not get factorization (which requires a single Pomeron contribution),
and one violates terribly binary scaling for $AA$ scattering, as I am
going to discuss in Sec. \ref{=======spoils factorization=======}. 

The solution to the problem is related to the treatment of saturation,
as I discuss in a very qualitative fashion in Sec. \ref{=======intro-saturation=======},
and in detail in Sec. \ref{=======saturation-recover-factorization=======}.

\section{Some introductory remarks about saturation and its relation with
energy sharing \label{=======intro-saturation=======}}

The above sketched ``rigorous parallel scattering scenario'' is
an elegant way to introduce unbiased parallel scattering, but in the
end, it does not work, one violates factorization (and binary scaling
in $AA$). So something is still missing. 

\begin{figure}[h]
\centering{}\includegraphics[scale=0.22]
{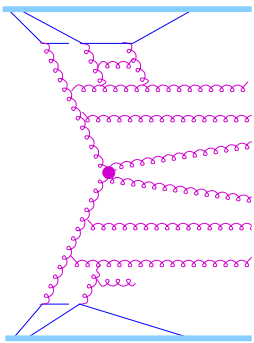}
\caption{Nonlinear effects: ladders which evolve first independently and in
parallel, finally fuse. \label{non-linear-effects}}
\end{figure}

There is actually another important issue in high-energy scattering:
with increasing energy, partons with very small momentum fractions
$x\ll1$ become increasingly important, since the PDFs at small $x$
become large. This means that the parton density becomes large, and
therefore the linear DGLAP evolution scheme is no longer valid, and
nonlinear evolution takes over, considering explicitly gluon-gluon
fusion. These phenomena are known as ``small x physics'' or ``saturation''
\cite{Gribov:1983ivg,McLerran:1993ni,McLerran:1993ka,kov95,kov96,kov97,kov97a,jal97,jal97a,kov98,kra98,jal99a,jal99b,jal99},
the main effect being a screening of low transverse momentum ($p_{t}$)
particle production (below a ``saturation scale''). Saturation effects
are expected to be even stronger in nucleus-nucleus collisions \cite{McLerran:1993ni,McLerran:1993ka},
simply because parton ladders emitted from different nucleons may fuse.
At high energies, the diagrams for each scattering actually look more
like the one shown in Fig. \ref{non-linear-effects}
(I do not consider, for simplicity, time-like parton emissions, but
in the real EPOS4 simulations, they are of course taken care of). 
At least for scatterings carrying a large value of $x^{+}x^{-}$,
one expects ``nonlinear effects'', which means that two ladders which
evolve first independently and in parallel, finally fuse. And only
after that is the (linear) DGLAP evolution realized. 

\begin{figure}[h]
\centering{}\includegraphics[scale=0.22]
{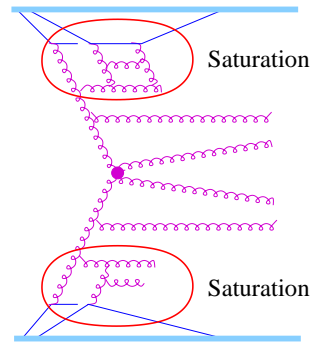}
\caption{Nonlinear effects (inside the red ellipses) are ``summarized'' in
the form of saturation scales. \label{saturation-one-pom}}
\end{figure}

\begin{figure}[h]
\centering{}\includegraphics[scale=0.22]
{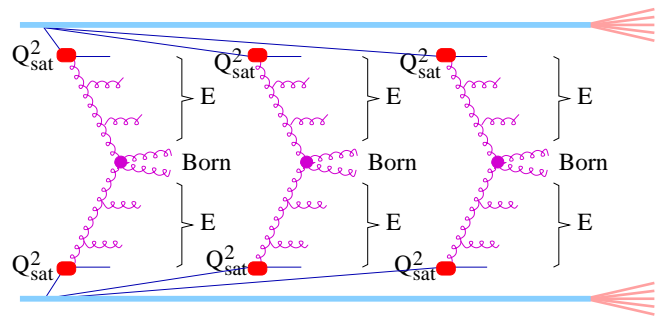}
\caption{Rigorous parallel scattering scenario, for $n=3$ parallel scatterings,
including non-linear effects via saturation scales. The red symbols
should remind one that the parts of the diagram representing nonlinear
effects are replaced by simply using saturation scales. \label{saturation-three-pom}}
\end{figure}

As mentioned above, such non-linear effects lead to strong destructive
interference, which may be summarized in terms of a saturation scale
\cite{McLerran:1993ni,McLerran:1993ka}. This is the motivation to treat
these ``saturation phenomena'' not explicitly, but by introducing
a saturation scale as the lower limit of the virtualities for the DGLAP
evolutions, as sketched in Fig. \ref{saturation-one-pom}.
So the diagrams inside the red ellipses are replaced by two scales
$Q_{\mathrm{sat,proj}}^{2}$ and $Q_{\mathrm{sat,targ}}^{2}$, and
in $pp$ scattering the two are equal. So the final version of the ``rigorous
parallel scattering scenario'' in EPOS4 is sketched in Fig. \ref{saturation-three-pom}.
One
still has DGLAP evolution, for each of the scatterings, but one introduces
saturation scales. 
Most importantly, as discussed in great detail in Sec. \ref{=======saturation-recover-factorization=======},
these scales are not constants, they depend on the number
of scatterings, and they depend as well on $x^{+}$ and $x^{-}$.
A smart choice of these dependencies allows finally to recover factorization
and binary scaling.
\begin{figure}[h]
\centering{}\includegraphics[scale=0.35]
{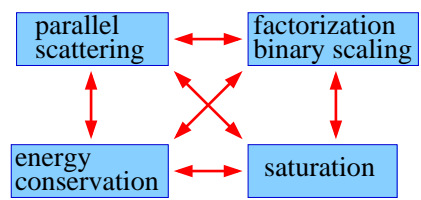}
\caption{Factorization, energy conservation, parallel scattering, and saturation:
four concepts that are deeply connected. \label{four-concepts}}
\end{figure}
One understands that there is a (so far unknown) very strong relation
between factorization, energy conservation (or better energy sharing),
parallel scattering, and saturation, see Fig. \ref{four-concepts}.
Let me summarize the reasoning for this statement: 
\begin{itemize}
\item at high energies, multiple scatterings must happen in parallel, and there
is nothing like a sequence or an ordering of elementary collisions;
\item ignoring energy sharing (as in the GR approach), factorization and
binary scaling are obtained (see Sec. \ref{=======factorization-without-energy-conservation=======});
\item implementing energy sharing in the sense of a ``rigorous parallel
scattering scenario'', not only do the technical difficulties increase
enormously, but there are conceptual problems: it spoils factorization
and binary scaling (see Sec. \ref{=======spoils factorization=======});
\item the only way out (it seems) is that one introduces saturation scales in a
particular way, which recovers factorization and binary scaling (see
Sec. \ref{=======saturation-recover-factorization=======}). 
\end{itemize}
Having solved the ``factorization and binary scaling problem'',
one may consider the ``low $p_{t}$ domain''---like studying high multiplicity
(collective) phenomena (see Sec. \ref{=======multiplicity-dependencies=======})---within 
a framework that has (finally) been proven to be compatible
with the factorization approach, allowing one to do ``high $p_{t}$ physics''
as well (as the generators that are based on factorization). In this sense
EPOS4 is meant to be a ``general purpose event generator''. \\

All this discussion about saturation and factorization is of fundamental
importance since both are considered to be very important issues,
but usually, they are discussed independently. But the message of this
paper is that they are connected, they affect each other (see Sec. \ref{=======saturation-recover-factorization=======}),
and they are just two aspects in a common approach.\\

Since saturation is so important in this approach, what about the other
models? The above-mentioned ``standard QCD generators'' (based on
factorization) do not explicitly deal with saturation (apart of a
constant low virtuality cutoff), but certain features have similar
effects. Let me consider $AA$ scattering in the Pythia/Angantyr model
\cite{Bierlich:2018xfw,Bierlich:2022ned}. As in EPOS4, there is first
the ``basic $AA$ model'' for $t=0$, and in a second step there are string interactions,
happening later. Concerning the basics $AA$ model, the total S-matrix
is given as product of sub-S-matrices. But in contrast to EPOS4, there
are no energy-momentum arguments, and therefore no energy sharing.
But one needs to introduce some kind of ``sequence'', i.e., one loops
over all the $NN$ scatterings, and treats the interaction in two different
ways: if a nucleon is already wounded (having had already a scattering
before) the current scattering is realized as diffractive scattering,
called secondary scattering; otherwise a ``normal'' scattering according
to Pythia happens, called primary scattering. The latter is in general
a multiple parton scattering process. Here, one needs to introduce
some ordering. The first sub-scattering is "normal", whereas
subsequent ones are not connected to projectile/target remnants as
the first one, but they are connected to the parton of the previous
sub-scatterings.

So EPOS4 and Pythia/Angantyr are fundamentally different, but in the
latter there are certain features that have similar effects as the
saturation scale in EPOS:

In Pythia/Angantyr one needs some ordering of $NN$ collisions in $AA$ scattering,
which is needed to distinguish primary and secondary scatterings.
This is necessary to avoid an overproduction of charged particles
in $AA$ collisions. In EPOS4, the same effect is obtained by treating
all $NN$ scatterings equally, but introducing the dynamical saturation
scale. 

Concerning multiple parton scatterings in $NN$ collisions, in Pythia/Angantyr
the first and the subsequent scatterings are treated differently
with regard to the color connections. This is needed to get the experimentally
observed increase of the mean transverse momentum with multiplicity.
In EPOS4, one treats all subscatterings equally, but one has a saturation
scale, which increases with multiplicity (as will be discussed later),
and this is the main mechanism that leads to the increase of the
mean transverse momentum with multiplicity. 

It is of course dangerous to generalize based on few examples, so
let me take the following statement as conjecture. I believe, based
on the work in this paper, compared to other approaches, that one
has two possibilities:
\begin{itemize}
\item either one considers subsequent sub-scatterings (parton-parton or nucleon-nucleon)
as strictly equal, with appropriate energy-sharing, which requires
a dynamical saturation scale as crucial element;
\item or one does not consider saturation (other than a simple cutoff in
the parton cascade), but one needs to distinguish between primary
and secondary scatterings (the first one and subsequent ones), for
both parton-parton and nucleon-nucleon collisions, which requires
some ordering.
\end{itemize}
This paper is meant to be an overview, with a minimum of technical
details. The latter can be found in separate publications, such as
\cite{werner:2023-epos4-heavy,werner:2023-epos4-smatrix,werner:2023-epos4-micro}.\\

After these introductory remarks, I will
\begin{itemize}
\item in Sec. \ref{======= EPOS4-S-matrix-approach =======}
present the EPOS4 formalism,
\item in Sec. \ref{=======factorization-without-energy-conservation=======}
show how factorization appears naturally if energy conservation is
dropped,
\item in Sec. \ref{=======deformation-function=======}
show how energy sharing deforms Pomeron energy distributions,
\item in Sec. \ref{=======spoils factorization=======}
show how deformed Pomeron energy distributions spoil factorization
and binary scaling in case of a ``naive'' Pomeron definition
\item in Sec. \ref{=======saturation-recover-factorization=======}
show how a dynamical saturation scale allows recovering factorization
and binary scaling, 
\item in Secs. \ref{=======EPOS4-factorization-mode =======}
to \ref{=======charmed-hadrons=======}
discuss results affected by saturation.
\end{itemize}
%%####################################################################
%%####################################################################

\section{EPOS4 S-matrix approach to realize parallel scatterings \label{======= EPOS4-S-matrix-approach =======}}

%%####################################################################
%%####################################################################

I first consider $pp$ scattering. An appropriate tool to
implement parallel scatterings is provided by S-matrix theory (see
\cite{Gribov:1967vfb,Gribov:1968jf,GribovLipatov:1972,Abramovsky:1973fm,Drescher:2000ha}),
where multiple parallel scatterings can be implemented in a simple
and transparent fashion. Factorization and binary scaling are not
``assumed'', they must come out. The S-matrix is by definition
the representation $S_{ij}=\left\langle i\right|\hat{S}\left|j\right\rangle $
of the scattering operator $\hat{S}$ using some basis of asymptotic
states, and the corresponding T-matrix is defined via $S_{fi}\!=\!\delta_{fi}+i(2\pi)^{4}\delta(p_{f}-p_{i})T_{fi}$).
Particularly important is the diagonal element, $T_{ii}$, representing
elastic scattering, where the asymptotic state $\left|i\right\rangle $
corresponds to two incoming protons. Assuming purely transverse momentum
transfer, one may Fourier transform $T_{ii}$ with respect to the
transverse momentum exchange $\vec{k}$ and in addition divide by
$2s$ to obtain some function $T(s,b)$, with the Mandelstam variable
$s$ and the impact parameter $b$, in the following simply named
``T-matrix'' $T$.

\begin{figure}[h]
\centering{}\includegraphics[scale=0.21]
{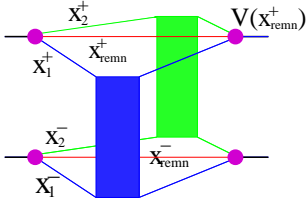}
\caption{Double scattering diagram. \label{two-boxes}}
\end{figure}

The EPOS4 S-matrix approach is based on the hypothesis that the T-matrix
$T$ can be written as a sum of products of ``elementary'' T-matrices
$T_{\mathrm{Pom}}^{(m)}$, the latter ones representing parton-parton
scattering by exchanging a ``Pomeron'' (without specifying its nature
for the moment), as 
\begin{equation}
iT=\sum_{n=1}^{\infty}\int dX\,\frac{1}{n!}\,V^{(+)}\!\times\!\left\{ iT_{\mathrm{Pom}}^{(1)}\times...\times iT_{\mathrm{Pom}}^{(n)}\right\} \!\times\!V^{(-)}\!\times\!\delta,\label{multiple-tmatrix-pp}
\end{equation}

\noindent with the ``vertices'' $V^{(\pm)}$ representing the connection
to the projectile and target remnants. The symbol $X$ stands for
all integration variables, to be specified in the following. In Fig.
\ref{two-boxes},
I show a graphical representation of a double scattering ($n=2$),
where the blue and green boxes are the elementary T-matrices $T_{\mathrm{Pom}}^{(m)}$,
representing parton-parton scattering, and the magenta dots are the
vertices $V^{(\pm)}$. The elementary T-matrices are characterized
by the light-cone momentum fractions $x_{i}^{\pm}$of the incoming
partons, in addition to $s$ and the impact parameter $b$, so one 
has
\begin{equation}
T_{\mathrm{Pom}}^{(i)}=T_{\mathrm{Pom}}(x_{i}^{+},x_{i}^{-},s,b).
\end{equation}
The precise content of the Pomerons (boxes) and the functional dependencies
on these variables will be discussed later, the general discussion
in this section does not depend on these details. The vertices depend
on the light-cone momentum fractions of the remnants, $x_{\mathrm{remn}}^{+}$
(projectile side) or $x_{\mathrm{remn}}^{-}$ (target side), i.e.,
\begin{equation}
V^{(\pm)}=V(x_{\mathrm{remn}}^{\pm}),\;
\end{equation}
with a simple functional form (power law) of $V$. The ``$\delta$''
in Eq. (\ref{multiple-tmatrix-pp}) stands for 
\begin{equation}
\delta(1-\sum_{i=1}^{n}x_{i}^{+}-x_{\mathrm{remn}}^{+})\,\delta(1-\sum_{i=1}^{n}x_{i}^{-}-x_{\mathrm{remn}}^{-}),
\end{equation}
to assure energy-momentum conservation, which will be crucial for
the discussions in this paper. The integration$\int dX$ amounts to
integrating over all light-cone momentum fractions. Each term (for
$n>1$) in the sum of Eq. (\ref{multiple-tmatrix-pp}) represents
multiple scatterings happening in parallel, as it should.

The generalization of the multiple parallel scattering picture towards
nucleus-nucleus ($AA$) collisions (including proton-nucleus) is trivial,
one simply writes a product of $pp$ expressions, 
\begin{align}
iT\!\! & =\!\!\sum_{n_{1}...n_{AB}}\int\!\!dX\,\,\prod_{i=1}^{A}V^{(+i)}\:\times\:\prod_{j=1}^{B}V^{(-j)}\label{multiple-tmatrix-aa}\\
 & \times\prod_{k=1}^{AB}\Bigg\{\frac{1}{n_{k}!}\left\{ iT_{\mathrm{Pom}}^{(k,1)}\times\!\!...\!\!\times iT_{\mathrm{Pom}}^{(k,n_{k})}\right\} \Bigg\}\times\delta\nonumber 
\end{align}
for colliding two nuclei with mass numbers $A$ and $B$, with at
least one $n_{k}>0$. Here, one has one vertex $V^{(m)}$ per remnant,
and a sum of products of elementary T-matrices 
\begin{equation}
T_{\mathrm{Pom}}^{(k,{\nu})}=T_{\mathrm{Pom}}(x_{k{\nu}}^{+},x_{k{\nu}}^{-},s,b_{k})\label{multiple-tmatrix-aa-1}
\end{equation}
per nucleon-nucleon pair $k$. The ``$\delta$'' in Eq. (\ref{multiple-tmatrix-aa})
stands for 
\begin{equation}
\prod_{i=1}^{A}\delta(1-\!\!\!\!\underset{\pi(k)=i}{\sum_{k,{\nu}}}\!\!x_{k{\nu}}^{+}\!-x_{\mathrm{remn},i}^{+})\prod_{j=1}^{B}\delta(1-\!\!\!\!\underset{\tau(k)=j}{\sum_{k,{\nu}}}\!\!x_{k{\nu}}^{-}\!-x_{\mathrm{remn},j}^{-}),\label{multiple-tmatrix-aa-2}
\end{equation}
where $\pi(k)\!=\!i$ amounts to summing Pomerons connected to projectile
$i$ and $\tau(k)\!=\!j$ to summing Pomerons connected to target
$j$. This formula does not mean at all a sequence of $pp$ collisions:
they are perfectly happening in parallel; the crucial ingredient is
the appearance of $\delta$ functions. The integration $\int dX$
here means integration over all light-cone momentum fractions and
over all transverse position of the nucleons. The generalization Eq.
(\ref{multiple-tmatrix-aa}) is conceptually trivial, but it should
be noted that one has (for big nuclei) 10000000 dimensional non-separable
integrals.

\begin{figure}[h]
\centering{}\includegraphics[scale=0.15]
{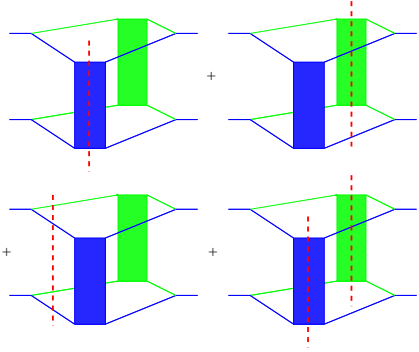}
\caption{Sum of all possible cuts of a two-Pomeron diagram. \label{two-boxes-cut}}
\end{figure}

So far I have discussed only elastic scattering for $pp$ and $AA$, the connection
with inelastic scattering provides the ``optical theorem'' (in $b$ representation),
which is at high energy given as 
\begin{equation}
\sigma_{\mathrm{tot}}=\int d^{2}b\,\mathrm{cut\,}T,\label{sigma-tot}
\end{equation}
with $\mathrm{cut}\,T=\mathrm{\frac{1}{i}disc}\,T$ (cut diagram),
with $\mathrm{disc}\,T$ being the s-channel discontinuity $T(s+i\epsilon)-T(s-i\epsilon)$.
So one needs to compute the ``cut'' of the complete diagram, $\mathrm{cut}\,T$;
for example, for $pp$, one needs to evaluate expressions like
\begin{equation}
\mathrm{cut}\Bigg\{ V^{(+)}\!\times\!\left\{ iT_{\mathrm{Pom}}^{(1)}\times...\times iT_{\mathrm{Pom}}^{(n)}\right\} \!\times\!V^{(-)}\Bigg\}.\label{cut-multi-Pomeron-diagram}
\end{equation}
Cutting a multi-Pomeron diagram corresponds to the sum of all possible
cuts \cite{Cutkosky60}, considering, in particular, all possibilities
of cutting or not any of the parallel Pomerons, so one has finally
sums of products with some fraction of the Pomerons being cut (``cut
Pomerons''), the others not (``uncut Pomerons''). I define $G$
to be the cut of a single Pomeron, 
\begin{equation}
G=G(x_{i}^{+},x_{i}^{-},s,b)=\mathrm{cut}\,T_{\mathrm{Pom}}(x_{i}^{+},x_{i}^{-},s,b).\label{definition-G}
\end{equation}
An uncut Pomeron is by definition the sum of two contributions, the
Pomeron being to the right or to the left of the cut, finally given
as $-G$ (see the discussion in the next section). Since cut and uncut
Pomerons have opposite signs, one gets big sums of positive and negative
terms, with plenty of interference and cancellations. In Fig. \ref{two-boxes-cut},
I show the sum of all possible cuts of a two-Pomeron diagram, not
considering remnants for simplicity. The uncut Pomerons represent
elastic scatterings, they are integrated out. So the expression for
the cut multi-Pomeron diagram Eq. (\ref{cut-multi-Pomeron-diagram})
is finally a product of ``$G$'' terms (and in addition the vertex
terms $V$), so $G$ is the fundamental building block of the approach.

\begin{figure}[h]
\begin{centering}
\hspace*{-1cm}%
\begin{minipage}[c]{0.2\columnwidth}%
Uncut \\
 diagram:%
\end{minipage}%
\noindent\begin{minipage}[c]{0.1\columnwidth}%
\noindent \begin{center}
\includegraphics[scale=0.2]
{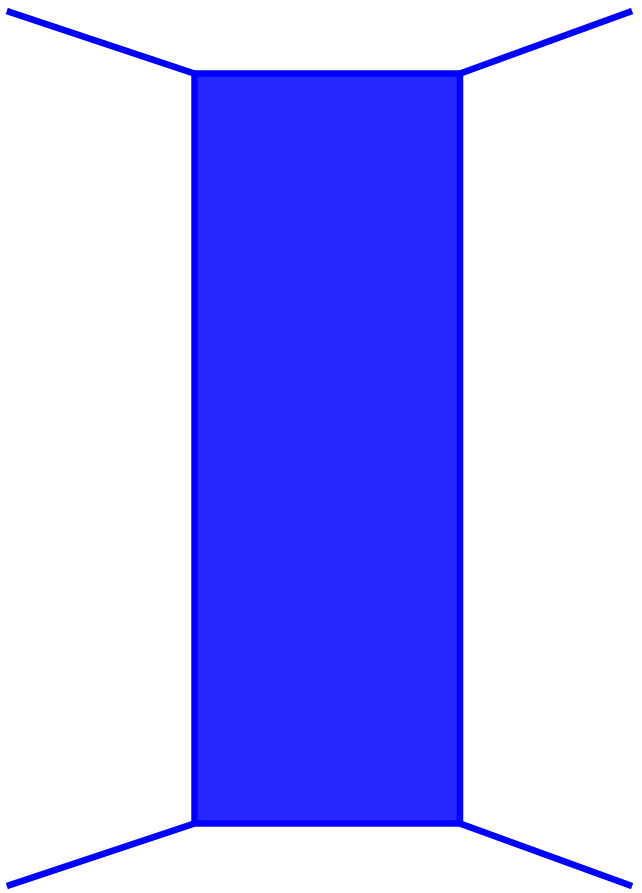} 
\par\end{center}%
\end{minipage}%
\noindent\begin{minipage}[c]{0.1\columnwidth}%
\noindent \begin{center}
\[
=
\]
\par\end{center}
~%
\end{minipage}%
\noindent\begin{minipage}[c]{0.1\columnwidth}%
\noindent \begin{center}
\includegraphics[scale=0.14]
{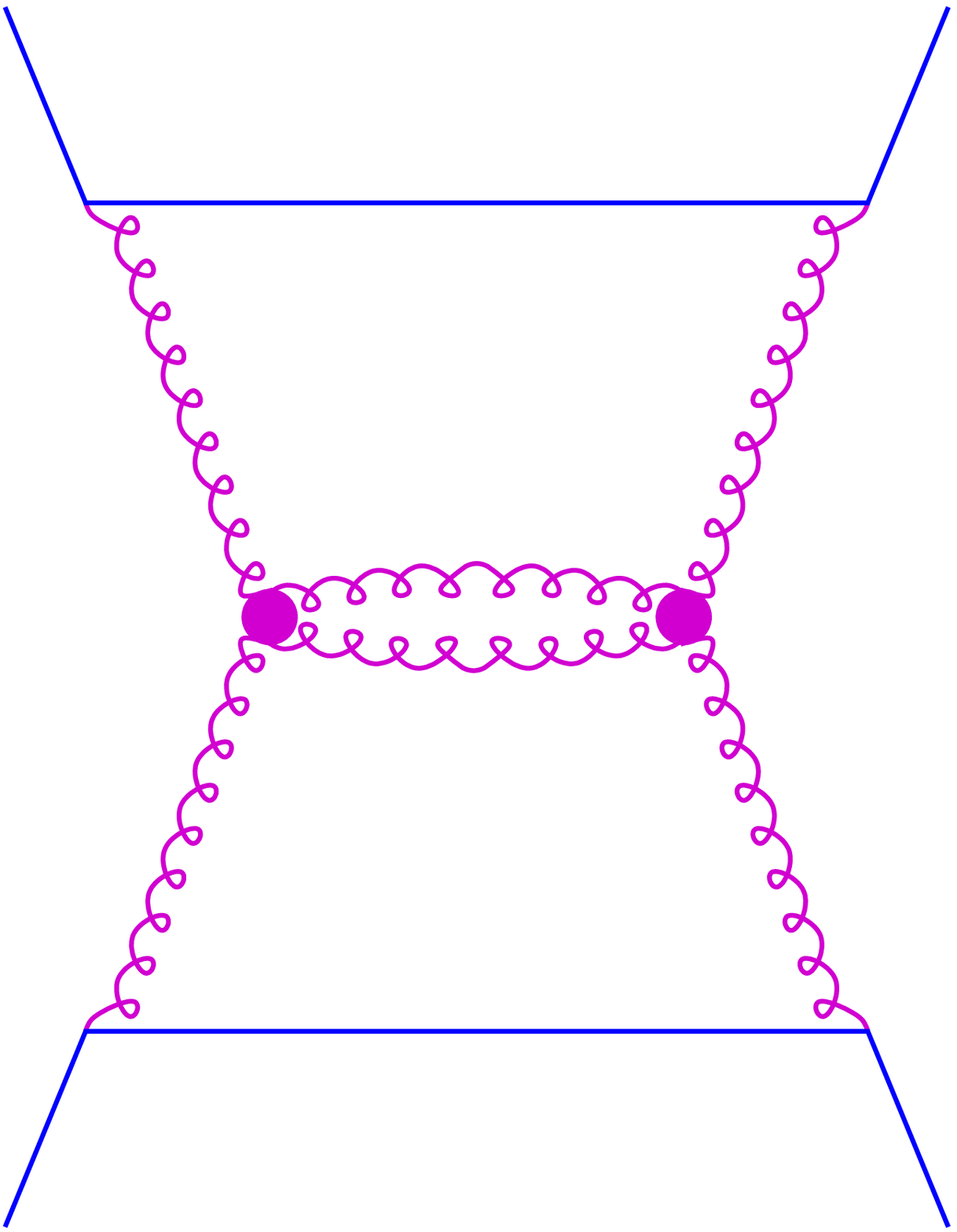} 
\par\end{center}%
\end{minipage}
\par\end{centering}
\noindent \centering{}\hspace*{-1cm}%
\begin{minipage}[c]{0.2\columnwidth}%
Cut \\
 diagram:%
\end{minipage}%
\noindent\begin{minipage}[c]{0.1\columnwidth}%
\noindent \begin{center}
\includegraphics[scale=0.2]
{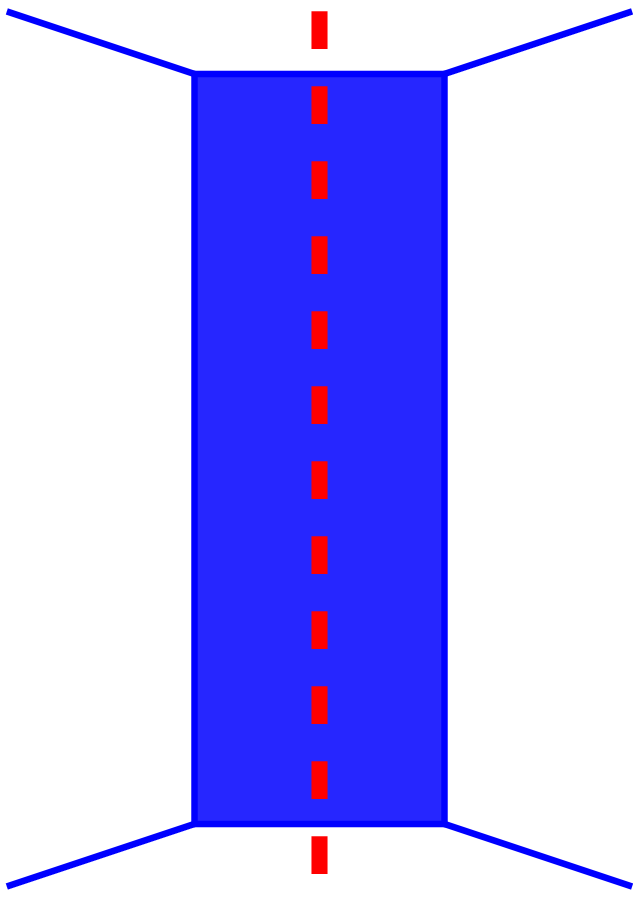} 
\par\end{center}%
\end{minipage}%
\noindent\begin{minipage}[c]{0.1\columnwidth}%
\noindent \begin{center}
\[
=
\]
\par\end{center}
~%
\end{minipage}%
\noindent\begin{minipage}[c]{0.1\columnwidth}%
\noindent \begin{center}
\includegraphics[scale=0.14]
{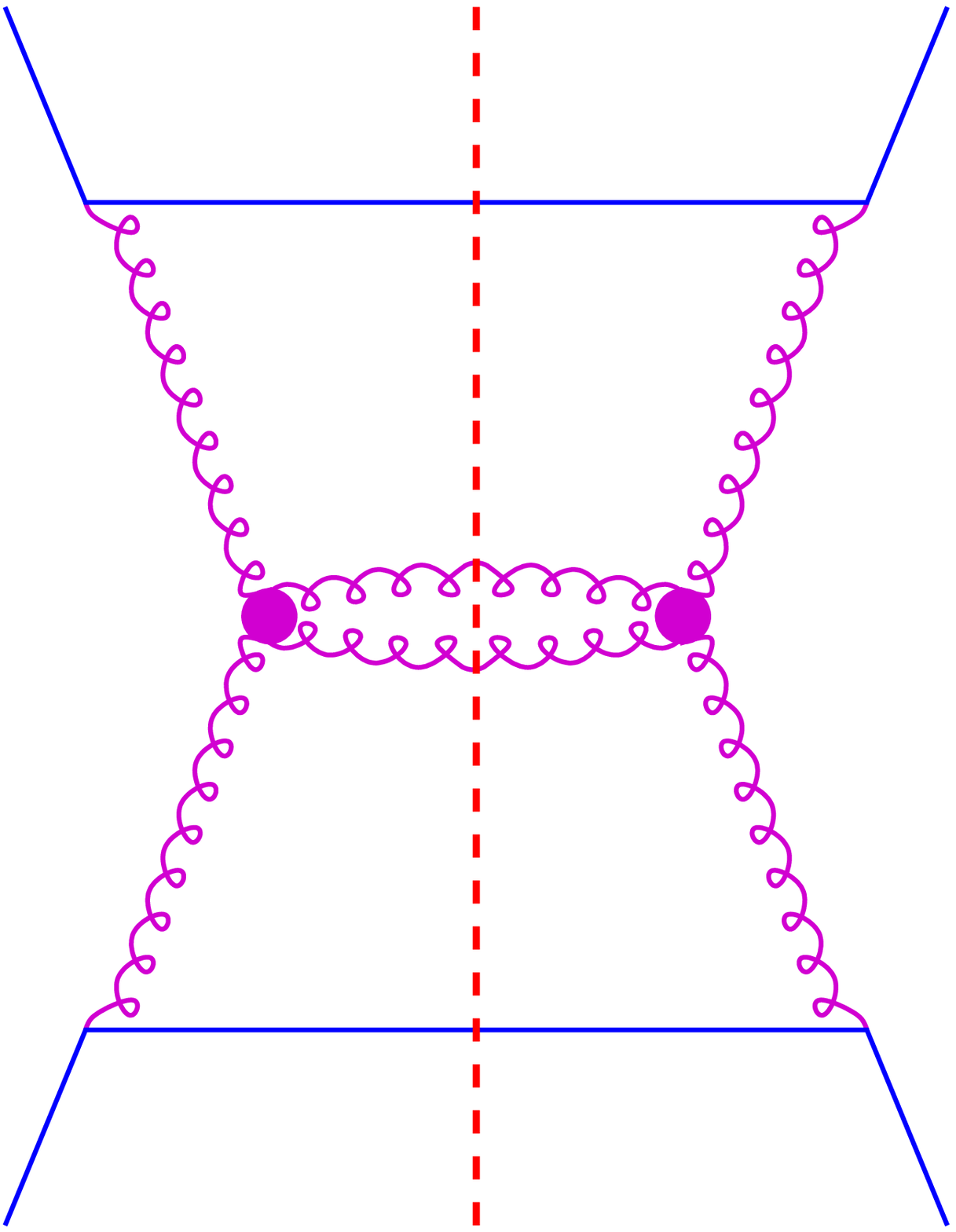} 
\par\end{center}%
\end{minipage}$\qquad$%
\noindent\begin{minipage}[c]{0.1\columnwidth}%
\noindent \begin{center}
\[
=
\]
\par\end{center}
~%
\end{minipage}%
\noindent\begin{minipage}[c]{0.1\columnwidth}%
\noindent \includegraphics[scale=0.14]
{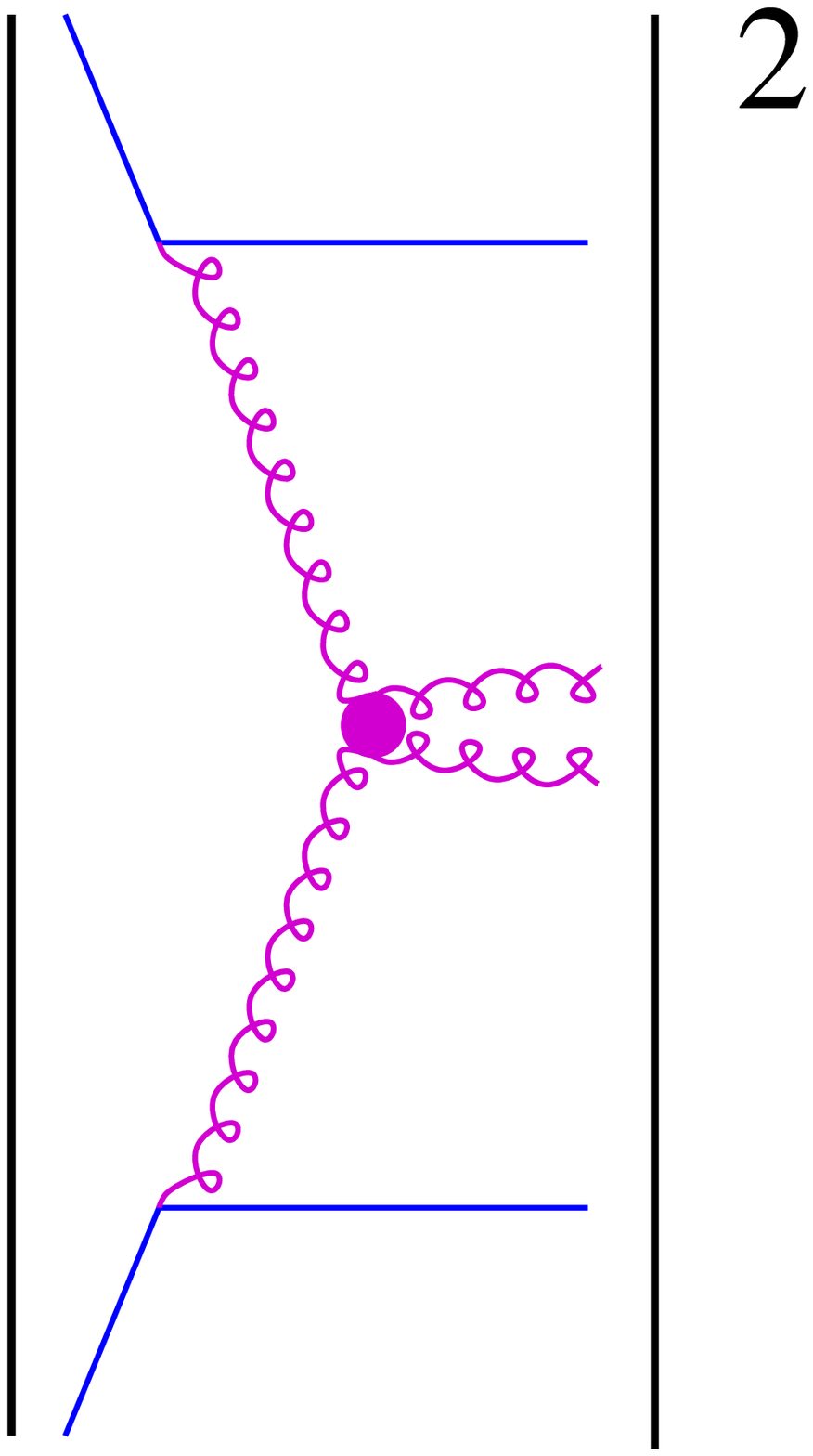}%
\end{minipage}
\caption{A simple example of an uncut and the corresponding cut diagram. \label{simple-example-cut}}
\end{figure}

Let me consider a simple example of a realization of a Pomeron (the
real one is much more complicated), namely a simple parton ladder
with two gluon and two quark ladder rungs, see Fig. \ref{simple-example-cut}.
The cut is represented by a vertical red dashed line. For a cut diagram,
the Feynman rules are modified in the sense that all elements to the
left of the cut are treated normally, for all elements to the right one
takes the complex conjugate of the normal result, and all propagators
crossing the cut line are replaced by a mass shell condition $\delta(p^{2}-m^{2})$.
The cut diagram corresponds therefore to an inelastic amplitude squared,
with all particles on the cut line being final on-shell particles.
This is true not only for this simple example but always. So the notion
of cut diagrams is very useful, in particular for multiple scattering
scenarios.

Concerning nuclear scatterings, the total cross section is still given
by Eq. (\ref{sigma-tot}), together with Eqs. (\ref{multiple-tmatrix-aa}-\ref{multiple-tmatrix-aa-2}),
which gives for a collision of two nuclei with mass numbers $A$ and
$B$:
\begin{align}
 & \sigma_{\mathrm{inel}}=\sum_{n_{1}...n_{AB}}\int\!\!dX\,\,\prod_{k=1}^{AB}\frac{1}{n_{k}!}\label{multiple-tmatrix-aa-3}\\
 & \times\!W_{AB}\!\left(\!\{x_{\mathrm{remn}\,i}^{+}\},\{x_{\mathrm{remn}\,j}^{-}\}\!\right)\!\prod_{k=1}^{AB}\prod_{{\nu}=1}^{n_{k}}G(\!x_{k{\nu}}^{+},x_{k{\nu}}^{-},s,b_{k}\!),\nonumber 
\end{align}
with at least one $n_k$ being nonzero. Here,  $W_{AB}$ contains all the vertices and the integration over
uncut Pomerons, with the symbol $\int dX$ explicitly given as
$\int d^{2}b \int dT_{AB}\int dX_{AB}$, with 
\begin{equation}
\int dT_{AB}=\int\prod_{i=1}^{A}d^{2}b_{i}^{A}\,T_{A}(b_{i}^{A})\prod_{j=1}^{B}d^{2}b_{j}^{B}\,T_{B}(b_{j}^{B}),\label{IdTAB}
\end{equation}
representing the nuclear geometry, with the nuclear thickness functions
$T_{A}(b)$ given as $\int dz\,\rho_{A}(\sqrt{b^{2}+z^{2}})$, with
$\rho_{A}$being the nuclear density, and with 
\begin{equation}
\int dX_{AB}=\int\,\prod_{k=1}^{AB}\prod_{{\nu}=1}^{n_{k}}dx_{k{\nu}}^{+}dx_{k{\nu}}^{-}.\label{IdXAB}
\end{equation}
The impact parameter is defined as $b_{k}=\left|\vec{b}+\vec{b}_{\pi(k)}^{A}-\vec{b}_{\tau(k)}^{B}\right|$,
where $\pi(k)$ and $\tau(k)$ refer to the projectile and the target
nucleons corresponding to pair $k$. There is no ``$\delta"$ term,
since here the remnant momentum fractions are no independent variables,
they are expressed in terms of the momentum fractions $x_{k{\nu}}^{\pm}$
as
\begin{equation}
x_{\mathrm{remn},i}^{+}=1-\!\underset{\pi(k)=i}{\sum_{k,{\nu}}}\!x_{k{\nu}}^{+},\quad x_{\mathrm{remn},j}^{-}=1-\!\underset{\tau(k)=j}{\sum_{k,{\nu}}}\!x_{k{\nu}}^{-}.
\end{equation}
For completeness, and since it is needed in Sec. \ref{=======spoils factorization=======},
let me note that $W_{AB}$ in Eq. (\ref{multiple-tmatrix-aa-3}) can
be written as \cite{werner:2023-epos4-smatrix}
\begin{align}
W_{AB} & =\prod_{i=1}^{A}\,V(x_{\mathrm{remn}\,i}^{+})\,\prod_{j=1}^{B}\,V(x_{\mathrm{remn}\,j}^{-})\label{vertex-function-W}\\
 & \times\prod_{k=1}^{AB}\exp\left(-\tilde{G}(x_{\mathrm{remn}\,\pi(k)}^{+}x_{\mathrm{remn}\,\tau(k)}^{-})\right),\nonumber 
\end{align}
with some known (simple) function $\tilde{G}$.\\

Let me close this section with some technical remarks, concerning the
impact parameter dependence and the energy dependence
of the T-matrices and of the $G$ functions:
\begin{itemize}
\item As discussed in \cite{werner:2023-epos4-heavy}, the (Mandelstam) $t$ dependence
of the original T-matrices is given (in all cases) as factors of the
form $\exp(R^{2}t)$, with parameters $R^{2}$. Considering purely
transverse momentum exchange, one has $t=-k_{\bot}^{2}$, and the
two-dimensional Fourier transform with respect to the transverse momentum
exchange $\vec{k}_{\bot}$ gives a factor $\exp\left(-b^{2}/(4R^{2})\right)$. 
\item In this paper, all the ``$G$'' and ``$T$'' expressions refer
to ``impact parameter representation''; so all $b$ dependencies
are simply Gaussian factors. Here, I do not specify the precise structure
of the Pomeron; this is done in very much detail in \cite{werner:2023-epos4-heavy},
where is shown that the ``real'' Pomerons are convolutions of several
parts (soft pre-evolution, hard part), but the $b$ dependencies are
always Gaussians, giving always a final $b$ dependence of the form
of a factor $\exp\left(-b^{2}/(4R^{2})\right)$. So the $b$ dependencies
are trivial, and easy to handle. In the following, I will
not write the $b$ dependencies explicitly.
\item All T-matrices and $G$ functions depend on $s$, with $s$
referring in all cases to the nucleon-nucleon center-of-mass squared
energy, because using the explicit arguments $x^{+}$ and $x^{-}$,
the transverse mass of a Pomeron is $x^{+}x^{-}s$. In the
following, I will not write this $s$ dependency
explicitly.
\end{itemize}

\section{A simple case: factorization and binary scaling in a scenario without
energy conservation \label{=======factorization-without-energy-conservation=======}}

Before further developing the full EPOS4 S-matrix approach, in order
to understand the real importance of energy conservation (or energy
sharing), I will discuss in this section the S-matrix approach without
energy sharing.

\begin{figure}[h]
\centering{}\includegraphics[scale=0.09]
{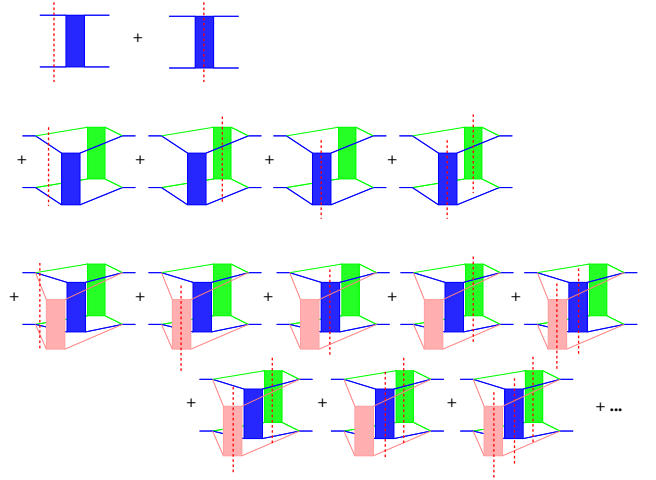}
\caption{All the diagrams which contribute to $\mathrm{cut}\,T$ in $pp$ scattering
up to order $n=3$. Red dashed lines refer to cuts. \label{many-boxes-cut}}
\end{figure}

I consider the general situation, where the precise structure of
Pomeron is not specified. All the diagrams which contribute to $\mathrm{cut}\,T$
(and therefore to the inelastic cross section) in $pp$ represent an
infinite series, composed of all possible cut and uncut Pomerons (boxes)
as shown in Fig. \ref{many-boxes-cut}
up to order $n=3$. However, here energy sharing will be dropped, which
is realized by removing the vertices and the $\delta$ term in Eq.
(\ref{multiple-tmatrix-pp}), so one has
\begin{equation}
iT=\sum_{n=1}^{\infty}\int dX\,\frac{1}{n!}\,\left\{ iT_{\mathrm{Pom}}^{(1)}\times...\times iT_{\mathrm{Pom}}^{(n)}\right\} \,.\label{multiple-tmatrix-pp-1}
\end{equation}
This simplifies things enormously, since with $\int dX=\int dX_{1}...dX_{n}$,
and defining $\tilde{T}_{\mathrm{Pom}}=\int dX_{i}\,T_{\mathrm{Pom}}^{(i)}$
(actually not depending on $i$), one gets 
\begin{equation}
iT=\sum_{n=1}^{\infty}\,\frac{1}{n!}\,\left\{ i\tilde{T}_{\mathrm{Pom}}\right\} ^{n}\,,\label{multiple-tmatrix-pp-1-1}
\end{equation}
which is precisely the expression used in the Gribov-Regge approach.
The sub-T-matrix $\tilde{T}_{\mathrm{Pom}}$ depends only on $s$
and $b$, as does the full S-matrix $T$ (in both cases I do not
write this dependence, for simplicity). Following Eqs. (\ref{sigma-tot}) and (\ref{cut-multi-Pomeron-diagram}),
including the subsequent discussion, and using Eq. (\ref{multiple-tmatrix-pp-1-1}),
one gets for the inelastic cross section 
\begin{equation}
\sigma_{\mathrm{inel}}=\int d^{2}b\,\mathrm{\sum_{n=1}^{\infty}\,\frac{1}{n!}\,\sum_{cuts}}\left\{ i\tilde{T}_{\mathrm{Pom}}\right\} ^{n},
\end{equation}
with at least one cut. One usually assumes the sub-T-matrix to be
purely imaginary, i.e. $\tilde{T}_{\mathrm{Pom}}=i\frac{a}{2}$, with
some real number $a$, and a factor $1\!/\!2$ for convenience. Then
one gets for the cut Pomeron $\mathrm{cut}\,\tilde{T}_{\mathrm{Pom}}=2\,\mathrm{Im}\,\tilde{T}_{\mathrm{Pom}}=a$.
Concerning the uncut Pomerons, one sums up the contributions where the
Pomeron is to the left or to the right of the cut, which gives $\{i\tilde{T}_{\mathrm{Pom}}\}+\{i\tilde{T}_{\mathrm{Pom}}\}^{*}=-a$.
So cut and uncut Pomerons have opposite signs, and one gets
\begin{equation}
\sigma_{\mathrm{inel}}=\int d^{2}b\,\mathrm{\sum_{n=1}^{\infty}\,\frac{1}{n!}\,\sum_{m=1}^{n}\left(\begin{array}{c}
n\\
m
\end{array}\right)}a^{m}(-a)^{n-m},
\end{equation}
where $m$ refers to the number of cut Pomerons.

\begin{figure}[h]
\centering{}\includegraphics[scale=0.09]
{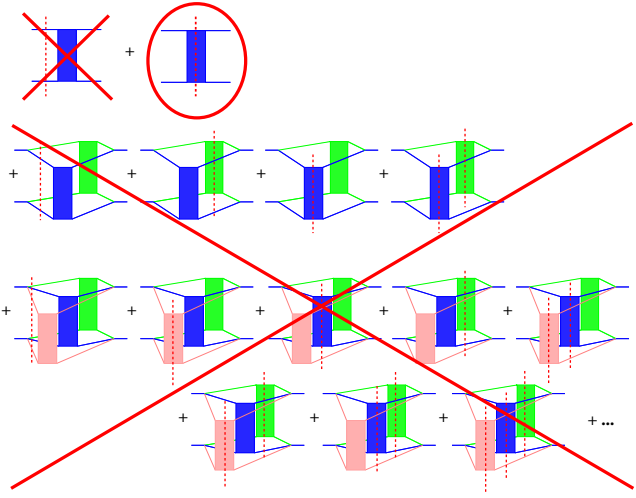}
\caption{Cancellations for inclusive cross sections, when energy-momentum conservation
is ignored. \label{many-boxes-cut-cancelled}}
\end{figure}

Let me consider inclusive cross sections, like jet cross sections,
where $m$-cut-Pomeron events contribute $m$ times more than single
Pomeron events, so one gets 
\begin{equation}
\sigma_{\mathrm{incl}}=\int d^{2}b\,\sum_{n=1}^{\infty}\,\frac{1}{n!}\,\left\{ \mathrm{\sum_{m=1}^{n}m\left(\begin{array}{c}
n\\
m
\end{array}\right)}a^{m}(-a)^{n-m}\right\} ,
\end{equation}
where the term in curly brackets represents the sum over all cuts.
\begin{comment}
\[
\mathrm{\sum_{m=0}^{n}m\left(\begin{array}{c}
n\\
m
\end{array}\right)}(-1)^{n-m}=\mathrm{\sum_{m=0}^{n}m\left(\begin{array}{c}
n\\
m
\end{array}\right)}x^{m}y{}^{n-m}=x\frac{d}{dx}\mathrm{\sum_{m=0}^{n}\left(\begin{array}{c}
n\\
m
\end{array}\right)}x^{m}y{}^{n-m}=x\frac{d}{dx}(x+y)^{n}=xn(x+y)^{n-1}=\left\{ \begin{array}{c}
1\,\,\mathrm{if}\,\,n=1\\
0\,\,\mathrm{if}\,\,n>1
\end{array}\right.
\]
\end{comment}
For a given number $n$ of Pomerons, an elementary calculation allows
to compute the sum over all possible cuts, and one finds an amazing
result:
\begin{equation}
\mathrm{\sum_{m=1}^{n}m\left(\begin{array}{c}
n\\
m
\end{array}\right)}a^{m}(-a)^{n-m}=a^{n}\times\left\{ \begin{array}{c}
1\,\,\mathrm{if}\,\,n=1\\
0\,\,\mathrm{if}\,\,n>1
\end{array}\right.,
\end{equation}
 known as AGK cancellations \cite{Abramovsky:1973fm}:
\begin{itemize}
\item For a given number $n>1$ of Pomerons, the sum of all cuts gives zero,
i.e.,one gets a complete cancellation. 
\item Only $n=1$ contributes, which corresponds
to the case of a single Pomeron.
\end{itemize}
Therefore, for inclusive cross sections and only for those,
only the single Pomeron events contribute, as indicated in Fig. \ref{many-boxes-cut-cancelled}.

\begin{figure}[h]
\centering{}%
\begin{minipage}[c]{0.55\columnwidth}%
(a)\\
\includegraphics[bb=20bp 0bp 360bp 205bp,clip,scale=0.35]
{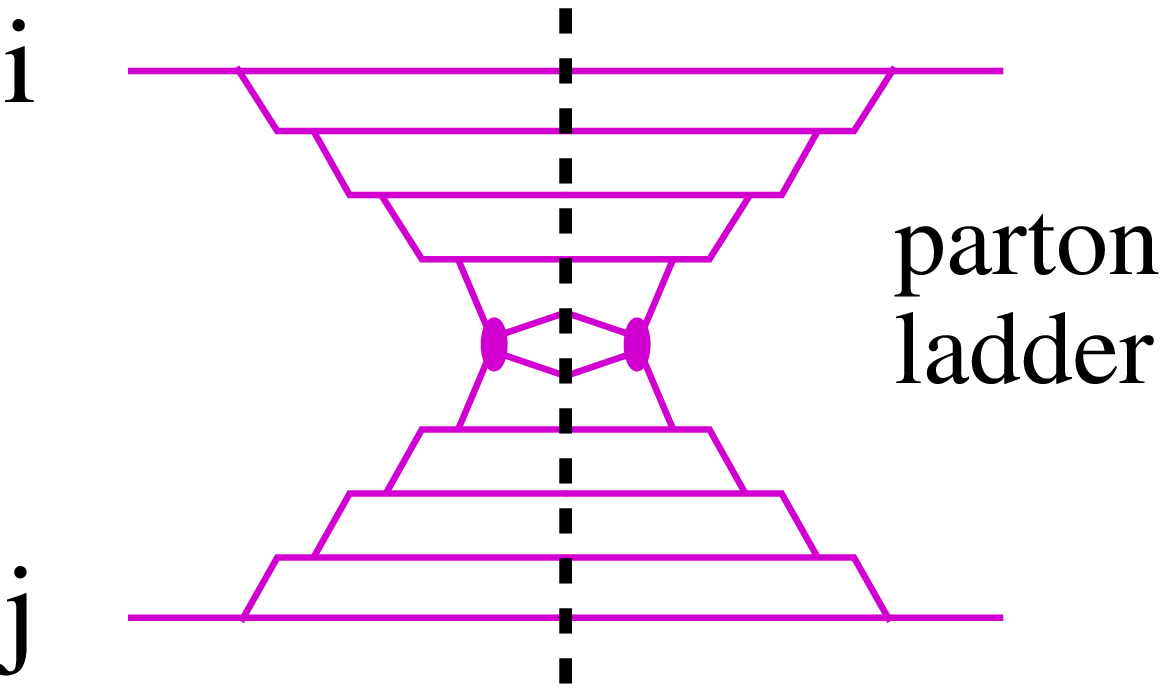}%
\end{minipage}%
\noindent\begin{minipage}[c]{0.21\columnwidth}%
(b)\\
\includegraphics[bb=20bp 0bp 160bp 205bp,clip,scale=0.35]
{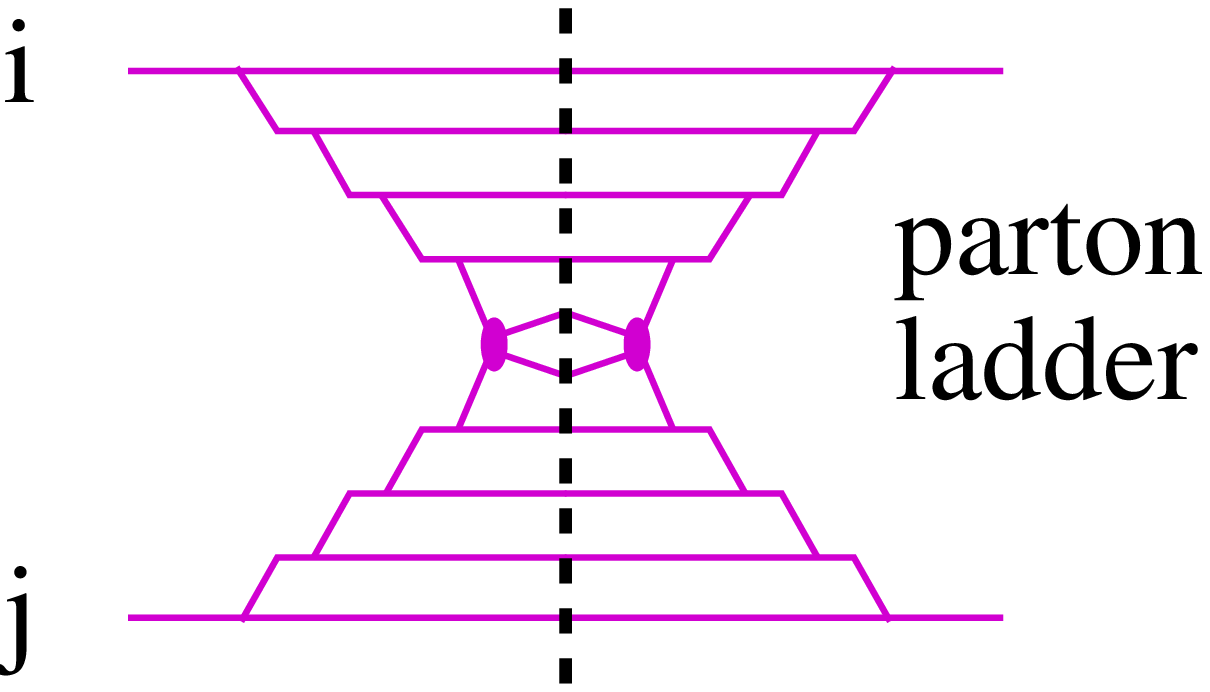}%
\end{minipage}%
\begin{minipage}[c]{0.15\columnwidth}%
~\smallskip{}

PDF\medskip{}

Born\medskip{}

PDF%
\end{minipage}
\caption{A single cut Pomeron (a) and the corresponding inelastic process (b).\label{single-pomeron-graph-1}}
\end{figure}

If a Pomeron is considered to be a parton ladder, a single cut Pomeron
looks like the graph in Fig. \ref{single-pomeron-graph-1}(a),
 with parton evolutions from both sides and a hard elementary parton-parton
scattering in the middle, the corresponding inelastic process is shown
in Fig. \ref{single-pomeron-graph-1}(b). So one can write the inclusive
$pp$ cross section as a convolution of two parton distribution functions
(PDFs) and a (calculable) elementary parton-parton cross section (Born
process),
\begin{equation}
\sigma_{\mathrm{incl}}=\int d^{2}b\,\int dx^{+}dx^{-}f_{\mathrm{PDF}}(x^{+})f_{\mathrm{PDF}}(x^{-})\sigma_{QCD}(x^{+}x^{-}s)\,,
\end{equation}
which amounts to factorization. So factorization here is the result
of a huge amount of cancellations. 

For completeness, also for inclusive cross sections in $AA$ scattering,
one observes this phenomenon of cancellations, such that finally only
a single Pomeron contributes. Colliding two nuclei with mass numbers
$A$ and $B$, the cross section turns out to be $AB$ times the
proton-proton cross section (see for example \cite{Drescher:2000ha}),
which is nothing but ``binary scaling''.\\

To summarize this section: 
\begin{itemize}
\item in a simplified picture, dropping energy conservation, one gets (easily)
factorization for inclusive cross sections in $pp$ scattering 
\item and binary scaling for inclusive cross sections in $AA$ scattering.
\end{itemize}
But dropping energy-momentum conservation is not really an acceptable
solution, in particular since the Monte Carlo procedures eventually
need the implementation of energy conservation, so one risks to introduce
inconsistencies, in the sense that the theoretical basis and the Monte
Carlo realization are not compatible.\\

In EPOS4, I insist on the compatibility of the theoretical basis
and the Monte Carlo realization (this is an evidence, but widely ignored),
so one must include energy conservation in the formulas representing
the theoretical basis (S-matrix theory). In Secs.
\ref{=======deformation-function=======} and
 \ref{=======spoils factorization=======}
I will discuss why energy conservation spoils factorization and binary
scaling, and in Sec. \ref{=======saturation-recover-factorization=======}
I discuss the solution of the problem, which depends strongly on
saturation and which leads to ``generalized AGK cancellations''.

\section{How energy sharing deforms Pomeron energy distributions \label{=======deformation-function=======}}

In this section, I come back to the full EPOS4 S-matrix formalism, including
energy sharing, and try to understand why and how energy sharing
affects Pomeron energy distributions, which will be crucial with respect
to factorization and binary scaling.

\begin{figure}[h]
\centering{}\includegraphics[scale=0.45]
{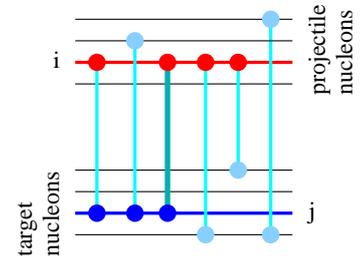}
\caption{A Pomeron connected to projectile nucleon \emph{i} and target nucleon
\emph{j,} together with other Pomerons connected to one (or both)
of these nucleons. \label{nconn-variable}}
\end{figure}

Let me consider a particular Pomeron in $AA$ collisions (including $pp$
as a special case), connected to projectile nucleon \emph{i} and target
nucleon \emph{j}; see Fig. \ref{nconn-variable}.
There might be other Pomerons, connected to one (or both) of these
nucleons. The corresponding Pomeron-nucleon connections are marked
as red and blue dots. It is obvious that the additional Pomerons connected
to the same nucleons $i$ and $j$ compete with each other: they have
to share the initial energy-momentum of the two nucleons. The more
Pomerons are connected, the less energy is available for one particular
Pomeron. 

To quantify this statement, I define the ``connection number''
\begin{equation}
N_{\mathrm{conn}}=\frac{N_{\mathrm{P}}+N_{\mathrm{T}}}{2},
\end{equation}
with $N_{\mathrm{P}}$ being the number of Pomerons connected to \emph{i,
}and with $N_{\mathrm{T}}$ being the number of Pomerons connected
to \emph{j }(the variable $N_{\mathrm{conn}}$ corresponds to half
of the number of red and blue points in Fig. \ref{nconn-variable}).
In the following, I will discuss the effect of energy sharing related
to the connection number. 

As discussed in Sec. \ref{======= EPOS4-S-matrix-approach =======},
the fundamental ``building block'' in EPOS4 is the cut single Pomeron
expression $G=\mathrm{cut}\,T_{\mathrm{Pom}}$. As shown in Eq. (\ref{multiple-tmatrix-aa-3}),
the inelastic cross section for the collision of two nuclei with mass
numbers $A$ and $B$ is given in terms of expressions
\begin{equation}
W_{AB}\!\left(\!\{x_{\mathrm{remn}\,i}^{+}\},\{x_{\mathrm{remn}\,j}^{-}\}\!\right)\!
\prod_{k=1}^{AB}\Bigg[ \frac{1}{n_k!} \prod_{{\nu}=1}^{n_{k}}G(x_{k{\nu}}^{+},x_{k{\nu}}^{-}) \Bigg] ,\label{multiple-G-expression}
\end{equation}
which represent particular configurations $\{n_{k}\}$ characterized
by $n_{k}$ cut Pomerons per nucleon-nucleon pair $k$ (with $k$
between 1 and $AB$). The indices $k,{\nu}$ refer to the ${\nu}{\mathrm{th}}$
Pomeron associated with the pair $k$. Due to energy-momentum conservation,
$1-x_{\mathrm{remn}\,i}^{+}$ is equal to the sum of all $x_{k{\nu}}^{+}$
with $k$ connected to projectile $i$ and $1-x_{\mathrm{remn}\,j}^{-}$
is equal to the sum of all $x_{k{\nu}}^{-}$ with $k$ connected
to target $j$. The expression Eq. (\ref{multiple-G-expression})
represents a multi-dimensional probability
distributions for the light-cone momentum fractions $x_{k{\nu}}^{\pm}$
for a given configuration $\{n_{k}\}$. 

Let me consider, for a given configuration $\{n_{k}\}$, a particular
Pomeron, which means a particular pair index $k$ and a particular
value ${\nu}$, with the associated momentum fractions $x_{k{\nu}}^{\pm}$.
Let $i$ and $j$ be the projectile and target the Pomeron is connected
to (see Fig. \ref{nconn-variable}). 

In the simplest case, one has $n_{k}=1$ (only one Pomeron associated
to pair $k$), and one has no other Pomeron connected to $i$ and
$j$, so one has $N_{\mathrm{conn}}=1$, the case of an isolated Pomeron.
Using the known form of $W_{AB}$ given in Eq. (\ref{vertex-function-W}),
one can see that the integration of Eq. (\ref{multiple-G-expression})
over all variables other than $x_{k{\nu}}^{\pm}$ gives up to a constant
the expression
\begin{equation}
W_{11}(x_{\mathrm{remn}\,i}^{+},x_{\mathrm{remn}\,j}^{-})G(x_{k{\nu}}^{+},x_{k{\nu}}^{-}).\label{single-G-expression}
\end{equation}
This is the probability distribution of $x_{k{\nu}}^{\pm}$ for the
case $N_{\mathrm{conn}}=1$, so I name it $f^{(1)}(x^{+},x^{-})$,
using simply $x^{\pm}$ instead of $x_{k{\nu}}^{\pm}$. Using the
energy-moment conservation relations, one gets for the $N_{\mathrm{conn}}=1$
probability distribution
\begin{equation}
f^{(1)}(x^{+},x^{-})=W'_{11}(1-x^{+},1-x^{-})\,G(x^{+},x^{-}),\label{2-dim-distri}
\end{equation}
with $W'_{AB}=c\,W{}_{AB}$ with a normalization constant $c$. Since
there may be more than one case with $N_{\mathrm{conn}}=1$, one averages
over them, and one averages over all configurations $\{n_{k}\}$. But
this does not change anything since they all have the same form, as
given in Eq. (\ref{2-dim-distri}).

This two-dimensional distribution Eq. (\ref{2-dim-distri}) allows one
to compute the distribution 
\begin{equation}
f^{(1)}(x_{\mathrm{PE}})=\int dy_{\mathrm{PE}}\,J\times f^{(1)}(x^{+},x^{-}),\label{1-dim-distri}
\end{equation}
being the probability distribution with respect to the ``Pomeron
energy fraction'', 
\begin{equation}
x_{\mathrm{PE}}=x^{+}x^{-}=\frac{M_{\mathrm{Pom}}^{2}}{s},\label{definition-p-PE}
\end{equation}
with $M_{\mathrm{Pom}}$ being the transverse mass of the Pomeron,
with $y_{\mathrm{PE}}=0.5\ln\frac{x^{+}}{x^{-}}$, and with $J$ being
the corresponding Jacobian determinant. Having very narrow $y_{PE}$
distributions, one may use $y_{PE}\approx0$, $x^{\pm}=\sqrt{x_{\mathrm{PE}}}e^{\pm y_{\mathrm{PE}}}$
$\approx\sqrt{x_{\mathrm{PE}}}$), and one gets
\begin{equation}
f^{(1)}(x_{\mathrm{PE}})\propto W'_{11}(1-x^{+},1-x^{-})G(x^{+},x^{-}).\label{1-dim-distri-1}
\end{equation}

Let me now consider a more complicated situation, corresponding to
$N_{\mathrm{conn}}>1$. The general formula for the probability distribution $f^{(N_{\mathrm{conn}})}(x^{+},x^{-})$
is given as
%%----------------------------------------------
\begin{align}
 &  f^{(N_{\mathrm{conn}})}(x^{+},x^{-})  = \sum_{k'=1}^{AB}\,\sum_{\{n_{k}\}\ne0}\,\sum_{{\nu'}=1}^{n_{k'}} 
   \delta_{N_{\mathrm{conn}}(k',{\nu'})}^{N_{\mathrm{conn}}}
   \label{sigma-incl-AB-2}\\
 & \qquad\quad \int dX\, \Big\{ P(K)\delta(x^{+}-x_{k'{\nu'}}^{+})\delta(x^{-}-x_{k'{\nu'}}^{-})\Big\},\nonumber 
\end{align}
where $\delta_{a}^{b}$ is the Kronecker delta, and where $P(K)$ is the expression Eq. (\ref{multiple-G-expression}) 
for particular multi-Pomeron configurations $K=\big\{\{n_{k}\},\{x_{k{\nu}}^{\pm}\}\big\}$, 
with given energy-momentum sharing. 
The symbol $\sum_{\{n_{k}\}\ne0}$ means summing over all possible choices of ${n_1,n_2,...,n_{AB}}$, 
excluding the case where all $n_k$ are zero.
The symbol $\int dX$ is explicitly given as $\int d^{2}b \int dT_{AB}\int dX_{AB}$, with 
$\int dT_{AB}$ and $\int dX_{AB}$ defined in Eqs. (\ref{IdTAB}) and (\ref{IdXAB}).
I only consider
Pomerons $(k',{\nu'})$ with connection number $N_{\mathrm{conn}}(k',{\nu'})$
equal to $N_{\mathrm{conn}}$.
%%----------------------------------------------
In principle one does  the same procedure as for the case $N_{\mathrm{conn}}=1$, namely for a given
configuration $\{n_{k}\}$, one chooses a particular pair index $k'$
and a particular value ${\nu'}$, with the associated momentum fractions
$x_{k'{\nu'}}^{\pm}$, which are replaced by $x^\pm$ after integrating over these variables, because of the delta functions.
Then one integrates Eq. (\ref{sigma-incl-AB-2})
over all variables other than $x_{k'{\nu'}}^{\pm}$, which is always
possibe. But in this case, these integration variables and the ``chosen
variables'' $x_{k'{\nu'}}^{\pm}$ (now $x^\pm$) can no longer be separated, and one
needs to do the integration numerically (via Monte Carlo, in practice).
Nevertheless, it is well defined, and
one gets the $x^{\pm}$ distribution $f^{(N_{\mathrm{conn}})}(x^{+},x^{-})$.
As for $N_{\mathrm{conn}}=1$, one integrates over $y_{\mathrm{PE}}$,
to get $f^{(N_{\mathrm{conn}})}(x_{\mathrm{PE}})$. In practice, one
defines event classes (according to multiplicity or impact parameter) and
computes the average $N_{\mathrm{conn}}$ values as well as the average
$x^{\pm}$ distributions per class, to finally get $f^{(\left\langle N_{\mathrm{conn}}\right\rangle )}(x_{\mathrm{PE}})$.
Then one takes the obtained distributions as the basis to compute $f^{(N_{\mathrm{conn}})}(x_{\mathrm{PE}})$
for arbitrary values of $N_{\mathrm{conn}}$ via interpolation. 

The distribution $f^{(N_{\mathrm{conn}})}(x_{\mathrm{PE}})$ or the
two-dimensional distribution $f^{(N_{\mathrm{conn}})}(x^{+},x^{-})$
are kind of ``master'' distributions for all kinds of ``inclusive
distributions'', for example the inclusive $p_{t}$ distribution
of partons, or of hadrons if one adds a fragmentation function. For
computing just inclusive spectra, the knowledge of $f$ is enough,
whereas otherwise the full calculations are needed, using Monte Carlo
methods based on Markov chains. In addition to $f^{(N_{\mathrm{conn}})}(x_{\mathrm{PE}})$
I also check the inclusive distribution of $y_{\mathrm{PE}}$, which
is narrow and strongly peaked at zero (note that the $x^{\pm}$ refer
to Pomeron momenta, not to those of outgoing partons). This is why
I concentrate in the following on $f^{(N_{\mathrm{conn}})}(x_{\mathrm{PE}})$.

Although $f^{(N_{\mathrm{conn}})}(x_{\mathrm{PE}})$ for $N_{\mathrm{conn}}>1$
cannot be calculated analytically, one has some idea of how it should
look like compared to $f^{(1)}(x_{\mathrm{PE}})$: the integrand of Eq. (\ref{sigma-incl-AB-2})
contains in addition to $G(x_{k'{\nu'}}^{+},x_{k'{\nu'}}^{-})$ other
$G$ terms and most importantly $W_{AB}(\{x_{\mathrm{remn}\,i}^{+}\},\{x_{\mathrm{remn}\,j}^{-}\})$,
contains factors of the form $(x_{\mathrm{remn}\,i}^{\pm})^{\alpha}$ with
$\alpha>0$, with arguments 
\begin{equation}
x_{\mathrm{remn}\,i/j}^{\pm}=1-x_{k'{\nu'}}^{\pm}-\sum{\!''}\,x_{k''{\nu''}}^{\pm},
\end{equation}
where $\sum''$ sums over all indices different from $k',{\nu'}$, being
connected to $i$ or $j$. Due to the additional term $\sum''x_{k''{\nu''}}^{\pm}$,
not present for $N_{\mathrm{conn}}=1$, one gets, compared to $N_{\mathrm{conn}}=1$,
a suppression of large values of $x_{k{\nu}}^{\pm}$. This is also
what one expects without any calculation: energy sharing involving more
than one Pomerons leads to a reduction of the energy of the Pomerons,
compared to the case of an isolated Pomeron. 

\begin{figure}[h]
\centering{}\includegraphics[bb=0bp 30bp 595bp 410bp,clip,scale=0.25]
{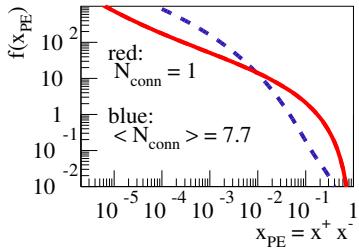}
\caption{Comparing $f^{(N_{\mathrm{conn}})}(x_{\mathrm{PE}})$ for the centrality
class 0-5\% with $f^{(1)}(x_{\mathrm{PE}})$, in PbPb collisions at
5.02 TeV. \label{r-deform-2-1}}
\end{figure}

\begin{figure}[h]
\centering{}\includegraphics[bb=0bp 30bp 595bp 642bp,clip,scale=0.25]
{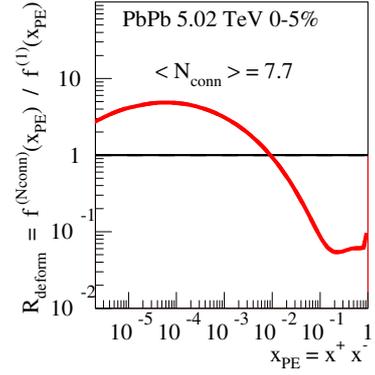}
\caption{Deformation function (see text) for the centrality class 0-5\% in
PbPb collisions at 5.02 TeV. \label{r-deform-2}}
\end{figure}

But let me be quantitative, and discuss the real calculations. In Fig.
\ref{r-deform-2-1},
I plot $f^{(1)}(x_{\mathrm{PE}})$ and $f^{(N_{\mathrm{conn}})}(x_{\mathrm{PE}})$
for the centrality class 0-5\% in PbPb collisions at 5.02 TeV with
an average value of $N_{\mathrm{conn}}$ of roughly 7.7, obtained
after a full EPOS4 simulation. One observes (as expected) for $N_{\mathrm{conn}}>1$
a ``deformation'' of the the $x_{\mathrm{PE}}$ distributions compared
to $f^{(1)}(x_{\mathrm{PE}})$, due to energy-momentum conservation.
Therefore I define the ratio 
\begin{equation}
R_{\mathrm{deform}}=R_{\mathrm{deform}}^{(N_{\mathrm{conn}})}(x_{\mathrm{PE}})=\frac{f^{(N_{\mathrm{conn}})}(x_{\mathrm{PE}})}{f^{(1)}(x_{\mathrm{PE}})},\label{R-deform}
\end{equation}
called the ``deformation function''. In Fig. \ref{r-deform-2},
I show the deformation function for the centrality class 0-5\% in
PbPb collisions at 5.02 TeV. The functional form is as one expects:
large $x_{\mathrm{PE}}$ values (close to unity) are suppressed, and small
$x_{\mathrm{PE}}$ values are increased. And the effect is big: the suppression
in the interval $[0.1,1]$ is roughly 1/20.\\

Let me quickly summarize the main results of this section: 
\begin{itemize}
\item Imposing energy sharing (as it should be) has a very important impact
on the distribution of Pomeron energies.
\item A useful variable to quantify the effect of energy sharing is the
connection number $N_{\mathrm{conn}}$, counting the number of ``other
Pomerons'' connected to the same remnants as a ``given Pomeron''.
$N_{\mathrm{conn}}=1$ represents an isolated Pomeron.
\item I define a variable $x_{\mathrm{PE}}=x^{+}x^{-}$, representing the
squared energy per Pomeron, and the corresponding probability distribution.
\item The probability distribution depends strongly on $N_{\mathrm{conn}}$,
so I use the notation $f^{(N_{\mathrm{conn}})}(x_{\mathrm{PE}})$.
The results for $N_{\mathrm{conn}}>1$ show a suppression at large
$x_{\mathrm{PE}}$, as a consequence of energy sharing. This must be
so, it is unavoidable, a fundamental feature.
\item I therefore define a ``deformation function'' $R_{\mathrm{deform}}$
as the ratio of $f^{(N_{\mathrm{conn}})}(x_{\mathrm{PE}})$ over
$f^{(1)}(x_{\mathrm{PE}})$, which drops below unity for large $x_{\mathrm{PE}}$. 
\end{itemize}

%%###############################################################
%%###############################################################

\section{How deformed Pomeron energy distributions spoil factorization and
binary scaling in case of a ``naive'' Pomeron definition \label{=======spoils factorization=======}}

%%###############################################################
%%###############################################################

In this section, the aim is to understand why and how energy sharing ruins
factorization and binary scaling. I showed in the last section
that energy sharing leads unavoidably to a ``deformation'' of the
Pomeron energy distribution $f^{(N_{\mathrm{conn}})}(x_{\mathrm{PE}})$
compared to the reference $f^{(1)}(x_{\mathrm{PE}})$, with $N_{\mathrm{conn}}$
being the connection number, counting the number of other Pomerons
connected to the same remnants as a given Pomeron, which leads to
the definition of a ``deformation function'' $R_{\mathrm{deform}}$
as the ratio of $f^{(N_{\mathrm{conn}})}(x_{\mathrm{PE}})$ over
$f^{(1)}(x_{\mathrm{PE}})$. In the following, I will show how this
deformation spoils factorization.

\begin{figure}[h]
\centering{}\includegraphics[scale=0.3]
{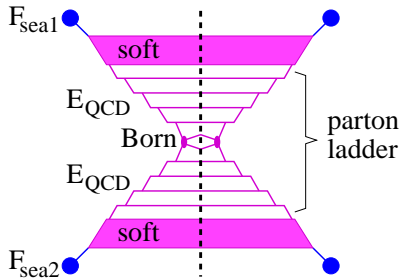}
\caption{The contribution $G_{\mathrm{QCD}}^{\mathrm{sea-sea}}$, which is  the convolution
$E_{\mathrm{soft}}\otimes E_{\mathrm{QCD}}\otimes\mathrm{Born}\otimes E_{\mathrm{QCD}}\otimes E_{\mathrm{soft}}$.
\label{G-sea-sea}}
\end{figure}

Actually ``the problem'' related to factorization and binary scaling
depends very much on the precise definition of $G$ in terms of QCD.
Sofar a Pomeron is a black box with all the QCD details hidden inside,
but now one needs to be more specific. To do so, one introduces in \cite{werner:2023-epos4-heavy}
(with many details and all necessary formulas) a ``QCD expression''
called $G_{\mathrm{QCD}}$ representing a QCD calculation of parton-parton
scattering. Every ``$G$ function'' which one uses, including $G_{\mathrm{QCD}}$,
is meant to be the cut of the Fourier transform of the T-matrix, divided
by $2s$. The term $G_{\mathrm{QCD}}$ is a sum of several contributions,
the most important one being the ``sea-sea'' contribution $G_{\mathrm{QCD}}^{\mathrm{sea-sea}}$;
see Fig. \ref{G-sea-sea}.  For a precise definition see \cite{werner:2023-epos4-heavy}.
The vertices $F_{\mathrm{sea}\,1}^{i}$ and $F_{\mathrm{sea}\,2}^{j}$
couple the parton ladder to the projectile and target nucleons. In
addition, one has three blocks: the two soft blocks and in between
a parton ladder, the latter being a DGLAP parton evolution from both
sides, with a pQCD Born process in the middle.
\begin{comment}
The corresponding diagram is shown in Fig. \ref{val-val-contrbution}.
\begin{figure}[h]
\centering{}\includegraphics[scale=0.22]
{Home/2/komm/conf/99aktuell/epos/ldmultaa3f}
\caption{Diagram associated to $G_{\mathrm{QCD}}$. \label{val-val-contrbution}}
\end{figure}
\end{comment}
I define parton evolution functions $E_{\mathrm{QCD}}$ obeying the
usual DGLAP equations, but in this case the evolution starts from a
parton, not from a nucleon, since a Pomeron corresponds to parton-parton
scattering. I compute and tabulate $E_{\mathrm{QCD}}$, and then
the convolution 
\begin{equation}
E_{\mathrm{soft}}\otimes E_{\mathrm{QCD}}\otimes\mathrm{Born}\otimes E_{\mathrm{QCD}}\otimes E_{\mathrm{soft}},
\end{equation}
with an elementary QCD cross section ``Born'' and a soft preevolution
$E_{\mathrm{soft}}$. In addition to ``sea-sea'', one has more contributions,
named ``val-val'', ``sea-val'', ``val-sea'', ``soft'', and
``psoft'', as discussed in great detail in \cite{werner:2023-epos4-heavy}.

Like $G$, the QCD expression $G_{\mathrm{QCD}}$ depends as well on
$x^{+}$ and on $x^{-}$, and in addition there is the crucial parameter
$Q_{0}^{2}$, which is the low virtuality cutoff in the DGLAP evolution,
so I use the notation $G_{\mathrm{QCD}}(Q_{0}^{2},\,x^{+},x^{-})$.
Whereas the cutoff is usually a constant of the order of $1\,\mathrm{GeV}^{2}$,
I consider it as a variable that may take any value, and I compute
and tabulate $G_{\mathrm{QCD}}(Q^{2},\,x^{+},x^{-})$ for large ranges
of discretized values of all arguments, such that $G_{\mathrm{QCD}}$
can be computed via interpolation for any choice of arguments.
After these preparations, the functional form of $G_{\mathrm{QCD}}(Q^{2},\,x^{+},x^{-})$
is known. Actually, $G$ and $G_{\mathrm{QCD}}$ also depend on $s$ and $b$, not written
explicitly as discussed earlier, so one should always consider ``for
given $s$ and $b$''.

\begin{figure}[h]
\centering{}\includegraphics[scale=0.35]
{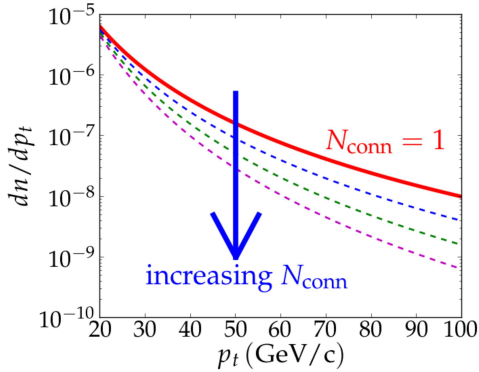}
\caption{Sketch of the suppression of parton yields at high $p_{t}$ with increasing
$N_{\mathrm{conn}}$. \label{facto}}
\end{figure}

What is the relation between $G$ (the Pomeron, the main building
block of the multiple scattering theory) and $G_{\mathrm{QCD}}$ (which
contains all the QCD part)? A first attempt might be (and this is what
was actually used in \cite{Drescher:2000ha}) to consider the two to be
equal, i.e.,
\begin{equation}
G(\,x^{+},x^{-})=G_{\mathrm{QCD}}(Q_{0}^{2},\,x^{+},x^{-}),
\end{equation}
with a constant $Q_{0}^{2}$. Then one gets for the Pomeron energy distribution
for an isolated Pomeron corresponding to $N_{\mathrm{conn}}=1$ [see
Eq. (\ref{1-dim-distri-1})]
\begin{equation}
f^{(1)}(x_{\mathrm{PE}})\propto W'_{11}(1-x^{+},1-x^{-})G_{\mathrm{QCD}}(Q_{0}^{2},x^{+},x^{-}),
\end{equation}
with the Pomeron energy variable $x_{\mathrm{PE}}=x^{+}x^{-}$, and using
$x^{\pm}=\sqrt{x_{\mathrm{PE}}}$). In case of $N_{\mathrm{conn}}>1$,
one has [see Eq. (\ref{R-deform})]
\begin{equation}
f^{(N_{\mathrm{conn}})}(x_{\mathrm{PE}})=R_{\mathrm{deform}}^{(N_{\mathrm{conn}})}(x_{\mathrm{PE}})\times f^{(1)}(x_{\mathrm{PE}}),
\end{equation}
which means that the $x_{\mathrm{PE}}$ distributions will get more
and more ``deformed'', in particular suppressed at large $x_{\mathrm{PE}}$.
This is a general feature and is unavoidable, a direct consequence of energy
sharing. What does this mean concerning transverse momentum ($p_{t}$)
distributions of the outgoing particles from the Born process? Here
one needs to consider the internal structure of $G_{\mathrm{QCD}}$,
first of all $G_{\mathrm{QCD}}^{\mathrm{sea-sea}}$ (actually similar
arguments hold for the other contributions). The important element
is the parton ladder, see Fig. \cite{werner:2023-epos4-heavy}, given
as a convolution 
\begin{equation}
E_{\mathrm{QCD}}\otimes\mathrm{Born}\otimes E_{\mathrm{QCD}}
\end{equation}
(for the formulas, see \cite{werner:2023-epos4-heavy}). The $p_{t}$
of the outgoing partons is related to the factorization scale $\mu_{\mathrm{F}}^{2}$
(one uses $\mu_{\mathrm{F}}^{2}=p_{t}^{2}$), which is the virtuality
of the partons entering the Born process. Large values of $p_{t}$
require large $\mu_{\mathrm{F}}^{2}$ and large squared energy $\hat{s}$
of the Born process, and this requires a large Pomeron squared energy,
and therefore a large value of $x_{\mathrm{PE}}$. The essential points are
\begin{itemize}
\item Large values of $p_{t}$ of the outgoing partons are strongly correlated
with large values of $x_{\mathrm{PE}}$.
\item A suppression of large $x_{\mathrm{PE}}$ values in $f^{(N_{\mathrm{conn}}>1)}(x_{\mathrm{PE}})$
compared to $f^{(1)}(x_{\mathrm{PE}})$ will therefore lead to a suppression
of large $p_{t}$ values in the case of $N_{\mathrm{conn}}>1$ compared
to $N_{\mathrm{conn}}=1$. 
\end{itemize}
In Fig. \ref{facto}, I sketch this situation of a suppression of
parton yields at high $p_{t}$ with increasing $N_{\mathrm{conn}}$.
Let me first discuss the consequences for $pp$ scattering. In order
to get factorization, as discussed in Sec. \ref{=======factorization-without-energy-conservation=======},
one would need something like AGK cancellations, such that the full
multiple scattering scenario is identical to the single Pomeron ($N_{\mathrm{conn}}=1$)
case, which means eventually 
\begin{equation}
\frac{dn^{(pp\,MB)}}{dp_{t}}=\frac{dn^{(N_{\mathrm{conn}}=1)}}{dp_{t}}\label{facto-condition}
\end{equation}
for the minimum bias (MB) inclusive particle production result. The
latter may be written as a superposition of the different contribution
for given values of $N_{\mathrm{conn}}$ [the latter being (for $pp$)
identical to the number of cut Pomerons] with the corresponding weights
$w^{(N_{\mathrm{conn}})}$ :
\begin{equation}
\frac{dn^{(pp\,MB)}}{dp_{t}}=\sum_{N_{\mathrm{conn}}=1}^{\infty}w^{(N_{\mathrm{conn}})}\,\frac{dn^{(N_{\mathrm{conn}})}}{dp_{t}},
\end{equation}
where the contributions $\frac{dn^{(N_{\mathrm{conn}})}}{dp_{t}}$
show with increasing $N_{\mathrm{conn}}$ more and more suppression
at large $p_{t}$(as indicated in Fig. \ref{facto}). The average
Pomeron number at say 7 TeV is around 2, so one has definitely an
important contribution from terms with $N_{\mathrm{conn}}>1$. This
means that also the MB result will be reduced at high $p_{t}$ compared
to $N_{\mathrm{conn}}=1$, and this means one cannot fulfill Eq. (\ref{facto-condition}),
so factorization is not achieved. 

The discussion for the scattering of two nuclei with mass numbers
$A$ and $B$ is similar to the $pp$ case. To assure binary scaling,
one expects
\begin{equation}
\frac{dn^{(AB\,MB)}}{dp_{t}}=AB\times\frac{dn^{(N_{\mathrm{conn}}=1)}}{dp_{t}},\label{facto-condition-1}
\end{equation}
for the minimum bias (MB) inclusive particle yields. For the contributions
for different values of $N_{\mathrm{conn}}$, one has a picture similar to that 
shown in Fig. \ref{facto}, and also here one concludes that the MB
inclusive yield will be reduced at high $p_{t}$ compared to $N_{\mathrm{conn}}=1$,
and therefore one cannot fulfill Eq. (\ref{facto-condition-1}),
and therefore binary scaling is violated.

Let me summarize this section:
\begin{itemize}
\item I consider here the case where the cut Pomeron $G$ is identical
to $G_{\mathrm{QCD}}$, the latter representing a pQCD result for
parton-parton scattering.
\item Considering the internal structure of $G_{\mathrm{QCD}}$, one concludes
that there is a strong correlation between the Pomeron energy variable
$x_{\mathrm{PE}}$ and $p_{t}$ of the outgoing partons (large $p_{t}$
corresponds to large $x_{\mathrm{PE}}$). 
\item Therefore the suppression of large $x_{\mathrm{PE}}$ with increasing
$N_{\mathrm{conn}}$ amounts to a suppression of large $p_{t}$, and
one can conclude a suppression of yields at large $p_{t}$ for minimum
bias results compared to $N_{\mathrm{conn}}=1$.
\item This means one cannot obey Eqs. (\ref{facto-condition}) and (\ref{facto-condition-1}),
which are necessary conditions for factorization ($pp$) and binary scaling
($AA$). 
\end{itemize}

%%###############################################################
%%###############################################################

\section{How saturation allows one to recover factorization and binary scaling
(Generalized AGK cancellations) \label{=======saturation-recover-factorization=======}}

%%###############################################################
%%###############################################################

\begin{figure}[h]
\centering{}%
\begin{minipage}[c]{0.2\columnwidth}%
\noindent \begin{center}
\includegraphics[scale=0.2]
{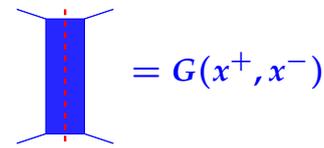} 
\par\end{center}%
\end{minipage}%
\begin{minipage}[c]{0.2\columnwidth}%
\noindent \begin{center}
\textcolor{blue}{\large{}{}{} 
\[
\boldsymbol{\boldsymbol{=G(x^{+},x^{-})}}
\]
}{\large{}{} } 
\par\end{center}%
\end{minipage}$\qquad\qquad$\caption{The cut Pomeron $G$. \label{box-equal-G}}
\end{figure}

In the following, I discuss the ``key issue'' of the EPOS4 approach,
namely the appropriate definition of $G(x^{+},x^{-})$, the cut Pomeron,
represented so far as a ``cut box'' as shown in Fig. \ref{box-equal-G},
and used earlier (see for example Figs. \ref{two-boxes-cut} and \ref{many-boxes-cut}),
to develop the multiple scattering scheme. The latter is actually
completely independent from the precise definition of $G$, which
is very useful, so one can investigate different options concerning
the internal structure of $G$.

I showed in the previous section that the ``naive'' assumption
\begin{equation}
G=G_{\mathrm{QCD}}\label{G=00003DGQCD}
\end{equation}
(which was also adopted in \cite{Drescher:2000ha} and \cite{Pierog_2015-EPOS-LHC})
completely spoils factorization and binary scaling. And from the discussion
in the previous section, it is known that this is a fundamental,
unavoidable problem, and not just a wrong parameter choice. So the assumption
Eq. (\ref{G=00003DGQCD}) seems to be simply wrong. 

There is another serious problem with Eq. (\ref{G=00003DGQCD}): as
discussed somewhat in the previous section (and in detail in \cite{werner:2023-epos4-heavy}),
the essential part of $G_{\mathrm{QCD}}$ is a cut parton ladder,
based on DGLAP parton evolutions. But as already discussed in the
introduction, this is certainly not the full story: with increasing
energy, partons with very small momentum fractions $x\ll1$ become
increasingly important, since the parton density becomes large, and
therefore the linear DGLAP evolution scheme is no longer valid and
nonlinear evolution takes over, considering explicitly gluon-gluon
fusion. These phenomena are known as ``small x physics'' or ``saturation''
\cite{Gribov:1983ivg,McLerran:1993ni,McLerran:1993ka,kov95,kov96,kov97,kov97a,jal97,jal97a,kov98,kra98,jal99a,jal99b,jal99}. 

\begin{figure}[h]
\centering{}\includegraphics[scale=0.22]
{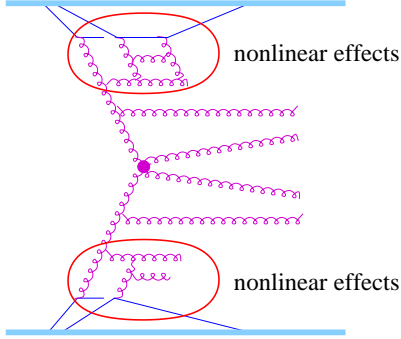}
\caption{Nonlinear effects (inside the red ellipses) also referred to as saturation
effects are ``summarized'' in the form of saturation scales, which
replace these non-linear parts. \label{saturation-one-pom-1}}
\end{figure}

At least for scatterings carrying a large value of $x^{+}x^{-}$,
one expects ``nonlinear effects'', which means that two ladders which
evolve first independently and in parallel, finally fuse. And only
after that is the (linear) DGLAP evolution realized. Such nonlinear
effects lead to strong destructive interference at low transverse
momentum ($p_{t}$), which may be summarized in terms of a saturation
scale \cite{McLerran:1993ni,McLerran:1993ka}. This suggests treating
these ``saturation phenomena'' not explicitly, but by introducing
a saturation scale as the lower limit of the virtualities for the DGLAP
evolutions, as sketched in Fig. \ref{saturation-one-pom-1}. 

So one has two problems:
\begin{itemize}
\item a wrong identity $G=G_{\mathrm{QCD}}$,
\item a missing treatment of saturation.
\end{itemize}
But fortunately, the two problems are connected, and there is an amazingly
simple solution that solves both problems. Instead of the ``naive''
assumption $G=G_{\mathrm{QCD}}$, 
one postulates:

\begin{equation}
G(x^{+},x^{-})=\frac{n}{R_{\mathrm{deform}}^{(N_{\mathrm{conn}})}(x_{\mathrm{PE}})}G_{\mathrm{QCD}}(Q_{\mathrm{sat}}^{2},x^{+},x^{-}),\label{fundamental-epos4-equation}
\end{equation}
with
\begin{equation}
 G \; \mathrm{independent}\; \mathrm{of} \;  N_{\mathrm{conn}}.
\end{equation}
Here, $R_{\mathrm{deform}}^{(N_{\mathrm{conn}})}(x_{\mathrm{PE}})$
is the deformation function discussed in Sec. \ref{=======deformation-function=======},
and $n$ is a constant, not depending on $x_{\mathrm{PE}}$. The independence
of $G$ on $N_{\mathrm{conn}}$ is absolutely crucial (as I will
show later) and to achieve this, one first parametrizes $G$ based
on the (very reasonable) assumption that $G$ has a ``Regge-pole
structure'' as $G\propto\alpha\:x_{\mathrm{PE}}\,^{\beta}$, where
the $s$ and $b$ dependences of $\alpha$ and $\beta$ are parametrized
with the parameters being fixed by comparing simulation results to
elementary experimental data, and then use Eq. (\ref{fundamental-epos4-equation})
to determine $Q_{\mathrm{sat}}^{2}$. In this way, $Q_{\mathrm{sat}}^{2}$
depends on $N_{\mathrm{conn}}$ and on $x^{\pm}$:
\begin{equation}
Q_{\mathrm{sat}}^{2}=Q_{\mathrm{sat}}^{2}(N_{\mathrm{conn}},x^{+},x^{-}),
\end{equation}
which means that this $Q_{\mathrm{sat}}^{2}$, being the low virtuality
cutoff for the DGLAP evolutions in $G_{\mathrm{QCD}}$, is not a constant,
but its value depends on the environment in terms of $N_{\mathrm{conn}}$
and on the energy of the Pomeron. I will refer to this as ``dynamical
saturation scale''.

\begin{figure}[h]
\centering{}\includegraphics[scale=0.35]
{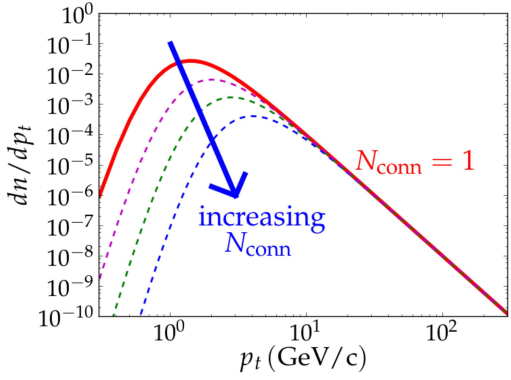}
\caption{Sketch of the suppression of low $p_{t}$ partons with increasing
$N_{\mathrm{conn}}$. \label{facto-1}}
\end{figure}

But why does Eq. (\ref{fundamental-epos4-equation}) work? One gets
for the Pomeron energy distribution for an isolated Pomeron, corresponding
to $N_{\mathrm{conn}}=1$ [see Eq. (\ref{1-dim-distri-1}), using
$x^{\pm}=\sqrt{x_{\mathrm{PE}}}\,$]
\begin{align}
 & f^{(1)}(x_{\mathrm{PE}})\propto W'_{11}(1-x^{+},1-x^{-})\nonumber \\
 & \qquad\times G_{\mathrm{QCD}}\left(Q_{\mathrm{sat}}^{2}(1,x^{+},x^{-}),x^{+},x^{-}\right),
\end{align}
where Eq. (\ref{fundamental-epos4-equation}) with $N_{\mathrm{conn}}=1$
was used to replace $G$. In the case of $N_{\mathrm{conn}}>1$, one has [see Eq.
(\ref{R-deform})]
\begin{align}
 & f^{(N_{\mathrm{conn}})}(x_{\mathrm{PE}})=R_{\mathrm{deform}}^{(N_{\mathrm{conn}})}(x_{\mathrm{PE}})\times f^{(1)}(x_{\mathrm{PE}})\\
 & \propto R_{\mathrm{deform}}^{(N_{\mathrm{conn}})}(x_{\mathrm{PE}})\times W'_{11}(1-x^{+},1-x^{-})G(x^{+},x^{-}).\label{cancel}
\end{align}
Here I will use again Eq. (\ref{fundamental-epos4-equation}) to
replace $G$, but with $N_{\mathrm{conn}}>1$ such that the $R_{\mathrm{deform}}^{(N_{\mathrm{conn}})}$
expressions cancel, and one gets
\begin{align}
 & f^{(N_{\mathrm{conn}})}(x_{\mathrm{PE}})\propto W'_{11}(1-x^{+},1-x^{-})\\
 & \qquad\times G_{\mathrm{QCD}}\left(Q_{\mathrm{sat}}^{2}(N_{\mathrm{conn}},x^{+},x^{-}),x^{+},x^{-}\right)
\end{align}
The crucial point here is the fact that thanks to Eq. (\ref{fundamental-epos4-equation})
and since $G$ does not depend on $N_{\mathrm{conn}}$, the $R_{\mathrm{deform}}^{(N_{\mathrm{conn}})}$
expressions disappear. Comparing now $f^{(N_{\mathrm{conn}})}(x_{\mathrm{PE}})$
and $f^{(1)}(x_{\mathrm{PE}})$, one finds

\begin{equation}
\frac{f^{(N_{\mathrm{conn}})}(x_{\mathrm{PE}})}{f^{(1)}(x_{\mathrm{PE}})}\propto\frac{G_{\mathrm{QCD}}\left(Q_{\mathrm{sat}}^{2}(N_{\mathrm{conn}},x^{+},x^{-}),\:x^{+},x^{-}\right)}{G_{\mathrm{QCD}}\left(Q_{\mathrm{sat}}^{2}(1,x^{+},x^{-}),\:x^{+},x^{-}\right)}\label{ratio-f-over-f}
\end{equation}
This equation is very interesting, it means that the $N_{\mathrm{conn}}$
dependence of $x_{\mathrm{PE}}$ distributions is guided by the saturation
scale, and nothing else. This is the only difference between the numerator
and the denominator. The Eq. (\ref{ratio-f-over-f}) also means that
the partonic structure is given by $G_{\mathrm{QCD}}$, and therefore
also the $p_{t}$ distribution of the outgoing partons is encoded
in the single Pomeron expression $G_{\mathrm{QCD}}$, for
any $N_{\mathrm{conn}}$. Only the saturation
scales $Q_{\mathrm{sat}}^{2}$ depend on $N_{\mathrm{conn}}$,
and these saturation scales suppress small $p_{t}$ particle production,
but do not affect high $p_{t}$ results, as sketched in Fig. \ref{facto-1}.

What does this mean concerning factorization? The minimum bias (MB)
inclusive parton yield may be written as a superposition of the different
contribution for given values of $N_{\mathrm{conn}}$ (for $pp$ identical
to the number of cut Pomerons) with weights $w^{(N_{\mathrm{conn}})}$:
\begin{equation}
\frac{dn^{(pp\,MB)}}{dp_{t}}=\sum_{N_{\mathrm{conn}}=1}^{\infty}w^{(N_{\mathrm{conn}})}\,\frac{dn^{(N_{\mathrm{conn}})}}{dp_{t}}.
\end{equation}
At large $p_{t},$ all contributions are equal, as indicated in Fig.
\ref{facto-1}, so one can replace $\frac{dn^{(N_{\mathrm{conn}})}}{dp_{t}}$
by $\frac{dn^{(1)}}{dp_{t}}$ and one gets
\begin{equation}
\frac{dn^{(pp\,MB)}}{dp_{t}}=\frac{dn^{(N_{\mathrm{conn}}=1)}}{dp_{t}}\;\mathrm{(for\,large\,p_{t})}\,,\label{AGK-pp}
\end{equation}
so only a single Pomeron contributes. This will allow one to define
parton distribution functions $f_{\mathrm{PDF}}$ and to compute cross
sections as convolutions $f_{\mathrm{PDF}}\otimes\mathrm{Born}\otimes f_{\mathrm{PDF}}$. 

The discussion for the scattering of two nuclei with mass numbers
$A$ and $B$ is similar to the $pp$ case, the Pomeron connections do not affect
high $p_{t}$, as shown in Fig.  \ref{facto-1}. The only difference compared to $pp$ scattering is the
fact that one has to sum over all possible nucleon-nucleon pairs,
which gives
\begin{equation}
\frac{dn^{(AB\,MB)}}{dp_{t}}=AB\times\frac{dn^{(N_{\mathrm{conn}}=1)}}{dp_{t}}\;\mathrm{(for\,large\,p_{t})\,,}\label{AGK-AB}
\end{equation}
for the minimum bias (MB) inclusive particle yields, which amounts
to binary scaling.\\

Equations (\ref{AGK-pp}) and (\ref{AGK-AB}) state the following:
\begin{itemize}
\item
The inclusive $pp$ cross section is equal to the one of a single
Pomeron contribution, and the inclusive cross section of the scattering
of two nuclei of mass numbers $A$ and $B$ is equal to $AB$
times the single Pomeron contribution, leading to factorization and
binary scaling. 
\item 
I refer to this as the ``generalized AGK theorem'',
valid at high $p_{t}$, in a scenario with energy sharing. 
\end{itemize}
One recalls that the classical AGK cancellations \cite{Abramovsky:1973fm}
are based on a scenario without energy sharing, as discussed in Sec.
\ref{=======factorization-without-energy-conservation=======}.
\medskip{}

Let me summarize this section:
\begin{itemize}
\item One tries to find the relation between $G$ (the multiple scattering
building block) and $G_{\mathrm{QCD}}$ which represents a QCD result
concerning single parton parton scattering.
\item Two problems are identified: (1) the naive expectation $G=G_{\mathrm{QCD}}$,
having been used so far, does not work, and (2) an appropriate treatment
of saturation is missing. 
\item Both problems are solved by postulating $G(x^{\pm})\propto G_{\mathrm{QCD}}(Q_{\mathrm{sat}}^{2},x^{\pm})/R_{\mathrm{deform}}$,
which means that a saturation scale, depending on the Pomeron connection
number $N_{\mathrm{conn}}$ and on $x^{\pm}$, replaces the virtuality
cutoff $Q_{0}^{2}$ usually used in DGLAP evolutions. In this way
one incorporates saturation.
\item A direct consequence of the above postulate is the fact that the Pomeron
energy distribution $f^{(N_{\mathrm{conn}})}(x_{\mathrm{PE}})$ is
for any value of $N_{\mathrm{conn}}$ given in terms of a single Pomeron
expression $G_{\mathrm{QCD}}$, with an only implicit  $N_{\mathrm{conn}}$ dependence
via $Q_{\mathrm{sat}}^{2}$. 
\item As a consequence, $N_{\mathrm{conn}}$ affects low $p_{t}$ (suppression)
but not high $p_{t}$, and one recovers factorization and binary scaling
(generalized AGK theorem).
\end{itemize}
As a final remark: within a rigorous parallel scattering scenario
(which seems mandatory), and respecting energy conservation
(which seems mandatory as well), the only way to not get in contradiction
with factorization and binary scaling seems to be the consideration
of saturation via $G=k\times G_{\mathrm{QCD}}(Q_{\mathrm{sat}}^{2})$
with $k$ being inversely proportional to the deformation function.
In this sense, parallel scattering, energy conservation, saturation, and
factorization are deeply connected.

%%###############################################################
%%###############################################################

\section{Remarks concerning deformation functions and saturation scales for
given event classes}

%%###############################################################
%%###############################################################

\begin{figure}[h]
\centering{}\includegraphics[bb=0bp 30bp 595bp 642bp,clip,scale=0.25]
{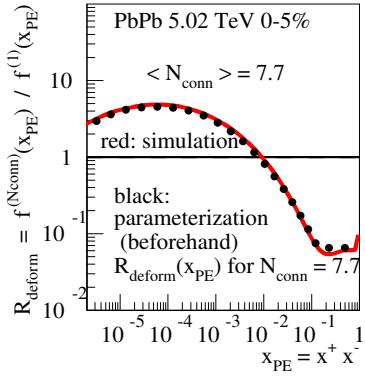}
\caption{Deformation function (see text) for the centrality class 0-5\% in
PbPb collisions at 5.02 TeV. \label{r-deform}}
\end{figure}

Let me come back to the deformation function $R_{\mathrm{deform}}^{(N_{\mathrm{conn}})}(x_{\mathrm{PE}})$,
which plays a fundamental role in the new approach. As explained earlier,
this function can be computed based on Monte Carlo simulations. But
to do so, one first needs to define the Pomerons.
This is done using a parametrization of $G$ in ``Regge pole form'' $\alpha s^{\beta}$,
and based on this, one computes  the deformation functions.
Then one uses  Eq. (\ref{fundamental-epos4-equation}), to do full simulations, and compare with data.
If needed, the initial parametrization of $G$, and as a consequence also the  deformation functions, 
are changed, and one repeats the procedure.
In practice, I have found a very simple functional form for the deformation
functions which accommodates all systems, centrality classes, and
energies \cite{werner:2023-epos4-smatrix}. I determine and tabulate the parameters,
and then use parametrized deformation functions. In Fig. \ref{r-deform},
I show as an example the deformation function for the centrality
class 0-5\% in PbPb collisions at 5.02 TeV, with an average value
of $N_{\mathrm{conn}}$ of roughly 7.7. I plot the parametrized
function (black dots) and the ``computed distribution'' which comes
out after a simulation (red curve). The two curves agree. 

One recalls that in the above iterative procedure, Eq. (\ref{fundamental-epos4-equation})
is used based on a $G$ already known and parametrized in ``Regge
pole form'' $\alpha s^{\beta}.$ As a historical side remark, including
saturation effects via a Regge pole form $\alpha s^{\beta}$ was
introduced first in EPOS1 \cite{Werner:2005-Parton-Lad-Split}, where
the term ``parton ladder splitting'' was used rather than ``saturation'',
but it refers to the same phenomenon. A dynamical saturation scale,
assuming some functional form of $Q_{\mathrm{sat}}^{2}(N_{\mathrm{conn}},x_{\mathrm{PE}})$,
was introduced first in EPOS3 \cite{Werner:2013-param-satur-scale}
for proton-lead scattering, where also for the first time the
expression ``saturation'' was used. Also in \cite{Werner:2013-param-satur-scale},
real simulation results show (Fig. 3) the suppression of parton yields
at high $p_{t}$ in case of the ``naive'' assumption $G=G_{\mathrm{QCD}}$,
and it is shown (Fig. 4) that the suppression can be avoided by introducing
a saturation scale. Finally in \cite{Pierog:2015-parameterized-geff}
it was proposed to use a parametrized $G$ as in \cite{Werner:2005-Parton-Lad-Split}
and use it to determine $Q_{\mathrm{sat}}^{2}(N_{\mathrm{conn}},x_{\mathrm{PE}})$.
However, at the time,  the role of the deformation was not yet understood,
but this is crucial to ensure the correct asymptotic behavior at large $x_{\mathrm{PE}}$ and large $p_t$ .

\begin{figure}[h]
\centering{}\includegraphics[bb=20bp 40bp 700bp 580bp,clip,scale=0.22]
{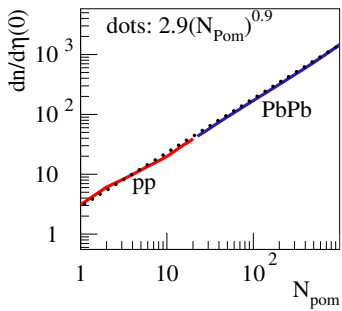}
\caption{The multiplicity $dn/d\eta(0)$ as a function of the Pomeron number
$N_{\mathrm{Pom}}$ for $pp$ (red line) and PbPb (blue line), together
with the dotted line representing the function $2.9(N_{\mathrm{Pom}})^{0.9}$.
\label{mult-vs-npom}}
\end{figure}

As was done for the case of central PbPb in Fig. \ref{r-deform}, one computes
the deformation functions always for event classes and associates
the obtained function to the mean $N_{\mathrm{conn}}$ for the corresponding
class. There are several ways to define event classes; one possibility
is to do it based on the number of cut Pomerons (or simply the Pomeron
number) $N_{\mathrm{Pom}}$, which is related to the multiplicity
$dn/d\eta(0)$. I consider simulations for $pp$ at 7 TeV and PbPb at
2.76 TeV (because I will later come back to these two systems to
compare simulation results with data). In Fig. \ref{mult-vs-npom},
I show the multiplicity $dn/d\eta(0)$ as a function of the Pomeron
number $N_{\mathrm{Pom}}$, for $pp$ at 7 TeV (red line) and PbPb at
2.76 TeV (blue line), together with the dotted line representing the
function $2.9(N_{\mathrm{Pom}})^{0.9}$, which provides a simple conversion
formula between these two quantities. This might be useful when I
analyze later observables as a function of $\left\langle dn/d\eta(0)\right\rangle $.
One gets a continuous curve when going from $pp$ to PbPb. 

\begin{figure}[h]
\centering{}

\hspace*{-0.3cm}\includegraphics[bb=20bp 40bp 642bp 570bp,clip,scale=0.22]
{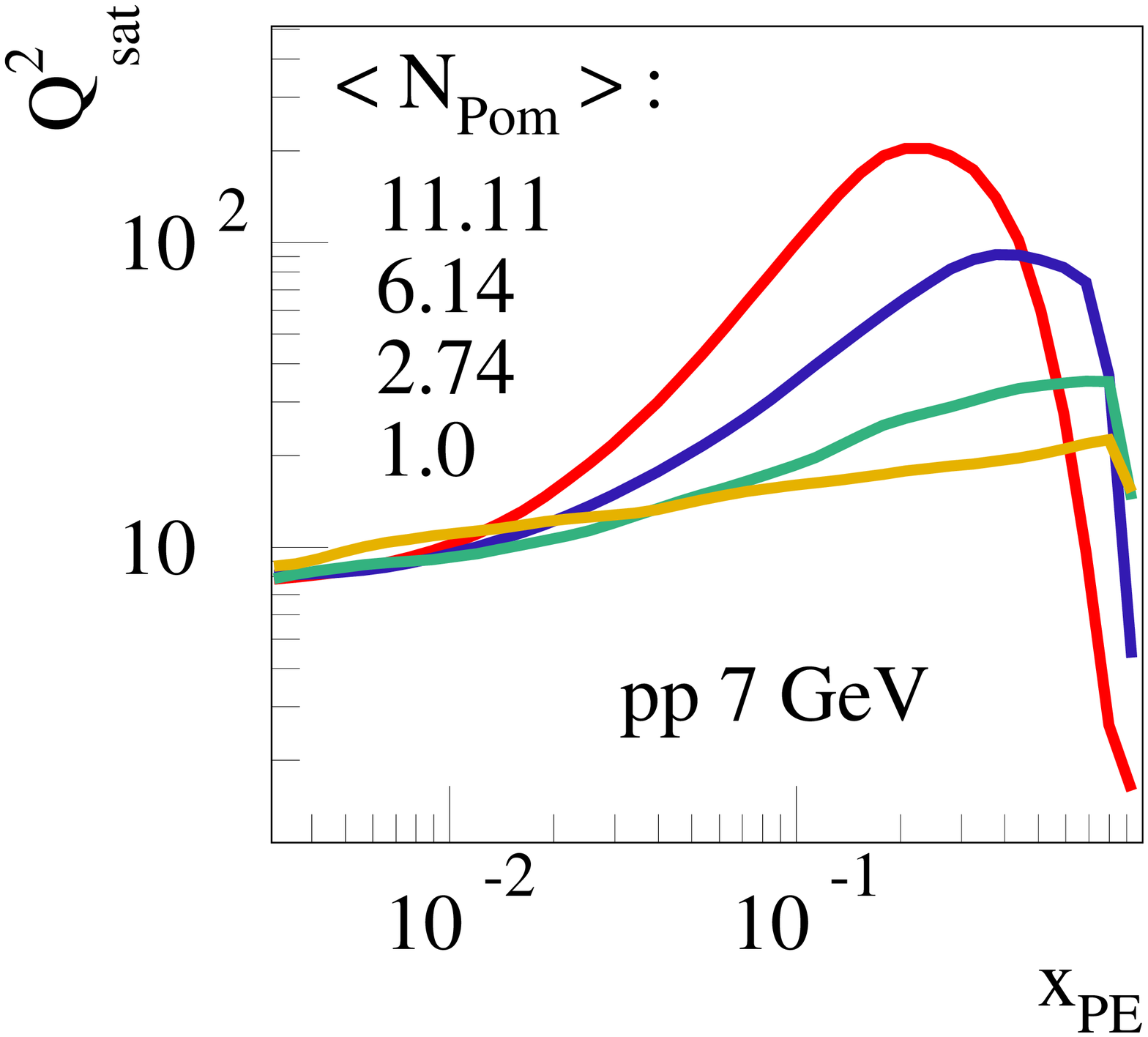}
\hspace*{-0.3cm}\includegraphics[bb=20bp 40bp 642bp 570bp,clip,scale=0.22]
{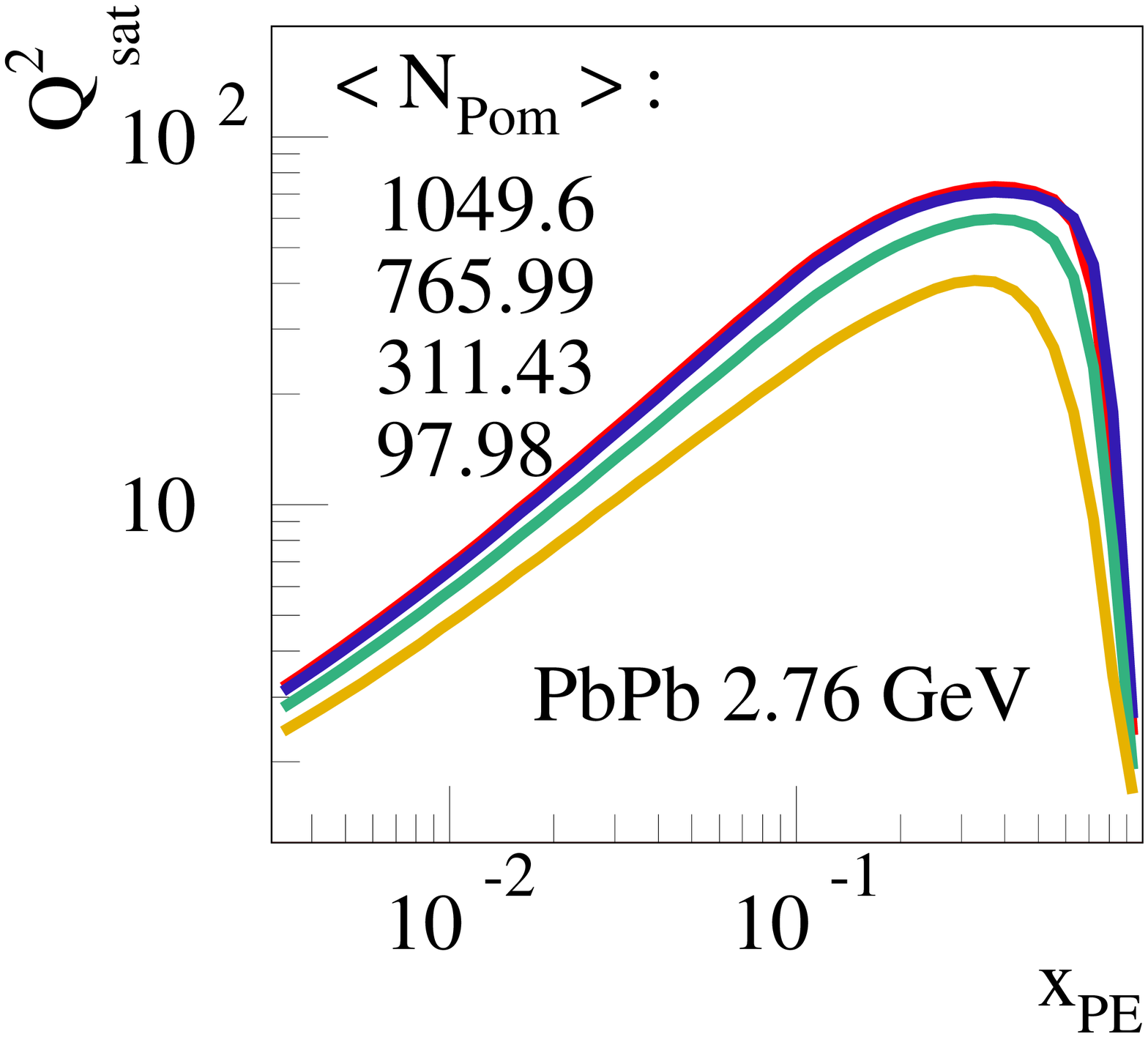}
\caption{The saturation scale $Q_{\mathrm{sat}}^{2}$ as a function of $x_{\mathrm{PE}}$,
for several $N_{\mathrm{Pom}}$ event classes. \label{q2sat-vs-xpe}}
\end{figure}

In Fig. \ref{q2sat-vs-xpe}, I plot the saturation scale $Q_{\mathrm{sat}}^{2}$
as a function of $x_{\mathrm{PE}}$, for several $N_{\mathrm{Pom}}$
event classes. The most striking result is the fact that
 in $pp$ the $Q_{\mathrm{sat}}^{2}$ values change very strongly with
$N_{\mathrm{Pom}}$, whereas for PbPb the variation is quite moderate,
and towards central collisions $Q_{\mathrm{sat}}^{2}$ even ``saturates''
(no variation anymore). This discussion will be important to understand
the results in Sec. \ref{=======multiplicity-dependencies=======}.

\begin{figure}[h]
\centering{}\includegraphics[bb=0bp 30bp 595bp 550bp,clip,scale=0.25]
{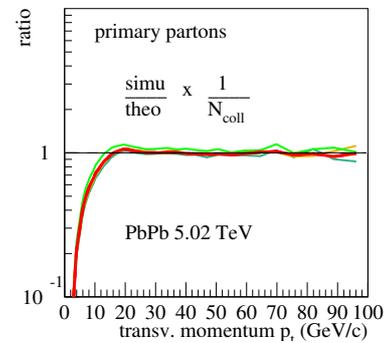}
\caption{The inclusive $p_{t}$ distribution of partons for a full simulation
(simu) divided by $N_{\mathrm{coll}}$ and the ``reference curve''
(theo), which is the corresponding distribution for a single Pomeron,
calculated analytically. I show results for minimum bias PbPb collisions
at 5.02 TeV (red curve) as well as results for different centrality
classes. \label{r-deform-1}}
\end{figure}

As a first result of EPOS4 simulations, I am going to show that
binary scaling really works in practice. In Fig. \ref{r-deform-1},
I show the inclusive $p_{t}$ distribution of partons for a full
simulation (simu) divided by $N_{\mathrm{coll}}$ and the ``reference
curve'' (theo), which is the corresponding distribution for a single
Pomeron, calculated analytically. I show results for minimum bias
PbPb collisions at 5.02 TeV (red curve) as well as results for different
centrality classes. One can see that the ratio is close to one for
large values of $p_{t}$, whereas low $p_{t}$ values are suppressed.
Also in $pp$, the full simulation over the ``reference curve'' (single
Pomeron) is close to unity at large $p_{t}$ and comparing $pp$ and
$AA$, one gets $R_{\mathrm{AA}}\approx1$.

An iterative procedure is employed that relies very much on experimental data: 
One starts with a parametrization of $G$ in ``Regge pole form'', 
already constrained by basic experimental data like energy dependence of cross sections.
Then one computes the deformation functions, which allows finally to determine  saturation scales,
which allows in this way including saturation effects. Based on $G_{\mathrm{QCD}}$,
with a (now) known saturation scale, one can generate partons, and then make very 
detailed comparisons with all kinds of data, and if needed redo the procedure with an improved parametrization of $G$.
So to some extent one has a data-driven method to obtain saturation scales, based on a fully self-consistent pQCD based multiple scattering framework, which is complementary to efforts of computing  saturation scales.

%%###############################################################
%%###############################################################

\section{EPOS4 factorization mode (single Pomeron) and EPOS4 PDFs \label{=======EPOS4-factorization-mode =======}}

%%###############################################################
%%###############################################################

\begin{figure}[h]
\centering{}\includegraphics[scale=0.3]
{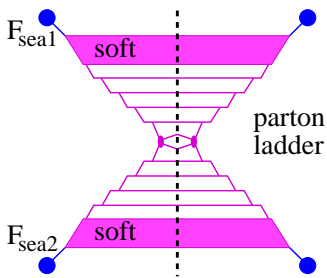}
\caption{Contribution $G_{\mathrm{QCD}}^{\mathrm{sea-sea}}$ to the cut single
Pomeron. \label{single-pomeron-graph}}
\end{figure}

Since in the case of inclusive spectra at large $p_{t}$ everything
can be reduced to the single Pomeron case (generalized AGK cancellations,
see Sec. \ref{=======saturation-recover-factorization=======}),
one may use ``a shortcut'' and compute inclusive particle production
simply by using a single Pomeron, without any need to use complicated
Monte Carlo procedures. This is referred to as ``EPOS4 factorization
mode''. In this case, one simply needs to evaluate the cut single Pomeron,
corresponding to $G_{\mathrm{QCD}}$, which is composed of several contributions;
see \cite{werner:2023-epos4-heavy}. The most important one is $G_{\mathrm{QCD}}^{\mathrm{sea-sea}}$;
see Fig. \ref{single-pomeron-graph}.
It is composed of a parton ladder with parton evolutions ($E_{\mathrm{QCD}}$)
from both sides and an elementary QCD Born process in the middle.
In addition, the QCD parton evolution is preceded
by a soft evolution ($E_{\mathrm{soft}}$). 
The vertices $F_{\mathrm{sea}\,1}$ and $F_{\mathrm{sea}2}$
couple the parton ladder to the projectile and target nucleons.
The complete expression
is a convolution of several elements $F_{\mathrm{sea}\,1}\otimes E_{\mathrm{soft}}\otimes E_{\mathrm{QCD}}\otimes\mathrm{Born}\otimes E_{\mathrm{QCD}}\otimes E_{\mathrm{soft}}\otimes F_{\mathrm{sea}2}$, 
which in addition needs to be convoluted with the vertices $V$ as $V\otimes ... \otimes V$. 
This expression may be regrouped in several ways. One possibility is to convolute
first the vertices, the soft evolution, and the QCD evolution on the
projectile side, representing the parton distribution function (PDF)
of the projectile, and correspondingly on the target side. The two
PDFs represent actually the upper and lower part of the graph in Fig.
\ref{single-pomeron-graph}, plus a vertex $V$, but excluding the Born process. 

So far I only consider the so-called ``sea-sea'' contribution $G_{\mathrm{QCD}}^{\mathrm{sea-sea}}$,
with a sea quark (after a soft evolution) being the first parton entering
the partonic cascade on both sides. But as shown in \cite{werner:2023-epos4-heavy}
there is, in addition, a ``val-val'' contribution, where valence quarks
enter the partonic cascade, and correspondingly ``val-sea'' and
``sea-val'' contributions. Since the parton distribution function
is just half of the Pomeron diagram, there are two contributions,
the ``sea'' and a ``val'' one. 
For a precise definition of the  PDFs, see \cite{werner:2023-epos4-heavy}.

One computes (and tabulates) the PDFs $f_{\mathrm{PDF}}^{k}(x,\mu_{\mathrm{F}}^{2}$),
with $x$ being the light-cone momentum fraction of the parton of
flavor $k$ entering the Born process, and $\mu_{\mathrm{F}}^{2}$
being the factorization scale. After this preparation, one may express
the di-jet cross section (where di-jet simply refers to the two outgoing
partons of the Born process) in terms of the PDFs, as 
\begin{align}
 & E_{3}E_{4}\frac{d^{6}\sigma_{\mathrm{dijet}}}{d^{3}p_{3}d^{3}p_{4}}=\sum_{kl}\int\!\!\int\!\!dx_{1}dx_{2}\,f_{\mathrm{PDF}}^{k}(x_{1},\mu_{\mathrm{F}}^{2})f_{\mathrm{PDF}}^{l}(x_{2},\mu_{\mathrm{F}}^{2})\nonumber \\
 & \qquad\qquad\qquad\frac{1}{32s\pi^{2}}\bar{\sum}|\mathcal{M}^{kl\to mn}|^{2}\delta^{4}(p_{1}+p_{2}-p_{3}-p_{4}),\label{differential-cross-section}
\end{align}
with $p_{1/2}$ and $p_{3/4}$ being the four-momenta of the incoming
and outgoing partons, and $\mathcal{M}^{kl\to mn}$ being the corresponding
matrix element. In order to get the complete expression corresponding
to Fig. \ref{single-pomeron-graph}, one needs to integrate $\int\frac{d^{3}p_{3}d^{3}p_{4}}{E_{3}E_{4}}$
over the differential cross section Eq. (\ref{differential-cross-section}),
whereas to obtain the inclusive jet (=parton) cross section one needs
to integrate $\frac{d^{3}p_{4}}{E_{4}}$. In any case, thanks to the
four-dimensional $\delta$ function, the remaining numerical integration
can be done, as discussed in detail in \cite{werner:2023-epos4-heavy}.
\begin{figure}[h]
\centering{}\includegraphics[bb=25bp 70bp 565bp 780bp,clip,scale=0.45]
{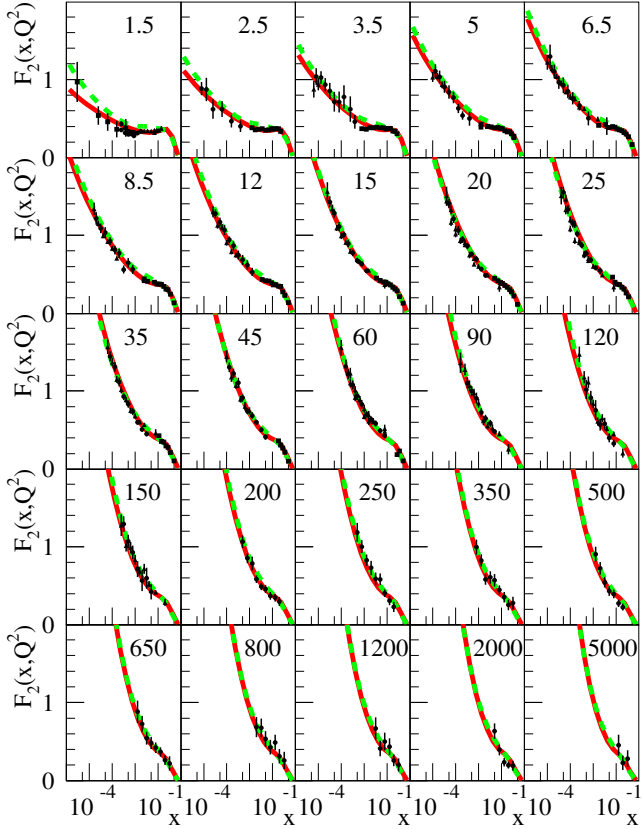}
\caption{$F_{2}$ as a function of $x$ for different values of $Q^{2}$, the
latter one indicated (in units of $\mathrm{GeV}^{2}$) in the upper
right corners of each subplot. The red curve refers to EPOS PDFs,
the green one to CTEQ PDFs, and the black points are data from ZEUS
and H1. \label{f2-epos}}
\end{figure}

At least the quark parton distribution functions can be tested and
compared with experimental data from deep inelastic electron-proton
scattering. The structure function $F_{2}$ is given as $F_{2}=\sum_{k}e_{k}^{2}\,x\,f_{\mathrm{PDF}}^{k}(x,Q^{2})$
with $x=x_{B}=Q^{2}/(2pq)$, with $p$ being the momentum of the proton,
$q$ the momentum of the exchanged photon, and $Q^{2}=-q^{2}$. In
Fig. \ref{f2-epos}, I plot $F_{2}$ as a function of $x$ for different
values of $Q^{2}$. The red curve refers to EPOS PDFs, the green one
to CTEQ PDFs \cite{Dulat_2016-CTEQ-PDF}, and the black points are
data from ZEUS \cite{ZEUS96} and H1 \cite{H1-94,H1-96a,H1-96b}.
The two PDFs give very similar results, and both are close to the
experimental data.

\begin{figure}[h]
\centering{}\includegraphics[bb=20bp 20bp 565bp 780bp,clip,scale=0.39]
{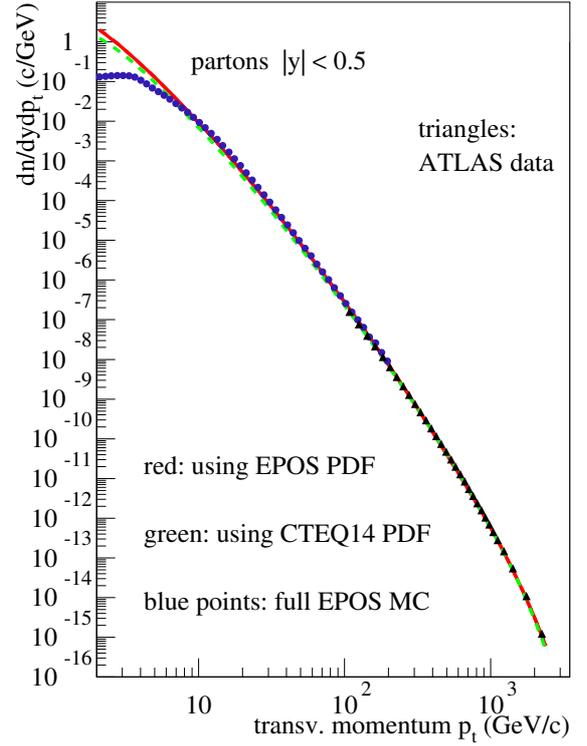}
\caption{Parton yield $dn/dp_{t}dy$ for $pp$ at 13 TeV. I show results based
on EPOS PDFs (red full line), CTEQ PDFs (green dashed line), the full
EPOS simulation (blue circles), and experimental data from ATLAS (black
triangles). \label{parton-pt-spectra}}
\end{figure}

Having checked the EPOS PDFs, I will use these functions to compute
the jet (parton) cross section, using Eq. (\ref{differential-cross-section}),
integrating out the momentum of the second parton and the azimuthal
angle of the first parton, for $pp$ at 13 TeV. I define the parton
yield $dn/dp_{t}dy$ as the cross section, divided by the inelastic
$pp$ cross section, showing the result in Fig. \ref{parton-pt-spectra}.
I show results based on EPOS PDFs (red full line), CTEQ PDFs \cite{Dulat_2016-CTEQ-PDF}
(green dashed line), the full EPOS simulation (blue circles), and
experimental data from ATLAS \cite{ATLAS:2017ble} (black triangles).
At large values of $p_{t}$, all the different distribution agree,
whereas at low $p_{t}$ the EPOS Monte Carlo simulation results (using
the full multiple scattering scenario) are significantly below the
PDF results, as expected due to screening effects.

%%###################################################################################
%%###################################################################################

\section{Full EPOS4 (core+corona, hydro, microcanonical decay): checking multiplicity
dependencies \label{=======multiplicity-dependencies=======}}

%%###################################################################################
%%###################################################################################

\noindent The ``factorization mode'' as discussed in the last section
is very useful to investigate inclusive cross sections at high $p_{t}$.
But this represents only a very small fraction of all possible applications,
and there are very interesting cases outside the applicability of
that approach. A prominent example, one of the highlights of the past
decade, concerns ``collective phenomena in small systems'',
following many discoveries showing that high-multiplicity $pp$ events
show very similar ``collective'' features as earlier observed in
heavy ion collisions \cite{CMS:2010ifv}. High multiplicity means
automatically ``multiple parton scattering'', and as discussed earlier,
this means that one has to employ the full parallel scattering machinery
developed earlier, based on S-matrix theory.

But this is not the full story. The S-matrix part concerns ``primary
scatterings'', happening instantaneously at $t=0$. As a result,
in the case of a large number of Pomerons, one has correspondingly
a large number of strings, which may overlap and ``fuse''.
In the EPOS framework, a core-corona procedure \cite{Werner:2007bf,Werner:2010aa,Werner:2013tya} is employed,
where the strings at a given proper time $\tau_{0}$ are first cut
into ``string segments'', which are separated into ``core'' and ``corona''
segments, depending on the energy loss of each segment when traversing
the ``matter'' composed of all the other segments. Corona segments
(per definition) can escape, whereas core segments lose all their
energy and constitute what is called the ``core'', which acts as an initial
condition for a hydrodynamic evolution \cite{Werner:2013tya}. The
evolution of the core ends whenever the energy density falls below
some critical value $\epsilon_{\mathrm{FO}}$, which marks the point
where the fluid ``decays'' into hadrons. It is not a switch from
fluid to particles; it is a sudden decay, called ``hadronization''.
\\

In EPOS4, as discussed in detail in \cite{werner:2023-epos4-micro},
a new procedure was developed concerning the energy-momentum flow through the ``freeze-out
(FO) hypersurface'' defined by $\epsilon_{\mathrm{FO}}$, which allows
defining an effective invariant mass, which decays according to microcanonical
phase space into hadrons, which are then Lorentz boosted according
to the flow velocities computed at the FO hypersurface. Also 
new and very efficient methods for the microcanonical procedure \cite{werner:2023-epos4-micro}
were developed.
Also in the full scheme, including primary and secondary interactions,
energy-momentum and flavors are conserved. All the technical details
about the new hadronization procedures can be found in \cite{werner:2023-epos4-micro};
the aim of this paper is to present an overview and some important
results. \\

To summarize the above discussion, one has 
\begin{itemize}
\item the ``full'' EPOS4 scheme, composed of 
\begin{itemize}
\item primary interactions, based on an S-matrix approach for parallel scatterings, 
\item secondary interactions, composed of 
\begin{itemize}
\item core-corona separation procedure, 
\item hydrodynamic evolution and microcanonical hadronization, 
\item hadronic afterburner (UrQMD \cite{Bas98,Ble99}). 
\end{itemize}
\end{itemize}
\end{itemize}
As an alternative, in order to better understand the different components,
I also consider 
\begin{itemize}
\item the ``core+corona'' (``co-co'') contribution, i.e. primary interactions
+ secondary interactions but \uline{without} hadronic afterburner; 
\item the ``core'' contribution, i.e. primary interactions + secondary
interactions but \uline{without} hadronic afterburner, only considering
core particles;
\item the ``corona'' contribution, i.e. primary interactions + secondary
interactions but \uline{without} hadronic afterburner, only considering
corona particles. 
\end{itemize}
\noindent One needs to exclude in these cases the hadronic afterburner,
because the latter affects both core and corona particles, so in the
full approach, the core and corona contributions are no longer visible.

\begin{figure}[h]
\centering{}\includegraphics[bb=20bp 50bp 700bp 580bp,clip,scale=0.27]
{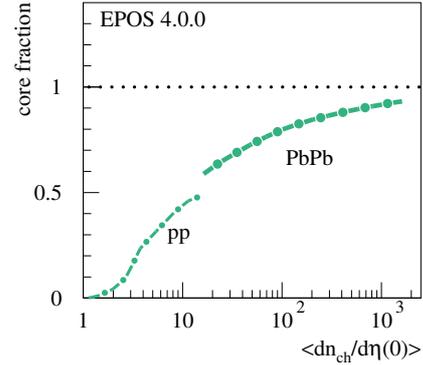}
\caption{The ``core fraction'' for $pp$ (thin line) and PbPb scattering (thick
line). \label{core-fraction}}
\end{figure}

In the following, I will present particle ratios, always relative
to pion yields, as well as mean $p_{t}$ results, for the different
contributions (``core'', ``corona'' etc), in $pp$ and PbPb collisions
at LHC energies. In all cases, the results depend strongly on the
relative weight of core to corona. It is clear that for low multiplicity
$pp$ scattering corona will dominate, whereas, for central PbPb collisions,
the core will dominate. To be more quantitative, I compute the ``core
fraction'', defined as the ratio of core to core+corona for pion
production (with pions being the most frequent particle species).
In Fig. \ref{core-fraction},
I show results for $pp$ (thin lines) and PbPb (thick lines), and one
sees an almost continuous curve, going from zero (for low multiplicity
$pp$) up to unity (for central PbPb). There is no overlap at intermediate
multiplicity because one is running out of statistics for both $pp$
and PbPb collisions. I consider actually $pp$ at 7 TeV and PbPb at
2.76 TeV, since the particle ratios in the following refer to these
energies.

\begin{figure}[h]
\hspace*{-0.4cm}\includegraphics[bb=20bp 50bp 642bp 570bp,clip,scale=0.28]
{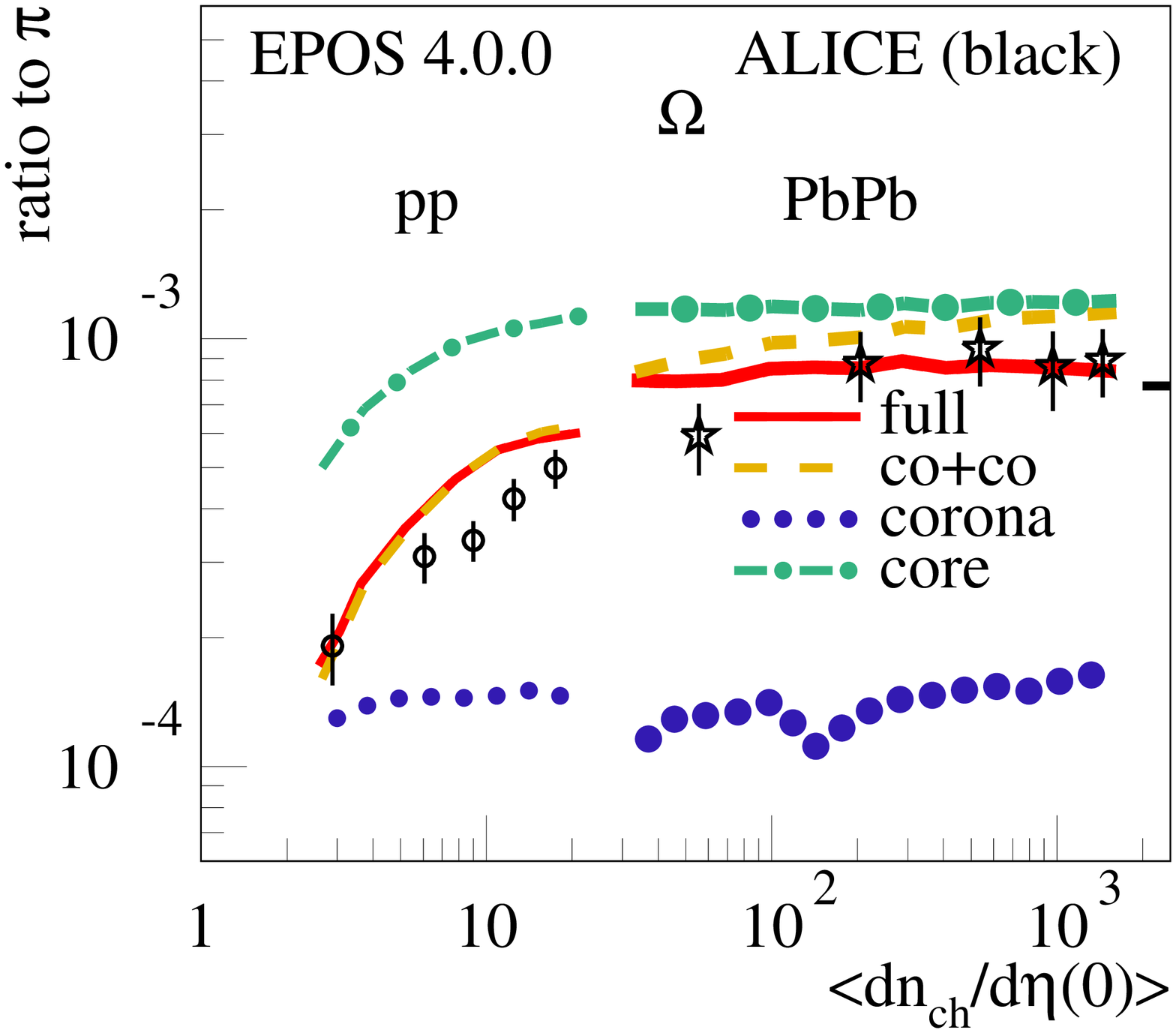}
\hspace*{-0.5cm}\includegraphics[bb=20bp 50bp 642bp 570bp,clip,scale=0.28]
{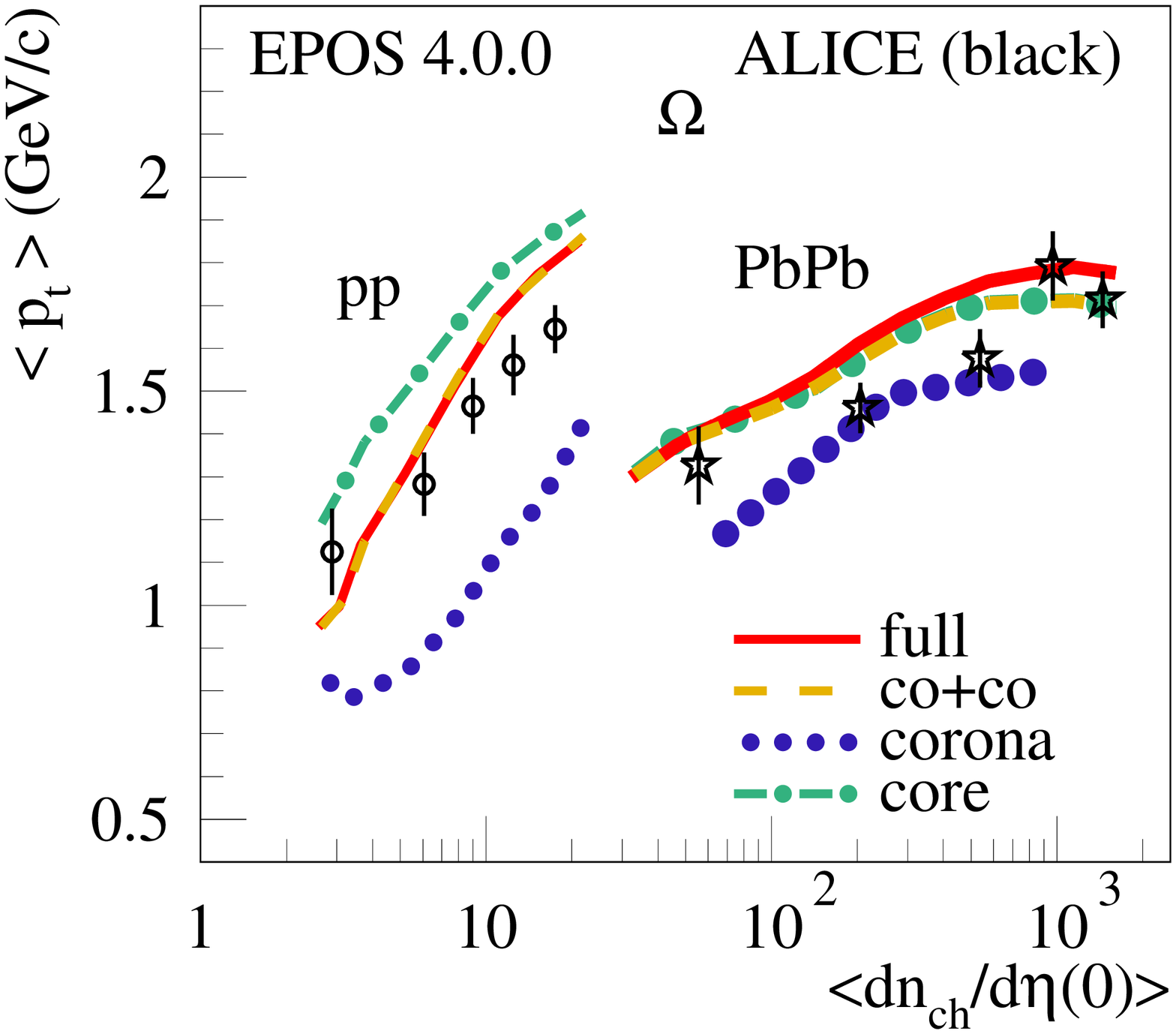}
\caption{Ratio of $\Omega$ over $\pi$ (upper panel) and average transverse
momentum (lower panel) versus multiplicity, for $pp$ at 7 TeV (thin lines)
and PbPb at 2.76 TeV (thick lines), compared to ALICE data. \label{core-corona-omega}}
\end{figure}

In Fig. \ref{core-corona-omega}(upper panel),
I plot the ratio of $\Omega$ baryon yields over $\pi$ yields versus
multiplicity. I show results for $pp$ at 7 TeV (thin lines) and PbPb
at 2.76 TeV (thick lines), compared to ALICE data \cite{ALICE:2016-pp-Ks-Lda-Xi-Oga,ALICE:2013-PbPb-Xi-Oga}.
The different line styles refer to different contributions: the yellow
dashed line refers to ``core+corona'' (``co-co''), i.e. primary
interactions + hydro but \uline{without} hadronic afterburner,
the blue dotted line refers to the ``corona'' and the green dashed-dotted
line refers to the ``core'' part. The red line is the ``full'' contribution,
i.e. core + corona + hadronic afterburner. One sees an almost flat line
for the corona contribution, similar for $pp$ and PbPb, which is understandable,
since ``corona'' means particle production from string fragmentation,
which does not depend on the system. One observes also a flat curve
for the ``core'' part at high multiplicity, which is again expected
since the core hadronization is determined by the freeze-out energy
density, which is system independent. However, when the system gets
very small, one gets a reduction of heavy particle production due to
the microcanonical procedure (with its energy and flavor conservation
constraints), whereas a grand canonical treatment would give a flat
curve down to small multiplicities. It is remarkable that the ``core''
curve is far above the ``corona'' one, which simply reflects the
fact that $\Omega$ production is much more suppressed in string decay,
compared to statistical (``thermal'') production. This explains
why the ``core+corona'' contribution increases by one order of magnitude
from low to high multiplicity, because simply the relative weight
of the core fraction increases from zero to unity. The effect from
hadronic rescattering (difference between ``full'' and ``co-co'')
is relatively small, some suppression due to baryon-antibaryon annihilation
can be seen.

Whereas the $\Omega$ over $\pi$ ratios are essentially smooth curves,
from $pp$ to PbPb, the situation changes completely when looking at
the average transverse momentum $\left\langle p_{t}\right\rangle $
versus multiplicity, as shown in Fig. \ref{core-corona-omega}(lower panel),
where I also show results for $pp$ (thin curves) and PbPb (thick curves),
for the different contributions. Here one sees (for all curves) a significant
discontinuity when going from $pp$ to PbPb. The ``corona'' contributions
are not flat (as the ratios), but they increase with multiplicity,
in the case of $pp$ being even more pronounced than for PbPb. This is a ``saturation
effect'': the saturation scale increases with multiplicity, which
means with increasing multiplicity the events get harder, producing
higher $p_{t}$. The situation is different for PbPb, where an increase
of multiplicity is mainly due to an increase of the number of active
nucleons, with a more modest increase of the saturation scale with
multiplicity. Also, the ``core'' curves increase strongly with multiplicity,
and here as well more pronounced in the case of $pp$, due to the fact
that one gets for high-multiplicity $pp$ high energy densities within
a small volume, leading to strong radial flow.
\begin{figure}[h]
\hspace*{-0.4cm}\includegraphics[bb=20bp 50bp 642bp 570bp,clip,scale=0.28]
{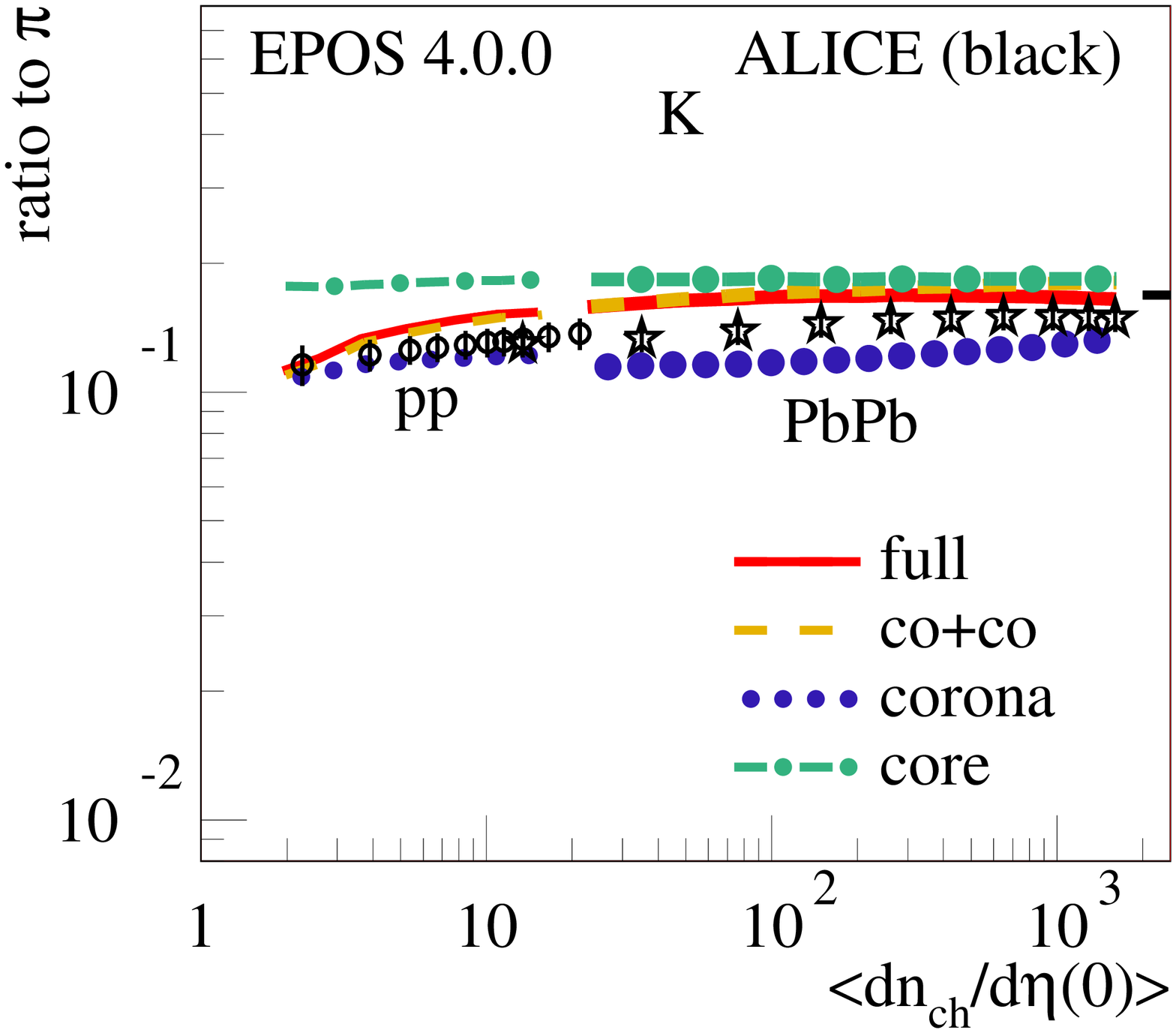}
\hspace*{-0.5cm}\includegraphics[bb=20bp 50bp 642bp 570bp,clip,scale=0.28]
{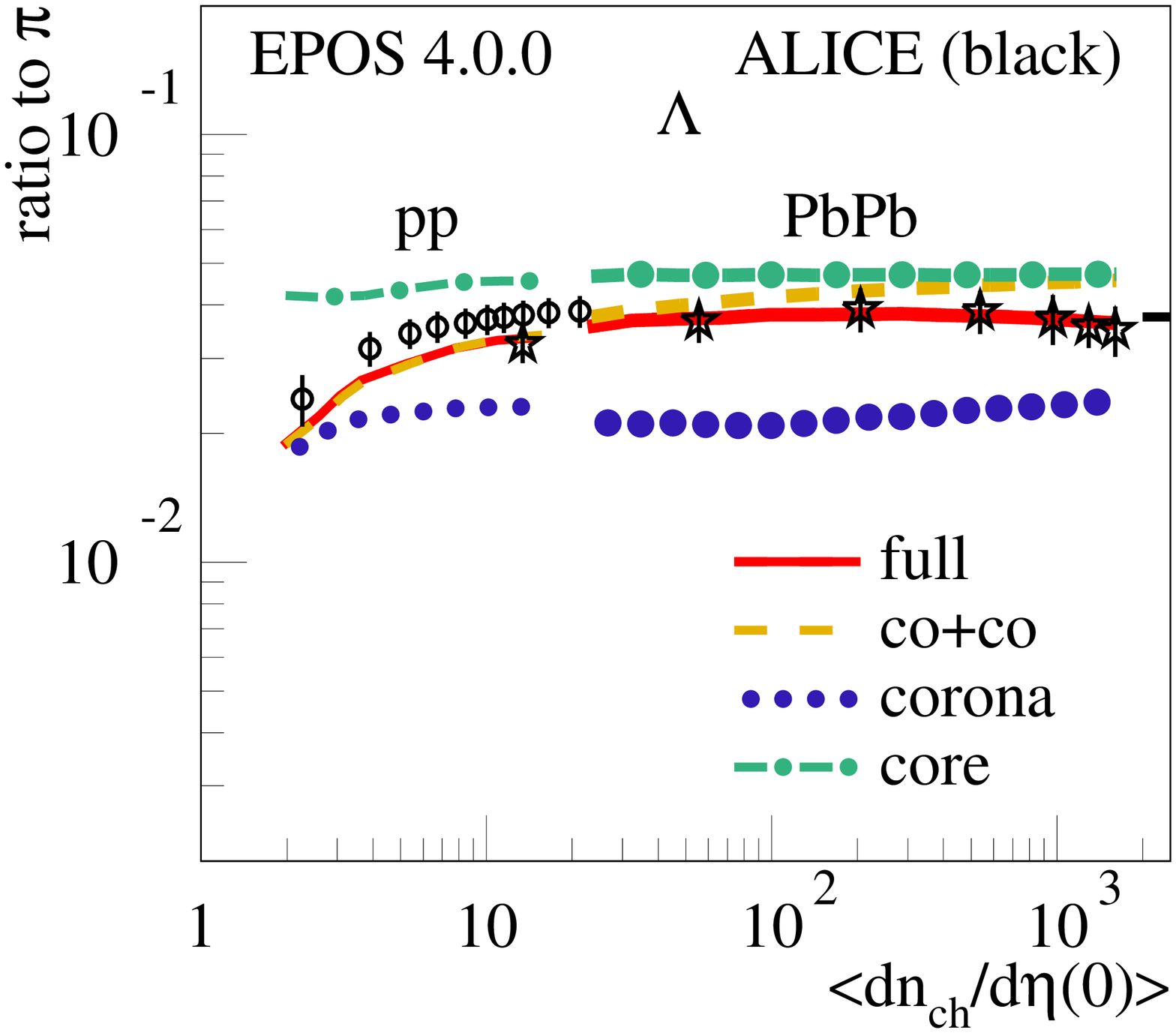}\\
 \hspace*{-0.4cm}\includegraphics[bb=20bp 50bp 642bp 570bp,clip,scale=0.28]
{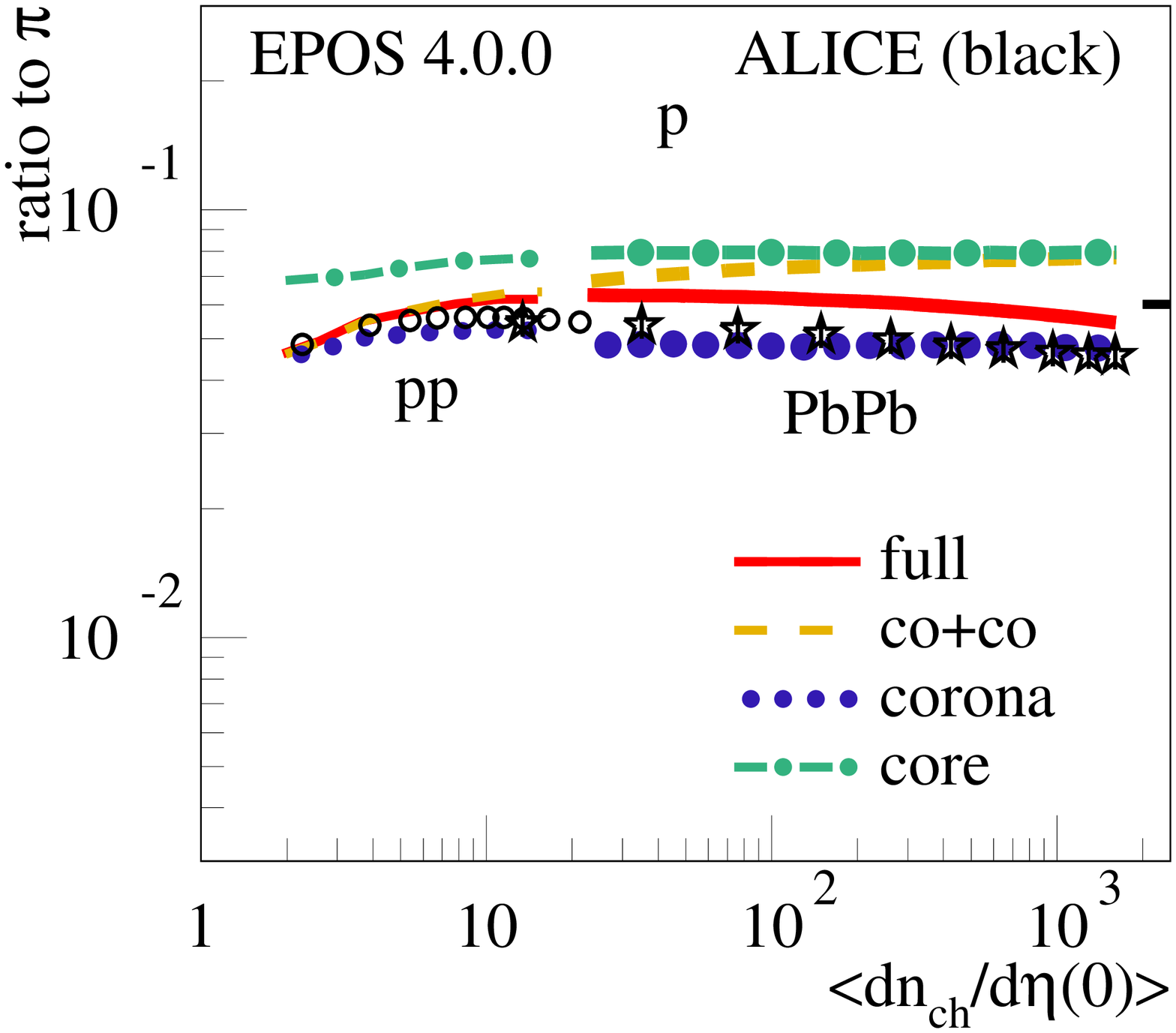}
\hspace*{-0.5cm}\includegraphics[bb=20bp 50bp 642bp 570bp,clip,scale=0.28]
{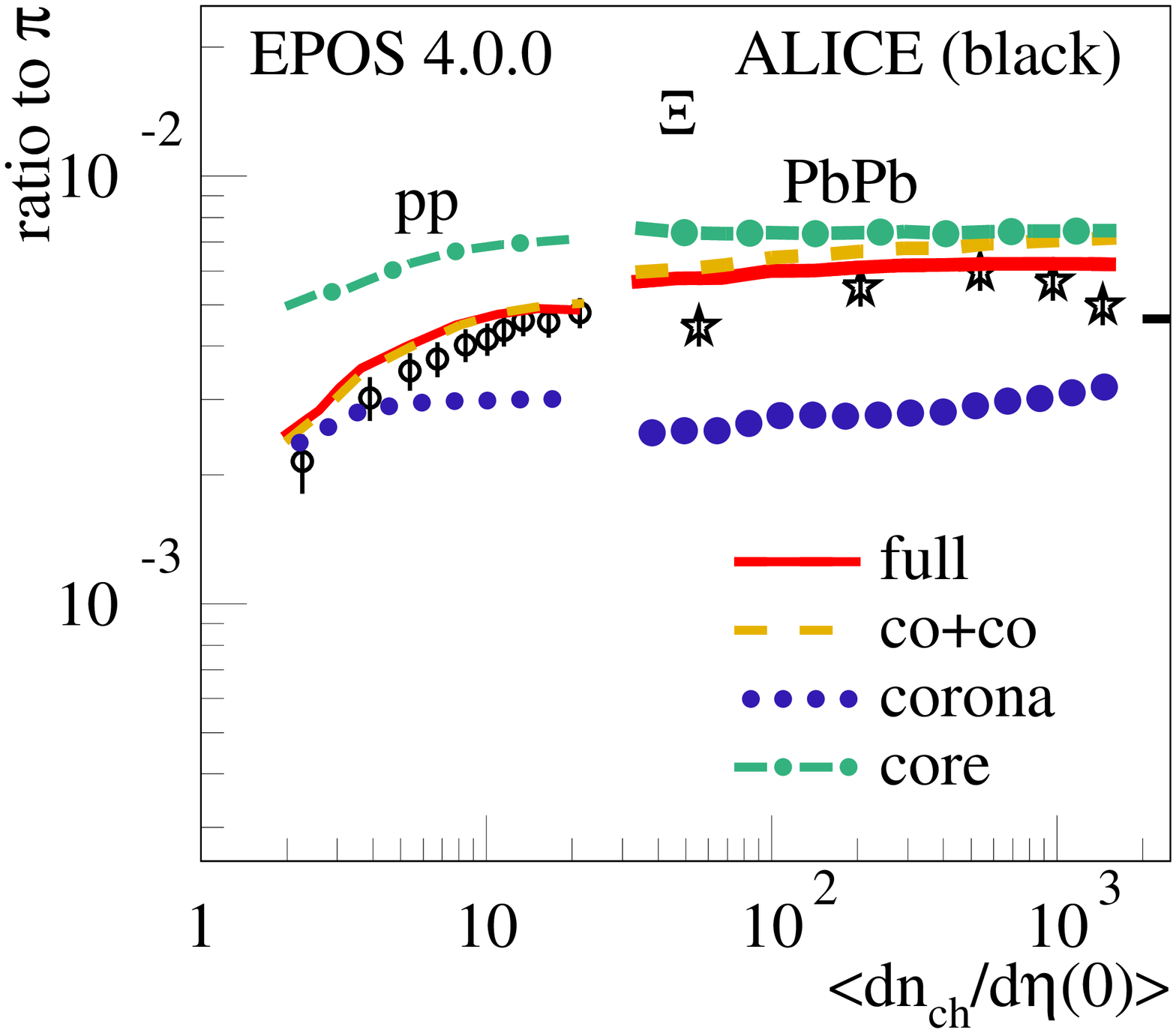}
\caption{As Fig. \ref{core-corona-omega}(upper panel), but for $K$, $p$, $\Lambda$,
$\Xi$. \label{core-corona-ratios}}
\end{figure}
\begin{figure}[h]
\hspace*{-0.4cm}\includegraphics[bb=20bp 50bp 642bp 570bp,clip,scale=0.28]
{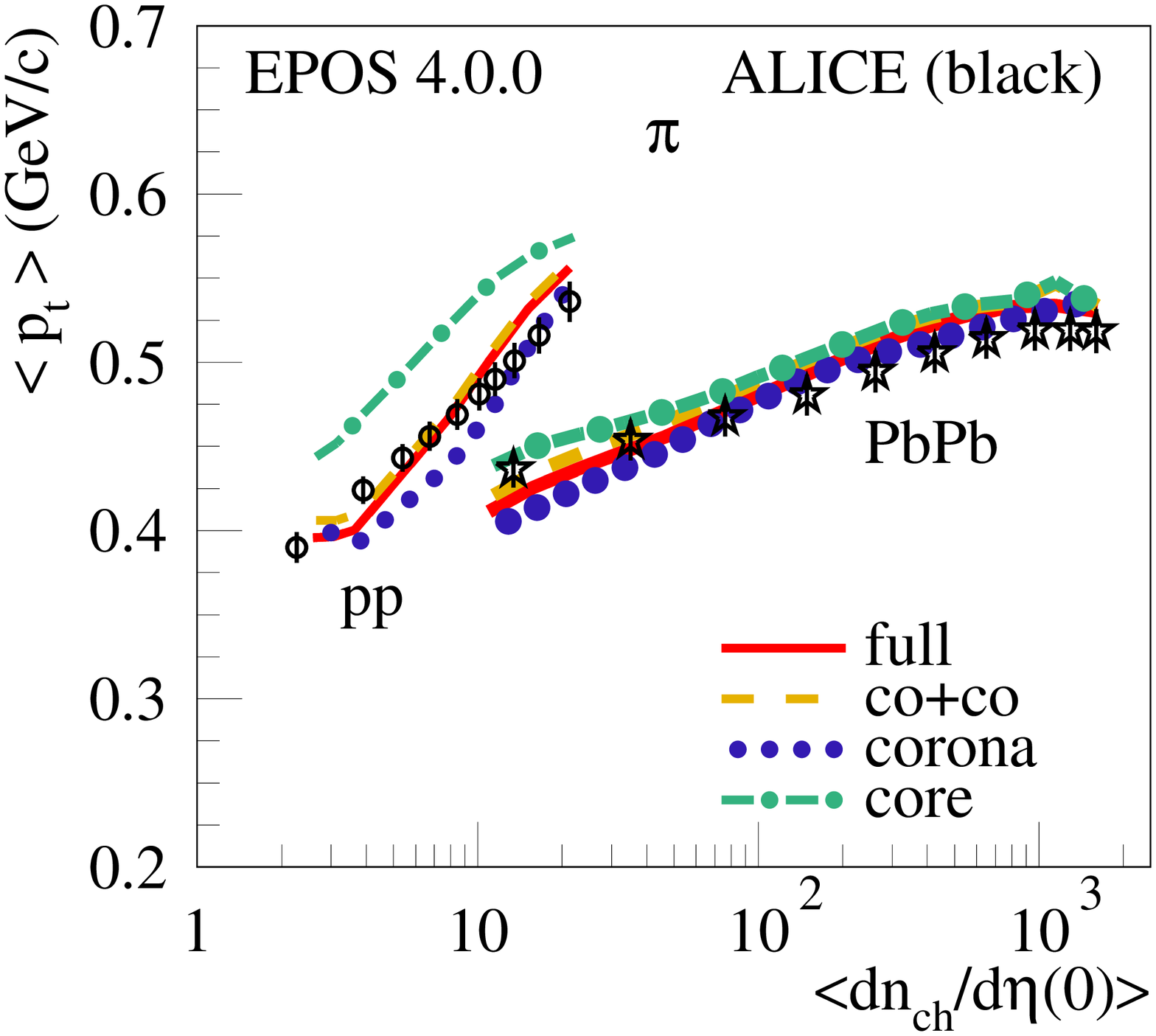}
\hspace*{-0.5cm}\includegraphics[bb=20bp 50bp 642bp 570bp,clip,scale=0.28]
{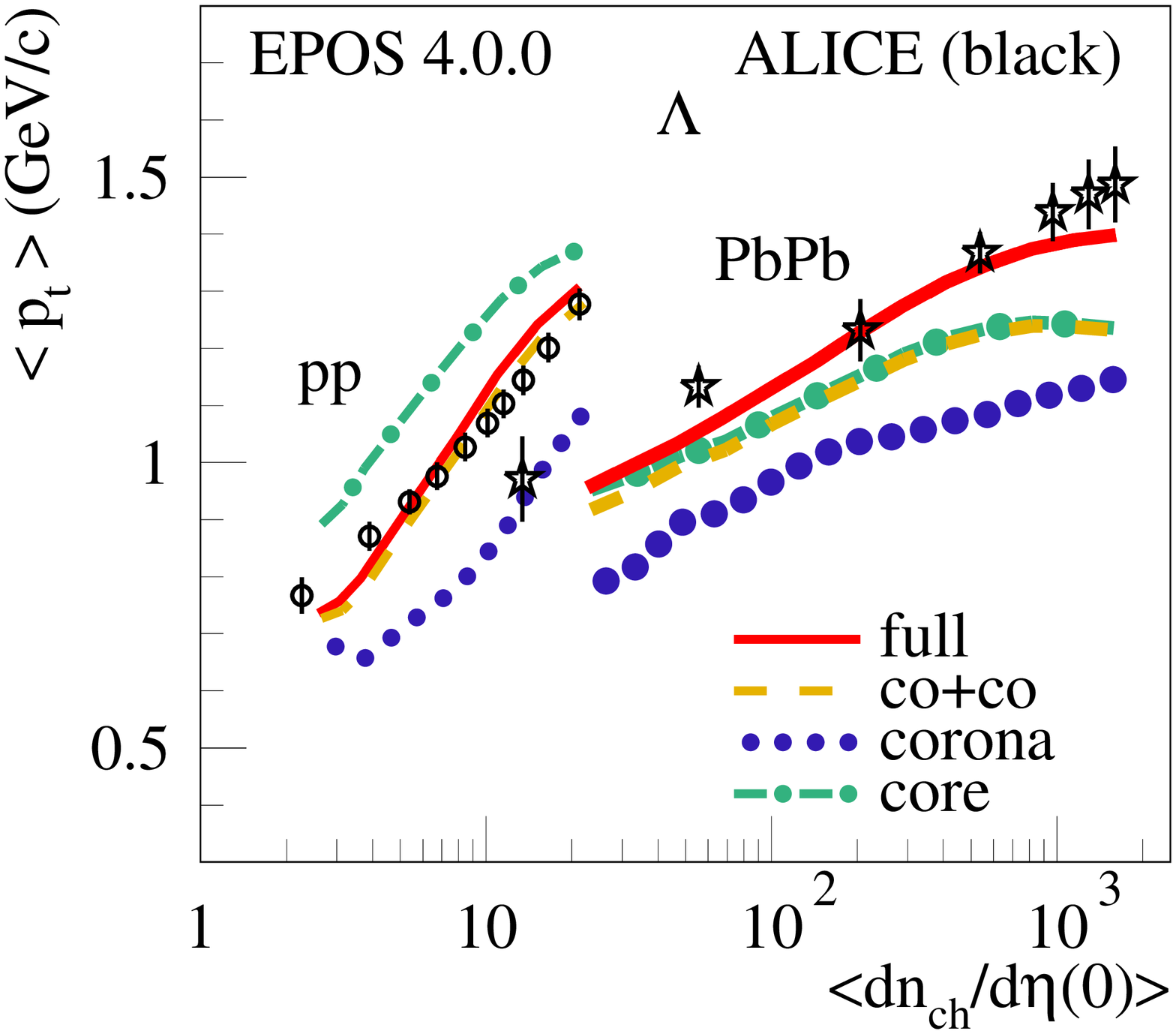}\\
 \hspace*{-0.4cm}\includegraphics[bb=20bp 50bp 642bp 570bp,clip,scale=0.28]
{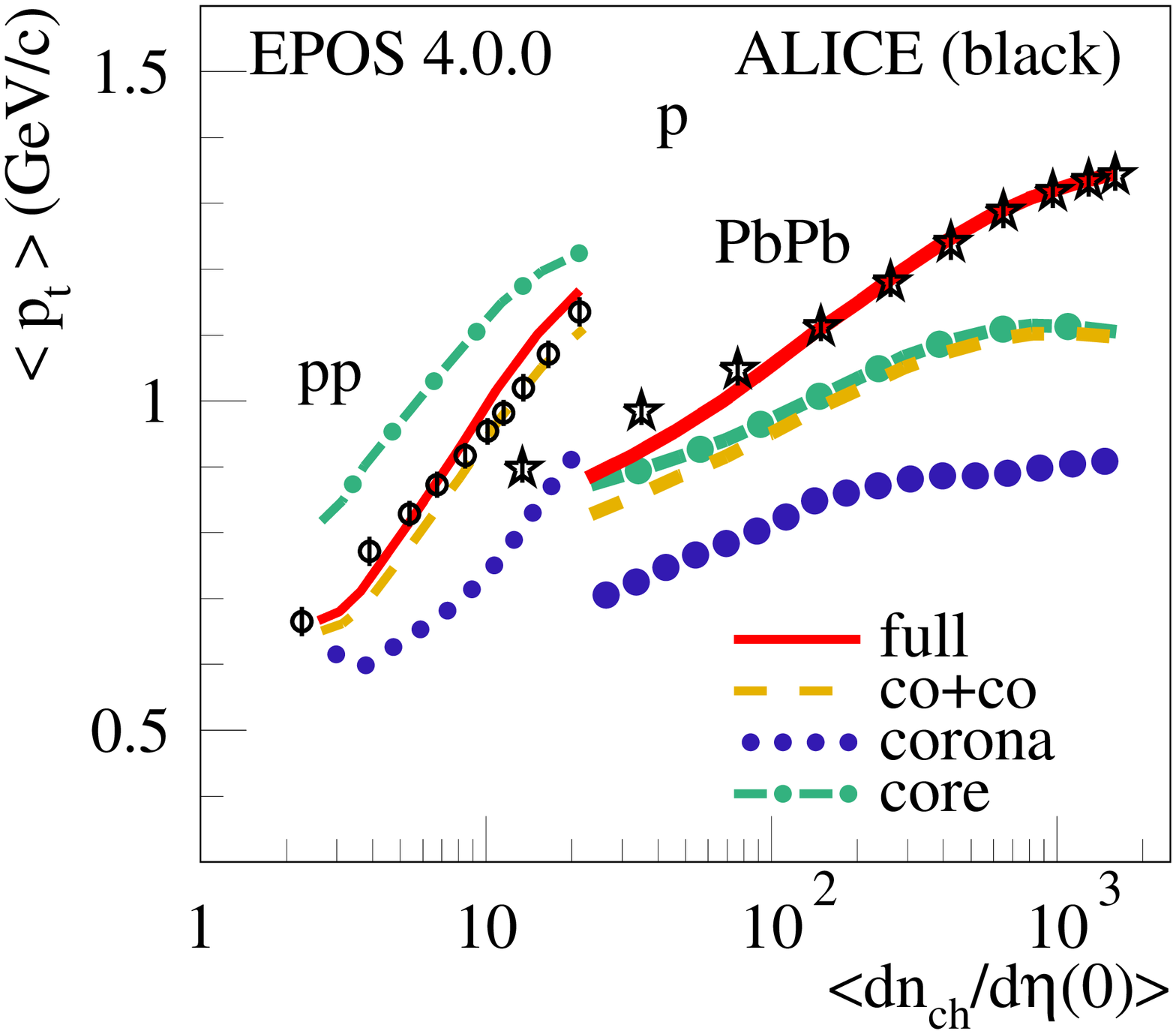}
\hspace*{-0.5cm}\includegraphics[bb=20bp 50bp 642bp 570bp,clip,scale=0.28]
{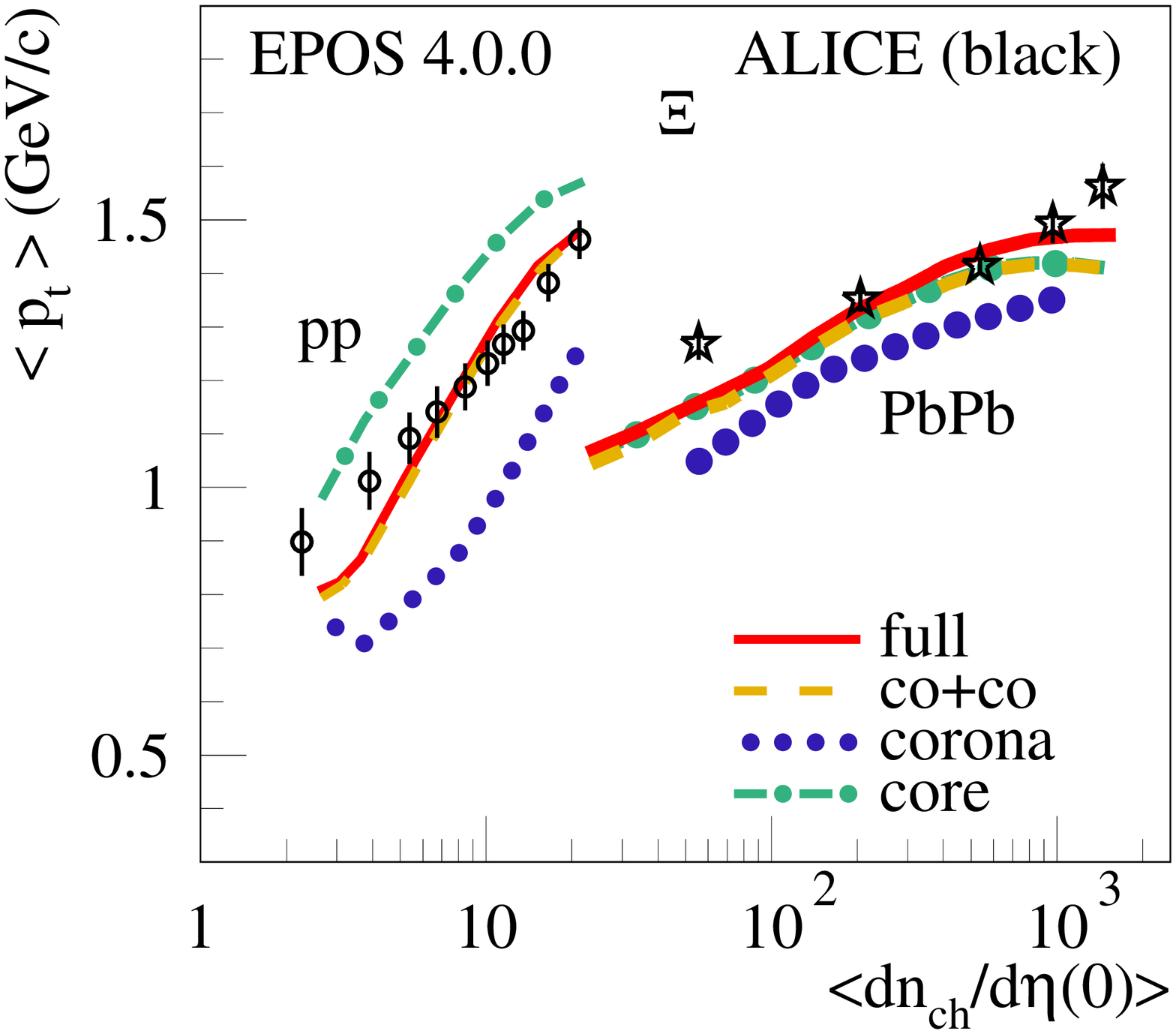}
\caption{As Fig. \ref{core-corona-omega}(lower panel), but for $K$, $p$, $\Lambda$,
$\Xi$. \label{core-corona-meanpt}}
\end{figure}
Again, the core+corona contribution is understood based on the continuous
increase of the core fraction from low to high multiplicity.

In Figs. \ref{core-corona-ratios} and \ref{core-corona-meanpt},
I show the multiplicity dependencies of ratios and mean $p_{t}$
for different hadrons, which are qualitatively similar to the $\Omega$
results, just the difference between the corona and the core curves
are smaller. The data are from ALICE \cite{ALICE:2013-PbPb-k-pi-p,ALICE:2013-PbPb-Ks-Lda,ALICE:2013-PbPb-Xi-Oga,ALICE:2015-pp-pi-K-p,ALICE:2016-pp-Ks-Lda-Xi-Oga}.

It is very useful (and necessary) to consider at the same time the
multiplicity dependence of particle ratios and of mean $p_{t}$ results,
since their behavior is completely different (the former is continuous,
the latter jumps). Despite these even qualitative differences between
the two observables, the physics issues behind these results is the
same, namely saturation, core-corona effects which mix flow (being
very strong) and non-flow, and microcanonical hadronization of the
core.

\begin{figure}[h]
\centering{}\includegraphics[bb=20bp 30bp 600bp 750bp,clip,scale=0.40]
{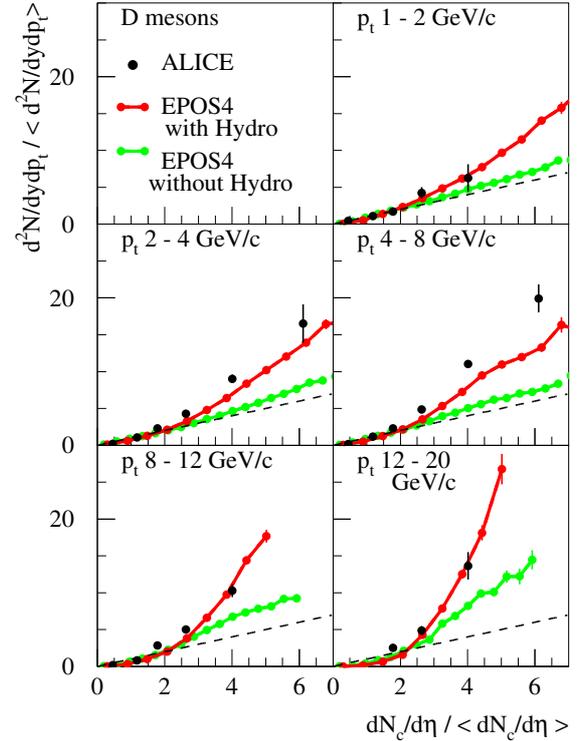}
\caption{Normalized $D$ meson multiplicity as a function of the normalized
charged particle multiplicity for different $p_{t}$ ranges in $pp$
scattering at 7 TeV. I show EPOS results with and without hydro,
compared to ALICE data. \label{charm-versus-charged}}
\end{figure}

Another very important and useful variable is the multiplicity dependence
of $D$ meson production, where D stands for the sum of $D^{0}$,
$D^{+}$, and $D^{*+}$. This is much more than just ``another particle'',
since the $D$ meson contains a charm quark, the latter one being
created exclusively in the parton ladder and not during fragmentation
or in the plasma. In Fig. \ref{charm-versus-charged},
I plot the normalized $D$ meson multiplicity $\frac{d^{2}N}{dydp_{t}}/<\!\!\frac{d^{2}N}{dydp_{t}}\!\!>$
as a function of the normalized charged particle multiplicity $\frac{d^{2}N_{c}}{dydp_{t}}/<\!\!\frac{d^{2}N_{c}}{dydp_{t}}\!\!>$
for different $p_{t}$ ranges in $pp$ scattering at 7 TeV, compared
to ALICE data \cite{ALICE:2015-pp-mult-charm}. It is interesting
to see in which way the simulations and the data deviate from the
reference curve, which is the dashed black line representing identical
multiplicity dependence for D mesons and charged particles. Considering
the EPOS results without hydro (green lines), for low $p_{t}$ (1-2GeV/c)
the curve is slightly above the reference, but with increasing $p_{t}$
the green curves get steeper, which is due to the fact that with increasing
multiplicity the saturation scale increases, and the events get harder,
producing more easily both high $p_{t}$ and charmed particles. Considering
EPOS with hydro (red curves), the increase compared to the green curves
is much stronger, which is due to the fact that ``turning on hydro''
will reduce the multiplicity (the available energy is partly transformed
into flow). The red curves are close to the experimental data, both
showing a much stronger increase compared to the reference curve,
with the effect getting bigger with increasing $p_{t}$. So one may
conclude this paragraph: to get these final results (the strong increase),
two phenomena are crucial, namely, saturation which makes high multiplicity
events harder, and the ``hydro effect'' which reduces multiplicity
and ``compresses'' the multiplicity axis.

Concerning earlier EPOS versions, there are no ``real publications''
concerning these multiplicity dependencies, only plots based on on
``preliminary versions'' shown at conferences or given to experimental
colleagues. But none of the preliminary versions were able to fit reasonably
well at the same time all the data shown in this section.

%%############################################################################
%%############################################################################

\section{Charmed hadrons \label{=======charmed-hadrons=======}}

%%############################################################################
%%############################################################################

Having already discussed the multiplicity dependence of charm production
in the last section, I will show here some basic charm results 
(a detailed discussion about charm production can be found 
in \cite{werner:2023-epos4-heavy}). 
I consider
here just primary interactions, no hydro and no hadronic cascade,
so the charm quarks originate from cut Pomerons, more precisely from
the parton ladder. Cut parton ladders correspond in general to two
chains of partons $q-g-...-g-\bar{q}$ identified as kinky strings,
with $q$ referring to light flavor quarks, and $g$ to gluons. The
Born process or branchings in the space-like or the time-like cascade
may lead to $Q\bar{Q}$ production, where $Q$ refers to ``heavy
flavor\textquotedblright{} (HF) quarks, i.e. charm or bottom. In this
case, one ends up with parton chains of the type $q-g-...-g-\bar{Q}$
and $Q-g-...-g-q$,
\begin{figure}[h]
\centering{}\includegraphics[bb=20bp 40bp 590bp 600bp,clip,scale=0.37]
{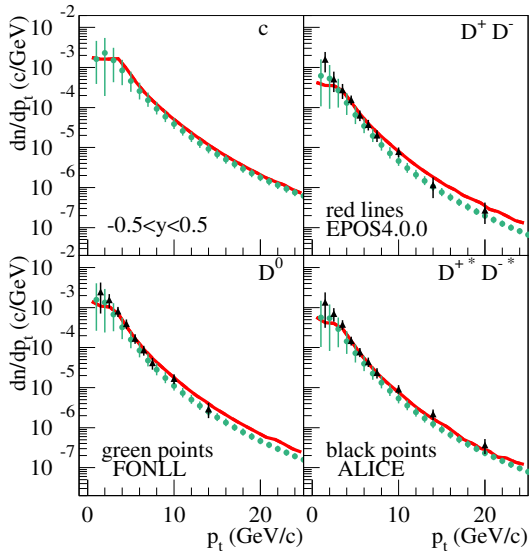}
\caption{Transverse momentum spectra of $c$ quarks and charmed mesons in $pp$
at 7 TeV. \label{dmeson-pt-spectra}}
\end{figure}
which will decay (among others) into HF hadrons. In Fig. \ref{dmeson-pt-spectra},
I show transverse momentum spectra of $c$ quarks (upper left), $D^{+}$,
$D^{-}$ mesons (upper right), $D^{0}$ mesons (lower left), and $D^{+*}$,
$D^{-*}$ mesons (lower right) in $pp$ collisions at 7 TeV. The red
lines refer to EPOS simulations, the green points to FONLL calculations
\cite{Cacciari:2012ny}, and the black points to ALICE data \cite{ALICE:2012-pp-charm-spectra}.
In Fig. \ref{charmed-baryons-pt},
\begin{figure}[h]
\centering{}\includegraphics[bb=40bp 60bp 590bp 560bp,clip,scale=0.37]
{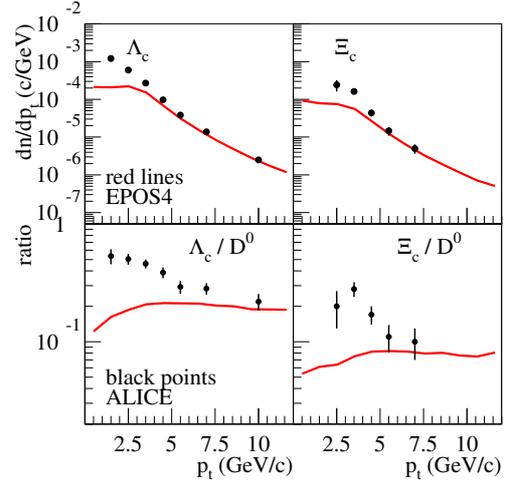}
\caption{Transverse momentum spectra of charmed baryons in $pp$ at 5.02 TeV..
\label{charmed-baryons-pt}}
\end{figure}
I plot transverse momentum spectra of $\Lambda_{c}$ and $\Xi_{c}$
baryons (upper panel) and their ratio with respect to $D^{0}$ mesons
in $pp$ collisions at 5 TeV. The red lines refer to EPOS simulations,
and the black points to ALICE data \cite{ALICE:2020-pp5-Lda_c,ALICE:2021-pp5-Xi_c}.
The production of charmed baryons is in principle straightforward,
they are also coming from $q-g-...-g-\bar{Q}$ and $Q-g-...-g-q$
strings (with $Q$ being a $c$ quark in this case). The only difference
compared to charmed meson production is the fact that here a diquark-antidiquark
breakup occurs, which results in an essentially flat baryon / meson
ratio, whereas the data show an increase towards small $p_{t}$. A
similar ``baryon/meson enhancement'' in the region around 2 - 6
GeV/c has already been observed in the light flavor sector, where
one possible explanation is collective flow.

\begin{figure}[h]
\centering{}\includegraphics[bb=20bp 20bp 590bp 680bp,clip,scale=0.4]
{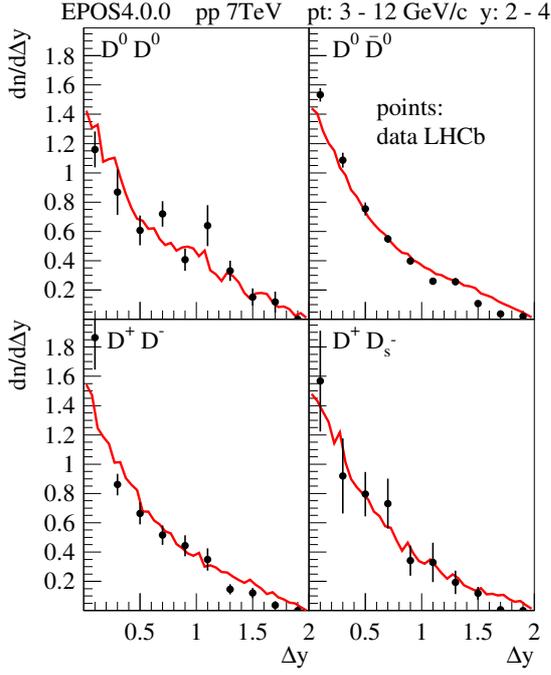}
\caption{Two hadron correlations for $D^{0}D^{0}$, $D^{0}\bar{D}^{0}$, $D^{+}D^{-}$,
and $D^{+}D_{s}^{-}$ as a function of the rapidity difference $\Delta y$
in $pp$ collisions at 7 TeV. Red lines represent EPOS4 simulations and
black dots data from LHCb. \label{correlation-rapidity}}
\end{figure}

\begin{figure}[h]
\centering{}\includegraphics[bb=20bp 20bp 590bp 680bp,clip,scale=0.4]
{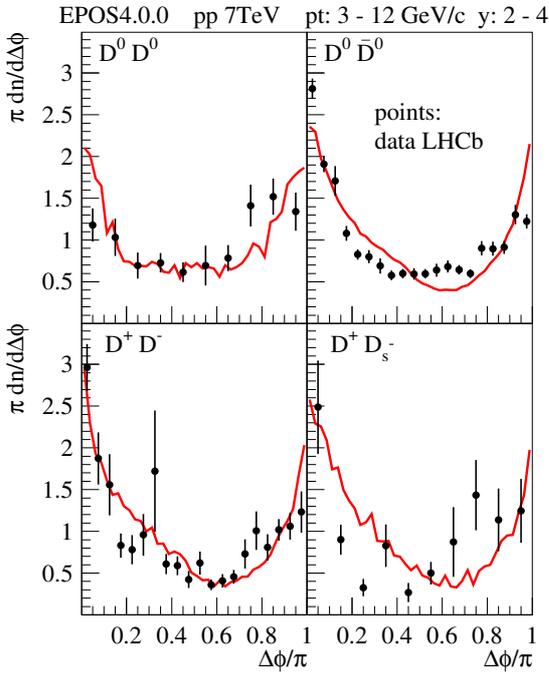}
\caption{Two hadron correlations for $D^{0}D^{0}$, $D^{0}\bar{D}^{0}$, $D^{+}D^{-}$,
and $D^{+}D_{s}^{-}$ as a function of the azimuthal angle difference
$\Delta\phi$ in $pp$ collisions at 7 TeV. Red lines represent EPOS4
simulations and black dots data from LHCb. \label{correlation-pt}}
\end{figure}

Since charm is produced (as everything else) in an event-by-event
manner, one produces for each charm quark the corresponding charm antiquark,
and depending on the production details they have characteristic correlations,
which are also visible in $D$ meson pair correlations. In Fig. \ref{correlation-rapidity},
I show two-hadron correlations for $D^{0}D^{0}$ (upper left plot),
$D^{0}\bar{D}^{0}$ (upper right), $D^{+}D^{-}$ (lower left), and
$D^{+}D_{s}^{-}$ (lower right) as a function of the rapidity difference
$\Delta y$ in $pp$ collisions at 7 TeV, with $p_{t}$ values between
3 and 12 GeV/c and rapidities between 2 and 4. Red lines represent
EPOS4 simulations and black dots data from LHCb \cite{LHCb:2012-double-charm}.
In Fig. \ref{correlation-pt},
I plot the correlations of these pairs as a function of the azimuthal
angle difference $\Delta\phi$, again compared to LHCb.

It should be noted that $D^{0}D^{0}$ represents a $c-c$ correlations,
whereas the three other combinations $D^{0}\bar{D}^{0}$, $D^{+}D^{-}$,
and $D^{+}D_{s}^{-}$ represent $c-\bar{c}$ correlations. For the
latter ones, the situation is quite simple: the $c$ and the $\bar{c}$
are always produced as pair from the same process, and therefore one
expects them to be close in rapidity, with a preference of $\Delta\phi=0$
(in case of a time-like $g\to c\bar{c}$), or at $\Delta\phi=\pi$
in case of a Born process $gg\to c\bar{c}$. This is precisely what
is seen: The rapidity correlations have maxima at $\Delta y=0$ and
then drop quickly, the $\Delta\phi$ correlations have maxima et $\Delta\phi=0$
and $\Delta\phi=\pi$, observed in both EPOS4 simulations and data.

Surprisingly, the $D^{0}D^{0}$ correlations (corresponding to a $c-c$
pair) look very similar, which suggests that also $c-c$ pairs originate
from the same process, like a timelike $g\to gg\to c\bar{c}\,c\bar{c}$
or a Born process $gg\to gg$ followed by $g\to c\bar{c}$, $g\to c\bar{c}$.

\begin{figure}[h]
\centering{}\includegraphics[bb=20bp 30bp 590bp 810bp,clip,scale=0.35]
{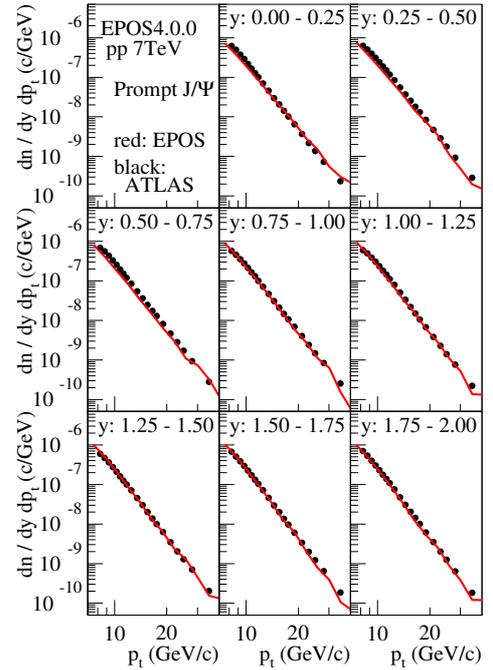}
\caption{Transverse momentum spectra of prompt $J/\Psi$ in $pp$ at 7 TeV. \label{prompt-jpsi-pt}}
\end{figure}

Since EPOS4 creates charm always in terms of $c-\bar{c}$ pairs, it
is quite tempting to look into the possibility to produce charmonium.
It is easy to implement the idea of the color evaporation model \cite{CEM:1980,CEM:1994,CEM:2016},
where charmonium is created with a certain probability in the case of
a $c-\bar{c}$ pairs being in the appropriate mass range. So one considers
all $c-\bar{c}$ pairs from the same Pomeron (fully evaluated, including
time-like emissions), and compute the invariant mass $M_{c\bar{c}}$.
Whenever this mass is less than the sum of two $D$ meson masses and
bigger than the $J/\Psi$ mass, the $c-\bar{c}$ pair is with a certain
probability $w_{J/\Psi}$ considered to be a $J/\Psi$. In Fig. \ref{prompt-jpsi-pt},
I plot prompt $J/\Psi$ (not coming from beauty decays) from EPOS4
simulations compared to ATLAS data \cite{ATLAS:2015-jpsi}.

\section{Summary \label{=======summary=======}}

I reported on new ideas, implemented in EPOS4, which provide a new
understanding of a deep connection between four basic concepts in
$pp$ and $AA$ collisions: rigorous parallel scattering, energy conservation,
factorization, and saturation. It is mandatory to treat multiple scatterings
in parallel, and a ``natural'' framework is S-matrix theory, with
an S-matrix being given as a product of several entities representing
individual scatterings referred to as Pomerons, and with energy-momentum
conservation being implemented in an unbiased fashion via $\delta$
functions without imposing any ordering of collisions (this is what
is meant by rigorous parallel scattering). The fundamental quantity
of the multiple scattering approach is the cut single Pomerons expression
$G$, representing inelastic parton-parton scattering. The fundamental
question discussed in this paper is how to relate $G$ to $G_{\mathrm{QCD}}(Q_{0}^{2})$,
where the latter refers to parton-parton scattering in the framework
of pQCD, having as basic elements parton evolutions with constant
virtuality cutoff $Q_{0}^{2}$ and a hard $2\to2$ elementary QCD
scattering. I refer to $G=G_{QCD}(Q_{0}^{2})$ as ``naive choice''. 

One recalls that factorization and binary scaling, often mentioned
in this paper, amount to reducing the inclusive cross sections for
$pp$ and $AA$ scattering to single Pomeron results, although the underlying
physical processes involve multiple parallel scatterings.

I showed in Sec. \ref{=======factorization-without-energy-conservation=======}
that neglecting energy conservation leads perfectly to factorization
and binary scaling. But in the Monte
Carlo procedures eventually one needs the implementation of energy
conservation, so one introduces inconsistencies, in the sense that
the theoretical basis and the Monte Carlo realization are not compatible.

On the other hand, as shown in Secs. \ref{=======deformation-function=======}
and \ref{=======spoils factorization=======},
considering energy conservation (or energy sharing) and using the
``naive choice'' $G=G_{QCD}(Q_{0}^{2})$, one completely spoils
factorization for hard processes, contradicted by data. I have shown
that the problem is due to a ``deformation'' of the inclusive energy
distribution of Pomerons connected to many other Pomerons, compared
to isolated Pomerons: the probability of carrying a large fraction of
the total energy is reduced, which is unavoidable. These deformations
can be quantified in terms of deformation functions $R_{\mathrm{deform}}$
depending on the number $N_{\mathrm{conn}}$ and the squared energy
fraction $x_{\mathrm{PE}}$.

In Sec. \ref{=======saturation-recover-factorization=======},
one takes note of two problems: (1) spoiling factorization when using
the naive choice $G=G_{QCD}$ in case of respecting energy conservation,
and (2) not considering saturation effects which are known to be important.
The solution of these two problems has been shown to be a dynamical
saturation scale $Q_{\mathrm{sat}}^{2}$, defined via $G=k\times G_{\mathrm{QCD}}(Q_{\mathrm{sat}}^{2})$
with $k$ being inversely proportional to the deformation function,
with a $G$ which must be independent of the connection number $N_{\mathrm{conn}}$.
In that case, even having multiple scattering, all inclusive $pp$ and
$AA$ cross sections are reduced to a single Pomeron result, but only for hard
processes as it should be. This is referred to as ``generalized AGK
cancellations'', which holds at large $p_{t}$, even in a scenario respecting
energy conservation. The dynamical saturation scale works, because
even a large number of parallel scatterings will not affect high $p_{t}$
particle production, it will only make the saturation scale big and
thus suppress small $p_{t}$ particle production.

Since in the new formalism, the full multiple scattering scenario
converges to the single Pomeron result in case of inclusive cross
sections (generalized AGK cancellations), one may use the single Pomeron
(or factorization) mode, based on EPOS parton distribution functions.
So one can now, with the same formalism, treat extremely high $p_{t}$
particle production in factorization mode, and as well collective
effects in high multiplicity events using the full simulation.

I discussed several examples, essentially multiplicity dependencies
(of particle ratios, mean $p_{t}$, charm production) which are very
strongly affected by the saturation issues discussed in this paper
and core-corona effects mixing flow (being very strong) and non-flow
contributions.

\subsection*{Acknowledgements}

I thank Tanguy Pierog for many contributions during the past decade
and in particular the important proposal (and first implementation)
in 2015 of using a Pomeron definition as $G\propto G_{QCD}(Q_{\mathrm{sat}}^{2})$
with a parametrized $G$. He also did a first attempt in 2020 to
take into account the ``deformation'', by using a correction factor
being a ratio of a ``theoretical'' $G_{\mathrm{theo}}$ and the
same function $G_{\mathrm{theo}}$ but with rescaled longitudinal
momentum arguments.

\end{document}